\documentclass[amssymb,amsmath,twocolumn,superscriptaddress,nofootinbib,floatfix]{revtex4-2}

\usepackage[deletedmarkup=xout,commentmarkup=uwave,commandnameprefix=ifneeded]{changes}
\definechangesauthor[name=Max, color=teal]{MI}

\usepackage{graphicx}
\usepackage{bm}
\usepackage{amssymb,amsmath}
\usepackage{mathrsfs}
\usepackage{latexsym}
\usepackage{color}
\usepackage{dcolumn}

\usepackage[colorlinks=true,citecolor=blue,urlcolor=blue]{hyperref}
\usepackage{multirow}
\usepackage{soul}

\usepackage{orcidlink}

\graphicspath{{./}{figures/}}

\allowdisplaybreaks

\newcommand{\CIT}{\affiliation{TAPIR, California Institute of Technology, Pasadena, CA 91125, USA}}
\newcommand{\CITLab}{\affiliation{LIGO Laboratory, California Institute of Technology, Pasadena, California 91125, USA}}
\newcommand{\CCA}{\affiliation{Center for Computational Astrophysics, Flatiron Institute, New York, NY 10010, USA}}

\newcommand{\UMassD}{\affiliation{Department of Mathematics,
    Center for Scientific Computing and Data Science Research,
    University of Massachusetts, Dartmouth, MA 02747, USA}}

\definecolor{kcmagenta}{rgb}{0.54, 0.17, 0.88}
\definecolor{shyellow}{rgb}{0.15625, 0.609375, 0.316406}
\definecolor{chorange}{rgb}{0.851, 0.372, 0.007}
\definecolor{tlteal}{rgb}{0,.55,.55}
\definecolor{jcpink}{rgb}{1.0, 0.0, 0.5}
\definecolor{mmgreen}{rgb}{0.0, 0.8, 0.6}
\definecolor{bbsalmon}{rgb}{1.0, 0.47, 0.42}
\definecolor{smgreen}{rgb}{0.26, 0.625, 0.277}

\definecolor{recomment}{rgb}{0.85, 0.1, 0.1}

\newcommand{\chieff}{\chi_{\textrm{eff}}}
\newcommand{\chip}{\chi_\mathrm{p}}

\newcommand{\Msun}{M_\odot}

\newcommand{\maxL}{\mathrm{max.}\,\mathcal{L}}

\newcommand{\Hz}{\mathrm{Hz}}
\newcommand{\tcut}{t_\mathrm{cut}}

\begin{document}

\title{Measuring spin precession from massive black hole binaries with gravitational waves: Insights from time-domain signal morphology}

\author{Simona J.~Miller~\orcidlink{0000-0001-5670-7046}} \CIT \CITLab 
\author{Maximiliano Isi~\orcidlink{0000-0001-8830-8672}} \CCA
\author{Katerina Chatziioannou~\orcidlink{0000-0002-5833-413X}} \CIT \CITLab 
\author{Vijay Varma~\orcidlink{0000-0002-9994-1761}}\UMassD
\author{Sophie Hourihane~\orcidlink{0000-0002-9152-0719}} \CIT \CITLab 
\date{\today}

\begin{abstract}
Robustly measuring binary black hole spins via gravitational waves is key to understanding these systems' astrophysical origins, but remains challenging---especially for high-mass systems, whose signals are short and dominated by the merger.
Nonetheless, events like GW190521 show that strong spin precession can indeed be gleaned from high-mass systems.
In this work, we track how spin precession imprints on simulated high-mass binary black hole signals cycle-by-cycle using time-domain inference.
We investigate a suite of signals, all with the same spins and (near-unity) mass ratio---yielding identical spin evolution---but different signal-to-noise ratios, total masses, and extrinsic angles, all of which affect the observed waveform morphology.
We truncate each signal at various times and infer source parameters using only the data before or after each cutoff.
The resultant posterior allows us to identify which time segments of each signal inform its spin precession constraints.
We find that at a sufficiently high post-peak signal-to-noise ratio (SNR, $\rho\sim 20$), spin precession can be constrained by the {\sc NRSur7dq4} waveform model when just the post-peak data (i.e., ringdown) are visible.
Similarly, at a large enough pre-cutoff SNR ($\rho\sim 10$), spin precession can be constrained using only pre-peak data (i.e., inspiral); this occurs for signals with detector-frame total mass $\lesssim 100 \Msun$ at GW190521's full-signal SNR.
Finally, we vary the inclination, polarization, and phase angles, finding that their configuration need not be fine-tuned to measure spin precession, even for very high-mass and short signals with two to three observable cycles. 
We do not find that the same morphological features consistently drive precession constraints: in some signals, precession inference hinges on the relationship between a loud merger and quiet pre-merger cycle, as was the case for GW190521, but this is not generically true. 
Our studies enable a morphological understanding of precession that can help bolster confidence in precession inference in the presence of signal or noise systematics.
\end{abstract}

\maketitle

\section{Introduction}\label{sec:intro}

Black hole (BH) spin is a key astrophysical observable, promised to illuminate physics on multiple scales, from understanding of stellar interiors~\cite{Fuller:2019sxi} to the formation and evolution of binary black holes (BBHs) throughout cosmic history \cite{Mandel:2018hfr}. 
Measurable precession of the BBH orbital plane caused by misalignment between the spin and orbital angular momenta is particularly useful for distinguishing between different BBH evolutionary pathways \cite{Mandel:2018hfr,Mapelli:2018uds,Rodriguez:2016vmx,Gerosa:2017kvu,Farr:2017uvj,Zevin:2020gbd,Baibhav:2022qxm,Kalogera:1999tq,Gerosa:2018wbw,OShaughnessy:2017eks,Wysocki:2017isg,Steinle:2020xej,Callister:2020vyz,LIGOScientific:2020kqk,Farr:2017uvj,Rodriguez:2019huv,Zhang:2023fpp,Doctor:2019ruh,Kimball:2020opk,Payne:2024ywe,Vitale:2015tea,Stevenson:2017dlk,Talbot:2017yur}. 
Gravitational waves (GWs) from BH mergers observed by Advanced LIGO~\cite{aLIGO}, Advanced Virgo~\cite{aVirgo}, and KAGRA~\cite{KAGRA} are one of the few ways to measure the spins of stellar-mass BHs directly.

Spin precession is, however, notoriously difficult to measure from GW signals~\cite{Vitale:2014mka,Vitale:2016avz,Shaik:2019dym,Fairhurst:2019srr}.
Informative spin measurements are rare to begin with: most observations are consistent with negligible spin magnitudes, and contain no information about spin precession~\cite{GWTC2,GWTC2.1,GWTC3}. 
Moreover, among observed BBHs with interesting spin configurations, a few are contaminated by non-Gaussian noise transients (``glitches"~\cite{Davis:2022ird,LIGOScientific:2016gtq,LIGO:2021ppb}). 
The anti-aligned spins of GW191109\_010717~\cite{Udall:2024ovp} and the large spin precession of GW200129\_065458~\cite{Payne:2022spz}, for example, are both dependent on the assumed glitch model.
Measuring precessing spins is further complicated by waveform systematics: different models for the waveform can yield different in-plane spin measurements for the same signals~\cite{Varma:2021csh}.
Therefore, obtaining robust spin measurements can be even more challenging than other BBH properties like mass, not only because the imprint is weaker but also because it is more subtle to model.

We submit that a detailed understanding of \textit{signal morphology} (i.e., strain over time) can lead to more robust spin inference and interpretation.
Connecting astrophysical parameters with waveform phenomenology---or how these parameters imprint cycle-by-cycle onto observable GW signals---bolsters the measurements' robustness.
For instance, if spin precession consistently imprints a distinct waveform feature, identifying that feature in a new GW event would strengthen confidence in the measurement and clarify potential systematics.
Additionally, consider the highly probable situation in which we detect a GW that overlaps with a glitch or its residual~\cite{Davis:2022ird}:
with \textit{a priori} knowledge of how and where precession imprints on GWs, we can better identify whether or not that glitch can mimic precession. This knowledge is especially powerful in the case where the imprint of precession is subtle and can be easily mimicked by non-Gaussianities in data.
In this way, signal morphology is a tool to help safeguard against spurious spin measurements caused by data quality issues or waveform systematics.
A similar idea of dividing data into time or frequency chunks to compare how data and simulations accumulate signal-to-noise ratio (SNR) is used in GW searches~\cite{Allen:2004gu,Usman:2015kfa,Davis:2020nyf} and glitch mitigation~\cite{Udall:2024ovp}.
Working in the time domain allows us to draw a connection between the waveform morphology and the dynamics of the source, leading to better intuitions.

The link between waveform morphology and system parameters is especially challenging for high-mass (``heavy") BBHs with redshifted (i.e., detector-frame) component masses $\gtrsim 50 + 50\,\Msun$.
These systems merge at the low-frequency edge of the sensitivity band for current ground-based interferometers~\cite{O3-sensitivity}, resulting in a short-duration detectable GW signal dominated by the merger phase of coalescence.
In such cases, up to seventeen BBH parameters\footnote{Fifteen parameters describe a non-eccentric BBH, and an additional two are necessary if the orbit is eccentric.} must be inferred from just a few observed waveform cycles.
If a cycle overlaps with a glitch, a large portion of the signal is compromised.
Moreover, the observational signature of BBH parameters on the merger phase remains poorly understood.
Unlike in the inspiral which is described by, e.g., the post-Newtonian formalism~\cite{Blanchet:2013haa,Nagar:2018zoe}, spin is not analytically tractable in the merger~\cite{Kang:2025nio}.
Precession is particularly difficult to measure, as the precession timescale is comparable to the duration of the signal itself~\cite{Gerosa:2015tea}. 
A sizable fraction of the most recent catalog of GW transients~\cite{GWTC3} is consistent with being high-mass, and therefore plagued by these issues.
Although the typical BBH detected with GWs is consistent with having  detector-frame masses of less than $40+40\,\Msun$, generating an observable inspiral, about \textit{one third}\footnote{Nuance is necessary when interpreting the significance of high-mass BBH merger detections. This back-of-the-envelope estimate uses the standard astrophysical probability significance threshold~\cite{GWTC3} and not the more stringent thresholds used when searching for intermediate mass BHs~\cite{LVK_IMBH}.}
of the BBHs in the third GW transient catalog~\cite{GWTC3} have a median inferred detector-frame total mass $M\gtrsim100\,\Msun$, which corresponds to a merger at ${\sim}150$\,Hz and only ${\sim}5$ observable cycles in-band.

Investigations with numerical relativity (NR) and other simulated data (e.g., with NR surrogates and ringdown models) suggest that, despite observational challenges, information about spin precession \textit{does} exist in heavy BBH signals~\cite{OShaughnessy:2012iol, Biscoveanu:2021nvg, Hughes:2019zmt,Finch:2021iip,Kamaretsos:2012bs,Hamilton:2023znn,Xu:2022zza,Zhu:2023fnf}.
However, studies of simulated high-mass GW signals, such as \citet{Biscoveanu:2021nvg} and \citet{Xu:2022zza}, find that the observability of spin precession in these systems is contingent on many factors: the specific spin configuration of the system, the angle at which the BBH source is oriented with respect to the Earth, and the Gaussian noise realization. 
In some cases, even changing spin tilt angles by as little as order one-tenth of a radian can affect
whether or not spin precession is observed, cf. Fig.~8 of~\citet{Biscoveanu:2021nvg}.
Generally, it is not always understood why precession inference is informative or not in any specific signal.
The relationship between the input (parameters) and output (strain) of waveform models and the \textit{measurability} of these parameters via said strain remains murky. 

In \citet{Miller:2023ncs} we used \textit{time-domain inference}~\cite{Isi:2021iql,Isi:2020tac} to understand the morphological imprint of spin precession in one particular signal: GW190521~\cite{GW190521_detection, GW190521_astro}, the most massive and highly precessing BBH confidently observed to date.\footnote{GW190521's status as most massive has since been eclipsed by GW231123\_135430~\cite{GW231123}.}
We dissected GW190521's data in the time domain and performed parameter estimation independently on different temporal sub sets of data, i.e., before/after different ``cutoff times." 
With this method, we localized the segment of data driving the previously elusive measurement of spin precession, under the assumption of a quasicircular orbit: \textit{the constraint hinges on a short and quiet portion of the pre-merger data that is dampened out with respect to the signal's loud merger}. 
Looking into the system's source dynamics, the relative suppression of this pre-merger cycle corresponds to the tilt of the binary's direction of maximal GW emission away from the Earth as the orbital plane precesses. 
If the system were \textit{not} precessing (and we assume the orbit is non-eccentric), such a change in emission direction would not occur, leaving the final pre-merger amplitude louder than it was for GW190521.

Here, we expand upon \citet{Miller:2023ncs} and use \textit{simulated} data to further investigate the signal morphology conditions that underpin measurements of spin precession in high-mass systems.
Like past studies, our motivating question is: should it be surprising that out of $\mathcal{O}(100)$ detections the most massive BBH is also the most highly precessing, i.e., does it hinge on a fine-tuned configuration? 
Furthermore, when precession \textit{is} constrained, does our \textit{interpretation} of the morphological origin of the GW190521 measurement hold generically true, i.e., do the relative strengths of the pre-merger and merger data always drive the measurement?
This paper is the first in a series of two throughout which we address these questions. 
In this first paper, we start from the ``best fit'' (maximum likelihood; $\maxL$) parameters of GW190521~\cite{Miller:2023ncs} and consider a variety of heavy, highly-precessing signals.
In the second paper~\cite{Miller:2025}, we investigate effects that could \textit{mimic} precession in high-mass signals, namely eccentricity~\cite{Romero-Shaw:2020thy} and intermediate mass ratio~\cite{Nitz:2020mga}.

By tweaking GW190521’s morphology along five different axes in parameter space, we examine how features in the signal impact the confidence of the spin precession measurement for both the full and restricted subsets of data, i.e., the inspiral vs.~ringdown.
Throughout this work, we define the ``ringdown" as the data after the signal's peak, and the ``inspiral" as the data before the signal's peak, where the peak refers to the waveform amplitude integrated over the celestial sphere~\cite{Blackman:2017pcm}.
We keep the six spin degrees of freedom and mass ratio fixed, meaning all of our simulated signals have identical spin evolution.
The effect of spin magnitudes and angles on inference in high-mass signals has been explored in other works~\cite{Biscoveanu:2021nvg,Xu:2022zza}.
We vary the system's SNR, adjust merger frequency by changing the total mass at a fixed SNR, and explore extrinsic angular parameters (inclination, polarization,and phase).
Each of these variations affects the \textit{observed} waveform morphology and precession measurability in different ways.
We find that: 
\begin{itemize}
    \item Spin precession is measurable later into and past the merger regime of the signal as the SNR increases. At a sufficiently  high SNR, precession can be constrained using ringdown data alone, with informative measurements persisting for start times up to $20\,M$ after the signal's peak.
    \item Decreasing the total mass, and thereby increasing the duration of the detectable inspiral, makes the signal more informative about precession.
    At sufficiently low masses ($M\lesssim100\,\Msun$), precession can be constrained using the inspiral data alone at GW190521's SNR.
    \item A fine-tuned extrinsic angular configuration is not necessary to measure spin precession, meaning that precession is constrained away from the prior for many different combinations of the inclination, polarization, and phase angles. 
    The inclination and phase angles do, however, drastically affect measurability, ranging from signals which have no information to being more informative than the real GW190521 data. 
    The polarization angle does not affect precession inference for the configurations we study. 
    These trends hold true for the full and restricted durations of data.
    \item The relationship between loud merger and quiet final pre-merger does not always drive spin precession constraints, as it did for GW190521~\cite{Miller:2023ncs}.
    In some signals, inference does hinge on the relative amplitude of the pre-merger cycle, but generically, the informativeness and the ratio of the signal's pre-merger strain peak to overall maximal amplitude are not negatively correlated. 
\end{itemize}

The remainder of the paper is organized as follows.
We briefly describe our methodology in Sec.~\ref{sec:methods}.
In Sec.~\ref{sec:maxL_injection} we present inference on the $\maxL$ waveform for GW190521 from \citet{Miller:2023ncs}.
Section~\ref{sec:results} presents results from changing different parameters of the $\maxL$ waveform, with Secs.~\ref{subsec:SNR}, \ref{subsec:total-mass}, and \ref{subsec:phase} corresponding to the first three bullets above.
Finally, we conclude in Sec.~\ref{sec:conclusion}.

\section{Methods}\label{sec:methods}

In this section, we establish spin notation (Sec.~\ref{subsec:spins}), and then describe our analysis approach, which can be broadly outlined in three steps.
First, we inject GW190521-like signals and conduct parameter estimation in the time domain before and after various cutoff times in the data.
The methods for generating and performing inference on simulated signals in the time domain are described in Sec.~\ref{subsec:TD_inf}.
Then, we assess the informativeness of posteriors across different time segments by computing the Jensen-Shannon divergence between each posterior and its prior, as described in Sec.~\ref{subsec:JSD}.
Finally, we trace the informativeness of the posteriors as a function of cutoff time to localize what features in the waveform may be driving the measurements of spin precession and other astrophysical phenomena.
Throughout, we look at the whitened time-domain strain and SNR, obtained through the procedures described in Sec.~\ref{subsec:whitening}.

\subsection{Spin parametrization}
\label{subsec:spins}

At leading order, BH spin imprints on GWs from merging BBHs in two ways. 
The strongest effect---modulation of the signal duration---comes from the spin components \textit{parallel} to the binary's orbital angular momentum (``aligned spin"). 
These are parametrized via the effective spin $\chieff$ which is approximately conserved in the inspiral~\cite{Racine:2008qv,Ajith:2009bn},
\begin{equation}
    \chieff = \frac{\chi_1 \cos\theta_1 + q \, \chi_2 \cos\theta_2}{1+q} \in (-1,1)\,,
\label{eqn:chieff}
\end{equation}
where $\chi_i$ are spin magnitudes, $\cos\theta_i$ are the polar angles between the spins and the orbital angular momentum, and $q\equiv m_2/m_1$ is the binary's mass ratio. 
Subscript 1 corresponds to the primary (more massive) BH, while 2 corresponds to the secondary (less massive). 

Spin components \textit{misaligned} with the orbital angular momentum cause the binary's orbital plane to precess, which induces amplitude and phase modulation as the direction of maximum GW emission sweeps toward and away from the Earth~\cite{Apostolatos:1994,Kidder:1993}.
Although alternatives exist~\cite{Gerosa:2020aiw,Thomas:2020uqj}, we report spin precession constraints using the canonical ``effective precessing spin" parameter $\chip$~\cite{Schmidt:2010it,Schmidt:2012rh,Schmidt:2014iyl}, 
\begin{equation}
    \chip = \mathrm{max}\left[\chi_1 \sin\theta_1,
    \left(\frac{3+4q}{4+3q}\right) q \,\chi_2 \sin\theta_2\right]\in [0,1) \,\, ,
\label{eqn:chip}
\end{equation}
for ease of comparison to past work.
In \citet{Miller:2023ncs}, we showed that other precession parameters behave analogously to $\chip$ for GW190521. 
Spin inference is correlated with other parameters~\cite{Cutler:1994ys,Roulet:2022kot,Fairhurst:2023idl,Biscoveanu:2021nvg,Vitale:2014mka,Xu:2022zza,Baird:2012cu}. 
Thus, while we focus on $\chip$, we also examine $\chieff$, detector-frame total mass $M$, and mass ratio $q$.

Given that the spin vectors evolve over time due to precessional dynamics~\cite{Apostolatos:1994}, they must be quoted at a specific reference time or frequency. 
We define all spin quantities at $t=-100\, M$~\cite{Varma:2021csh,Varma:2019csw}.
For reference, in geometric units, $1\,\Msun = 4.93 \times 10^{-6}\,\rm{s}$ and hence the $\maxL$ point for GW190521 has $M = 258.7\,\Msun= 1.27 \times 10^{-3}\,\mathrm{s}$. 
Using units of mass allows us to compare systems with different total masses by referring to the analogous points in the orbital evolution (see Sec.~\ref{subsec:total-mass}) and facilitates comparison with ringdown analyses~\cite{Siegel:2023lxl}, which report results in units of detector-frame remnant mass.

\subsection{Analyzing signals in the time domain}
\label{subsec:TD_inf}

Inferring the parameters of GW sources is typically conducted in the frequency domain. 
Because detector noise is stationary, its covariance matrix becomes diagonal in the Fourier domain when periodic boundary conditions are imposed, enabling efficient likelihood calculations~\cite{LIGOScientific:2019hgc,Unser:1984}. 
However, if we want to isolate a specific segment of the signal in time, it may not be possible to enforce periodicity, meaning either non-trivial likelihood modifications must be performed in the Fourier domain~\cite[e.g.,][]{Capano:2021etf,Correia:2023ipz}, or the data must be analyzed directly in the time domain~\cite{Isi:2021iql}.
We use time-domain inference~\cite{Miller:2023ncs, Isi:2021iql}, based on methods originally derived for studying BH ringdowns~\cite{Isi:2019aib,Isi:2020tac,Carullo:2019flw}.
The code base used for this analysis has been released as {\tt tdinf}~\cite{tdinf}.

We use simulated GW signals with a zero noise realization, assuming the sensitivity of the LIGO and Virgo detectors at the time of GW190521~\cite{GW190521_gwosc}. 
We opt for zero noise because we are interested in studying the behavior of
simulated signals on average, and zero noise is equivalent to averaging
over Gaussian realizations~\cite{Nissanke:2009kt}.
The specific Gaussian noise realization will impact the posteriors~\cite{Biscoveanu:2021nvg,Xu:2022zza} in the standard way, i.e., by shifting the best-fit parameters~\cite{Vallisneri:2007ev}.
We confirm that we can reproduce the GW190521 inference with the $\maxL$
waveforms in zero noise (see Fig.~\ref{fig:real_vs_maxL}), unlike \citet{Xu:2022zza}.

We use the inspiral-merger-ringdown (IMR) numerical relativity surrogate waveform \textsc{NRSur7dq4} \cite{Varma:2019csw} for both injection and recovery.\footnote{
To generate {\sc NRSur7dq4} with parameters defined at $t=-100\,M$, see Sec.~\ref{subsec:spins}, we use a custom version of \textsc{LALSuite} with the flag {\tt fref=-1}~\cite{Varma:2021csh,Varma:2019csw,lalsuite}.}
For each signal, we select a series of cutoff times, truncate the data at each time, and analyze the data before (``pre-cutoff") and after (``post-cutoff") independently, as well as the full signal.
Consistent with \textsc{NRSur7dq4}, we define the time of coalescence ($t=0$) as the time of peak total waveform amplitude integrated over the celestial sphere~\cite{Blackman:2017pcm}; see Eq.~(5) of \citet{Varma:2019csw}.

During inference, we draw samples from the posterior of the binary total mass $M$, mass ratio $q$, spin magnitude $\chi_i$ and tilt angles $\theta_i$, azimuthal inter-spin angle $\phi_{12}$, azimuthal precession cone angle $\phi_{JL}$, luminosity distance $d_L$, and phase of coalescence $\varphi$. 
Geocenter time of coalescence $t_0$, right ascension $\alpha$, declination $\delta$, and polarization angle $\psi$ are fixed to their true values; this fixes the time of arrival in each detector.
\citet{Miller:2023ncs} showed that fixing these parameters did not affect conclusions for GW190521. 
We confirm that the same is true for simulated data by repeating select analyses while varying over all parameters.
Priors are the same as Table I in \citet{Miller:2023ncs}, except that we use a wider uniform total mass prior to encompass each simulated system's true total mass value. 
We simulate data from the three-detector network of LIGO Livingston (LLO), LIGO Hanford (LHO), and Virgo with a sampling rate of $2048\,\Hz$, assuming the GW190521 event power spectral densities publicly available on the Gravitational-wave Open Science Center \cite{opendata_O1O2,opendata_O3}. 
We analyze the frequency range of 11--1024\,Hz.
Our time domain inference code~\cite{tdinf} utilizes a Markov Chain Monte Carlo sampler implemented in Python with {\tt emcee}~\cite{emcee}. 
Each analysis runs with 512 chains for 50,000 steps.

At face value, there are parallels between our time-domain inference framework and traditional IMR consistency tests (IMRCTs)~\cite[e.g.,][]{Ghosh:2016qgn,Ghosh:2017gfp,Breschi:2019wki} used to test General Relativity~\cite{GWTC3_TGR}. 
However, while both involve conducting parameter estimation on different portions of a GW signal, the two have distinct methods and goals.
IMRCTs operate in the frequency domain---frequency-domain truncation is not equivalent to time-domain truncation beyond the stationary phase approximation, e.g., in the merger or when higher-order modes are present.
One could theoretically slice GW data in any way in the time-frequency plane and require consistency within the different segments, but not all bases provide equal insight.
Segmenting in the time domain allows us to draw connections to a BBH's source dynamics in a more direct way than other bases.
Moreover, IMRCTs are, as the name suggests, consistency tests: they compare the posterior from different parts of the signal to see if the data are consistent with each other. 
In this work, we are not concerned with the question of consistency, but rather in understanding how information is accumulated throughout a signal.

\subsection{Quantifying a parameter's ``measurability"}
\label{subsec:JSD}

We use the Jensen-Shannon divergence (JSD)~\cite{Menendez:1997,Endres:2003,Lin:1991} to quantify how different two posterior probability distributions are. 
The JSD is a statistical quantity for the similarity between two probability distributions, $P$ and $Q$. 
It is the symmetrized and smoothed version of the Kullback-Lieber (KL) divergence $D$~\cite{Kullback:1951,Kullback:1959}, such that
\begin{equation}
    \mathrm{JSD}(P||Q) = \frac{1}{2} \Big[ 
        D\Big(P \Big|\Big| \frac{P+Q}{2} \Big) + D \Big( Q \Big|\Big| \frac{P+Q}{2} \Big)
    \Big]\,,
\label{eqn:JSD}
\end{equation}
where the KL divergence is defined as follows:
\begin{equation} \label{eqn:KL}
    D(P||Q) = \int^{\infty}_{-\infty} p(x) \log_2 \frac{p(x)}{q(x)}  dx\,,
\end{equation}
where $p$ and $q$ are the probability densities of $P$ and $Q$.
The JSD is normalized with a minimum of 0 nats, corresponding to the two distributions $P$ and $Q$ being identical, and a maximum dependent on the base of the logarithm used to calculate $D(P||Q)$.
We use a base $2$, corresponding to a maximum of $\log(2) = 0.69$ nats.
Lower JSDs indicate distributions that are more similar.

To define how ``measurable" a parameter is, we follow Refs.~\cite{Xu:2022zza,Romero-Shaw:2020owr} and calculate the JSD between its one-dimensional marginal posterior and prior.
Correlated priors between two parameters can lead to spuriously strong one-dimensional JSDs, i.e, confidently measuring one may induce a prior-driven measurement in the other~\cite{Gangardt:2022ltd}; we further discuss this in Appendix~\ref{app:other_params_in_RD}.
The KL divergence is estimated using a kernel density estimate (KDE) over a finite number of samples; it is thus subject to Monte Carlo error.
As an order of magnitude approximation, when generating sets of $N=10,000$ random draws from the $\chip$ prior distribution, we find JSDs ranging $\mathcal{O}(10^{-4}-10^{-6})$. 
When using a smaller number of samples, $N=1,000$, the JSDs increase up to $\mathcal{O}(10^{-3})$. 
Figure~\ref{fig:mass_in_RD} in the Appendix shows examples of such JSD null distributions for $N=3,000$ samples.
As such, to be conservative when dealing with varying numbers of samples and differently shaped distributions (KDEs are sensitive to the tails of distributions), we define JSDs $\gtrsim 10^{-2}$ as certainly indicating a difference between two distributions, and JSDs $\lesssim 10^{-4}$ as indicating identical distributions. 

\begin{table*}
    \centering
    \renewcommand{\arraystretch}{1.5}
    \begin{tabular}{ >{\raggedright}p{0.5\columnwidth} c p{0.5\columnwidth}  p{0.5\columnwidth} }
        \hline\hline
         Parameter & Symbol & $\maxL$ value ($f_{\rm ref} = 11\,{\rm Hz}$) & $\maxL$ value ($t_{\rm ref} = -100\, M$) \\
         \hline
          Detector-frame total mass & $M$ & $258.74\,\Msun$ & $277.15 \,\Msun$ \\
          Mass ratio & $q$ &  0.97 & 0.86  \\
          
          Primary spin & $\vec \chi_1$ & (0.67, 0.04, 0.29) & ($-0.87$, $0.18$, $-0.39$) \\
          Secondary spin & $\vec \chi_2$ & ($-0.05$, $-0.76$, $-0.58$) &  ($-0.17$, $-0.75$, $0.59$)\\
          Effective spin & $\chieff$ & $-0.13$ & 0.06 \\
          Effective precessing spin & $\chip$ & 0.74 & 0.89\\
          Inclination angle & $\iota$ & 1.16\,rad & 1.11\,rad  \\
          Polarization angle & $\psi$ & 2.30\,rad  & 2.30\,rad\\
          Phase & $\varphi$ & 1.31\,rad & $-3.12$\,rad  \\
          Luminosity distance & $d_{L}$ & 3253.11\,MPc & 3190.22\,MPc \\
          Right ascension & $\alpha$ &  6.26\,rad & 6.26\,rad \\
          Declination  & $\delta$ &  $-1.15$\,rad & $-1.15$\,rad \\
          Coalescence GPS time  & $t_{0}$ & 1242442967.407715\,s & 1242442967.4064941\,s\\

        \hline\hline
        \end{tabular}
        \caption{The maximum likelihood ($\maxL$) parameters from two time-domain inference analyses conducted on GW190521 by \citet{Miller:2023ncs}: one where all parameters are defined at $f_{\rm ref} = 11\,\rm Hz$ and another at $t_{\rm ref} = -100\,M$. 
        The $f_{\rm ref} = 11\,\rm Hz$ parameters are used as the baseline for the results presented in Sec.~\ref{sec:maxL_injection}, Sec.~\ref{subsec:SNR}, and Sec.~\ref{subsec:phase}. 
        The $t_{\rm ref} = -100\,M$ parameters serve as the baseline for results in Sec.~\ref{subsec:total-mass} for consistency when looking at sources from different total masses. Since they are obtained from a sampling algorithm and not a dedicated ``peak finding'' algorithm both sets of parameters should be interpreted as ``the maximum likelihood sample,'' rather than the absolute maximum of the likelihood surface.}
    \label{tab:GW190521_maxL_params}
\end{table*}
%

\subsection{Whitening a signal and calculating its signal-to-noise ratio} 
\label{subsec:whitening}

We visualize GW signals with \textit{whitened} waveforms, a helpful representation of GW data which brings the noise to a white-Gaussian baseline so we can better judge what goes into the likelihood by eye.
In the Fourier domain, whitened waveforms $\hat h (f)$ are obtained by dividing the original waveform $h(f)$ by the square root of the noise power spectral density, $S_n(f)$:
\begin{equation}
    \hat h (f) = \sqrt{\frac{f_s}{2\,S_n(f)}} ~ h (f)\,,
\label{eqn:whitening}
\end{equation}
where $f_s$ is the sampling rate.
We derive the corresponding whitened time series $\hat{h}(t)$ by inverse Fourier transforming.
The scaling of Eq.~\eqref{eqn:whitening} ensures that the units of $\hat{h}(t)$ are standard deviations of the noise, a dimensionless quantity denoted by $\sigma$.

When dissecting signals in the time domain, we calculate the \textit{optimal network SNR} in different time windows, denoted $\rho$. 
For a signal $x$, the SNR in a single detector $\mathbb{D}$ is calculated in the time domain over a region of time $T$ by taking the inner product of the signal with itself:  
\begin{equation}
    \rho_{\mathbb{D}} = \langle x | x \rangle_{T}^{1/2}\,,
\label{eqn:SNR}
\end{equation}
where the time-domain inner product is defined as 
\begin{equation}
    \langle x | y \rangle_{T} = \sum^{N}_{i,j=0} x_i C^{-1}_{ij} y_j\,,
\end{equation}
where $x$ and $y$ are time series with $N$ time steps over the duration $T$, with index $0$ corresponding to the first point in the region of $T$. 
The covariance matrix $C_{ij}$ is derived from a given power spectral density following the procedure presented in Sec.~III of \citet{Isi:2021iql}.
The SNR for a network of detectors is then obtained by summing the single-detector SNRs in quadrature, 
$$ \rho = \Big( \sum_{\mathbb{D}} {\rho_\mathbb{D}}^2 \Big)^{1/2}\,\,.$$
Here, $\mathbb{D}\in\{\mathrm{LHO,LLO,Virgo}\}$.
Using this method we calculate the SNR over different time slices of the data, e.g., the SNR in a signal's inspiral versus ringdown.

\section{Analyzing the GW190521 maximum likelihood waveform}
\label{sec:maxL_injection}

%
\begin{figure*}
    \centering
    \includegraphics[width=\textwidth]{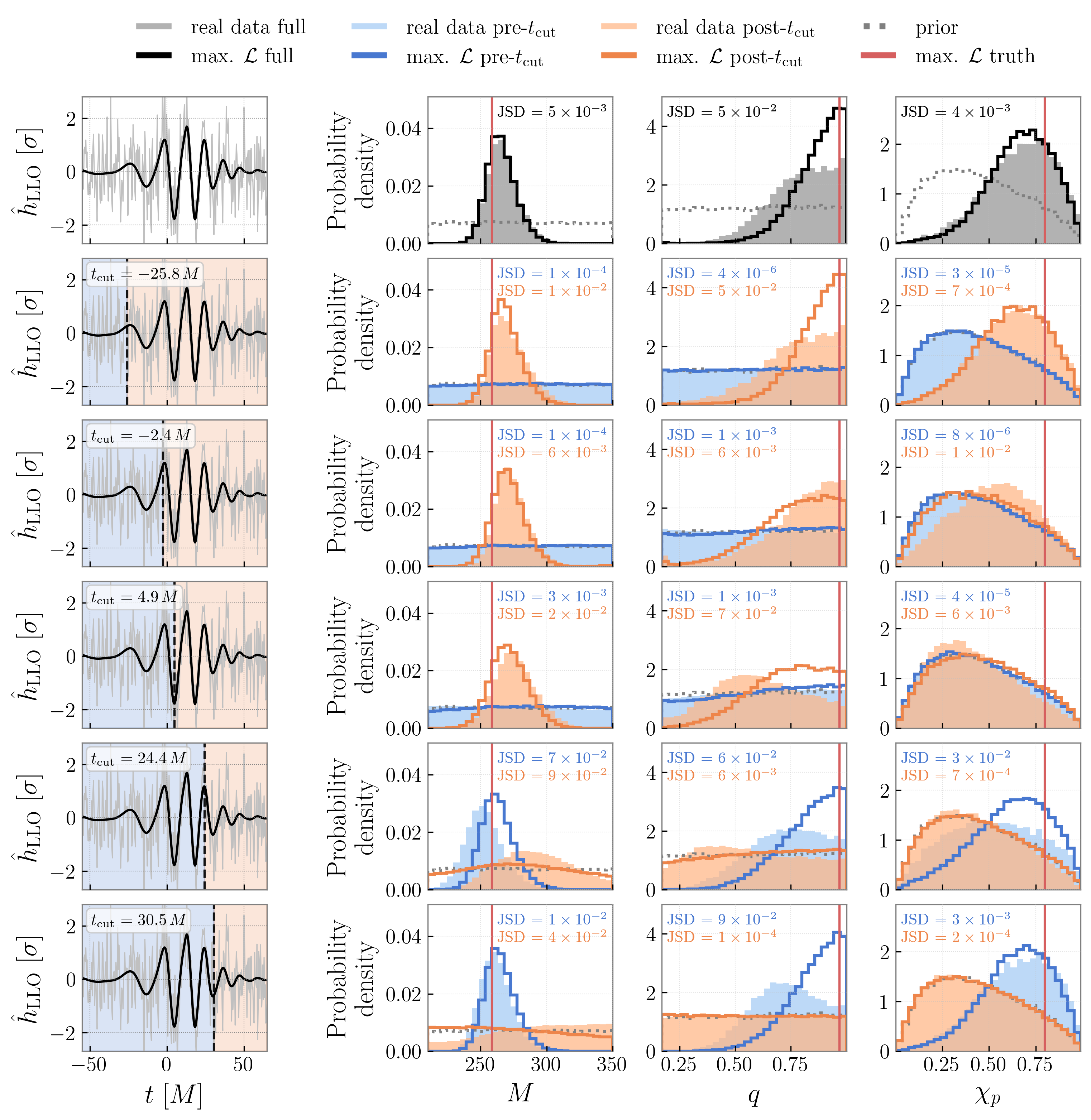}
    \caption{Evolution of posteriors for representative cutoff times for GW190521 and its $\maxL$ waveform, exemplifying that the $\maxL$ waveform behaves similarly to the real data, both for the full signals and as a function of cutoff time.
    Each row shows different segments of data analyzed:  the top row shows the full segment of real data (gray) and the full $\maxL$ waveform (black), while 
    subsequent rows show results from data before (blue) and after (orange) different cutoff times: $t_{\rm cut}= \{-25.6, -2.4, 4.9, 24.4,30.5\}\,M$ from top to bottom. 
    The lighter shaded histograms correspond to the real GW190521 data, while the darker empty histograms are for the $\maxL$ waveform. 
    JSDs between the two posteriors are given in inset text on each plot.
    \textit{First column:} The whitened $\maxL$ waveform from the full analysis (black) along with the whitened LLO data (gray). The blue/orange shaded regions highlight the data informing the same-color posteriors in the remaining columns, and the black-dashed vertical line is the cutoff time.
    \textit{Second through fourth columns:} Marginalized posteriors on the detector-frame total mass $M$, mass ratio $q$, and effective precessing spin $\chip$ inferred from each segment of data. Priors on each parameter are represented by the gray-dotted histograms and vertical red lines mark the true values for the $\maxL$ waveform (Table~\ref{tab:GW190521_maxL_params}).
    See Ref.~\cite{animation_figure01} for an animation of this figure, including more cutoff times.}
    \label{fig:real_vs_maxL}
\end{figure*}

The signals we consider are all based on the maximum likelihood waveform for GW190521 found in \citet{Miller:2023ncs}, with parameters listed in Table~\ref{tab:GW190521_maxL_params}. 
We herein call this the ``$\maxL$ waveform."
Before examining the impact of various system parameters by varying them one at a time in Sec.~\ref{sec:results}, we here confirm that the $\maxL$ waveform behaves similarly to the real event, both when analyzed in full and under time cutoffs.

Results from inference on the $\maxL$ waveform compared to the real data are shown in Fig.~\ref{fig:real_vs_maxL}.
We begin with the \textit{full-signal analysis} (top row), comprising the entire data segment.
In the first column, we plot the whitened GW190521 data (gray) superimposed with the $\maxL$ waveform (black) in LLO.%
\footnote{
We plot waveforms in LLO because it is the detector in which GW190521 is the loudest: the $\maxL$ optimal SNR is 10.8 in LLO, 8.6 in LHO, and 2.5 in Virgo.}
The remainder of the first row shows the posteriors from the real GW190521 data (gray filled histograms) and the $\maxL$ waveform (black empty histograms) for the total detector-frame mass $M$, mass ratio $q$, and effective precessing spin $\chip$.%
\footnote{We exclude the effective spin $\chieff$ from this figure because it is consistent with the prior for the full data and all time slices.
We present the evolution of $\chieff$ and other spin parameters in our accompanying data release~\cite{suppl_figs}. All spin parameters lose/gain information at the same cutoff times.}
The true $\maxL$ values are indicated by vertical red lines, with which all full posteriors are consistent. 
Posteriors for $M$ and $\chip$ are in high agreement with those from the real data, yielding JSDs of $\mathcal{O}(10^{-3})$ between them.%
\footnote{
The latter is notable: $\chip$ inference  from the $\maxL$ waveform is more consistent with the real data than previous studies, e.g.~Fig.~3 of \citet{Biscoveanu:2021nvg} or Fig.~2 of \citet{Hamilton:2023znn}.
We have verified that this difference comes down to the choice of ``$\maxL$'' waveform: while we use the parameters from \citet{Miller:2023ncs}, \citet{Biscoveanu:2021nvg} and  \citet{Hamilton:2023znn} use parameters from~\citet{GW190521_PE}, given in Table II of Ref.~\cite{Biscoveanu:2021nvg}.
When we simulate data based on that maximum likelihood waveform and perform inference in the time domain, we recover posteriors consistent with theirs (peaking at $\chip\sim0.6$). 
Thus the time-domain analysis of GW190521 happened to find a $\maxL$ point that yields a more similar posterior to the real data.
We attribute this to little more than a lucky chance in stochastic sampling.}
The $q$ posterior for the $\maxL$ waveform is slightly more informative than that from the real data, although they both peak at equal masses ($q=1$). 
Here, the JSD between the $\maxL$ and real data posteriors is an order of magnitude larger at $5 \times 10^{-2}$, just above our threshold for difference. 

We now turn to how inference progresses over time.
The second through final rows of Fig.~\ref{fig:real_vs_maxL} show the results from inference conducted on data before (blue) and after (orange) five representative cutoff times.
The dashed vertical line in the first column marks each cutoff, with blue and orange shaded regions representing pre- and post-cutoff data, respectively.
The filled (empty) histograms are posteriors from the analysis conducted on real data ($\maxL$ waveform).
The two analyses are again largely in agreement: JSDs between the analogous posteriors are $\lesssim \mathcal{O}(10^{-2})$.
Most relevantly, \textbf{\textit{inference of $\chip$ from the $\maxL$ waveform hinges on the relationship between the final pre-merger cycle and the post-peak data, just like for the real GW190521 event}}.
For both cases, the data loses the bulk of its precession information between approximately $-25\,M$ (second row) and $-2\,M$ (third row) corresponding to the exclusion of final pre-merger cycle. 

While our method is intrinsically data-driven, as it connects the measurability of parameters to observed waveform morphology, it can be useful to think about the underlying physical mechanisms that may generate the waveform morphology that drives our measurements.
In \citet{Miller:2023ncs}, we found that the suppressed final pre-merger cycle of GW190521 was driven by the orbital plane precessing away from our line of sight.
The same interpretation holds for the $\maxL$ signal, as analogous correlations between the inclination and $\chip$ posteriors in the full versus excluding-pre-merger data can be seen in our accompanying data release~\cite{suppl_figs}.

Though the behaviors of GW190521 and it's $\maxL$ waveform are qualitatively similar as a function of cutoff time, small differences remain. 
The $\maxL$ waveform loses information about $\chip$ slightly earlier than the real data in the post-cutoff analyses, and gains it back earlier in the pre-cutoff analyses.
As can be seen in orange histograms in the third row of
Fig.~\ref{fig:real_vs_maxL}, the $\maxL$ posterior for the post-$t=-2\,M$ analysis is near identical to the prior, while that from the real data is shifted to slightly higher values. 
In the pre-cutoff analyses, when the data before $t=24\,M$ are included, the $\maxL$ signal becomes informative again (fifth row; blue). 
For the real data, this does not happen until $30\,M$ (sixth row).
Additionally, the $M$ and $q$ posteriors behave marginally differently around the peak amplitude of the data: the post-cutoff $q$ posterior for the real data shifts to lower values earlier than the $\maxL$ waveform (fourth row), while the $M$ posterior for the real data retains post-cutoff information longer (fifth row). 
Generally, the $\maxL$ waveform is more informative on $q$ than the real data.

Perfect agreement between the $\maxL$ waveform and the real GW190521 signal is not expected, because they differ in noise realization and true signal parameters.%
\footnote{
In the high SNR regime where the likelihood becomes approximately Gaussian, full agreement in the posteriors from the data and its $\maxL$ simulation is expected~\cite{Cutler:1994ys,Maggiore:2007ulw,Vallisneri:2007ev}. 
This is not the case at GW190521's SNR.
}
The two are, however, guaranteed to at least be relatively consistent within the uncertainty implied by Gaussian noise as encoded in the posterior of the original analysis~\cite{Maggiore:2007ulw,Veitch:2008}.
The same is not necessarily true when analyzing \textit{subsets} of the data, especially under mismodeling of the signal and/or noise.
This is because the waveform that is the best fit (e.g., minimizes the residual) for a \textit{portion} of the data will in general not be the same waveform that minimizes the residual over a bigger window---a template may match well for part of the signal but not the rest.
For example, as shown in Fig.~2 of~\citet{Miller:2023ncs}, the $\maxL$ waveform found for GW190521's full data is not the same as that when just its merger-ringdown data are analyzed.
That the $\maxL$ signal behaves similarly to the real GW190521 makes it a worthy baseline to use in the remainder of this study.

We interpret the time-cut tests performed here and in the subsequent sections as nontrivial consistency checks of the interpretation of the real GW190521 data, in a similar vein to the $\chi^2$ tests performed in GW searches~\cite{Allen:2004gu,Usman:2015kfa,Davis:2020nyf} and glitch mitigation~\cite{Udall:2024ovp}.
Those consistency checks test how SNR is accumulated.
A waveform is split into frequency (or time) bins of equal optimal SNR. 
If given data are equivalent to the sum of that waveform and Gaussian noise, the matched-filter SNR in these bins should then follow a $\chi^2$ distribution.
If an unexpected amount of matched-filter SNR is accumulated in a bin, this is attributed to, e.g, a glitch. 
Our cutoff-time analyses are similar in spirit. 
Here, we use the accumulation of \textit{information} about a parameter (in the form of the JSD), rather than SNR, to see if a signal (in this case, the $\maxL$) is consistent with data (GW190521).

\section{Exploring Spin Measurability}
\label{sec:results}

%
\begin{figure*}[]
    \centering
    \includegraphics[width=\textwidth]{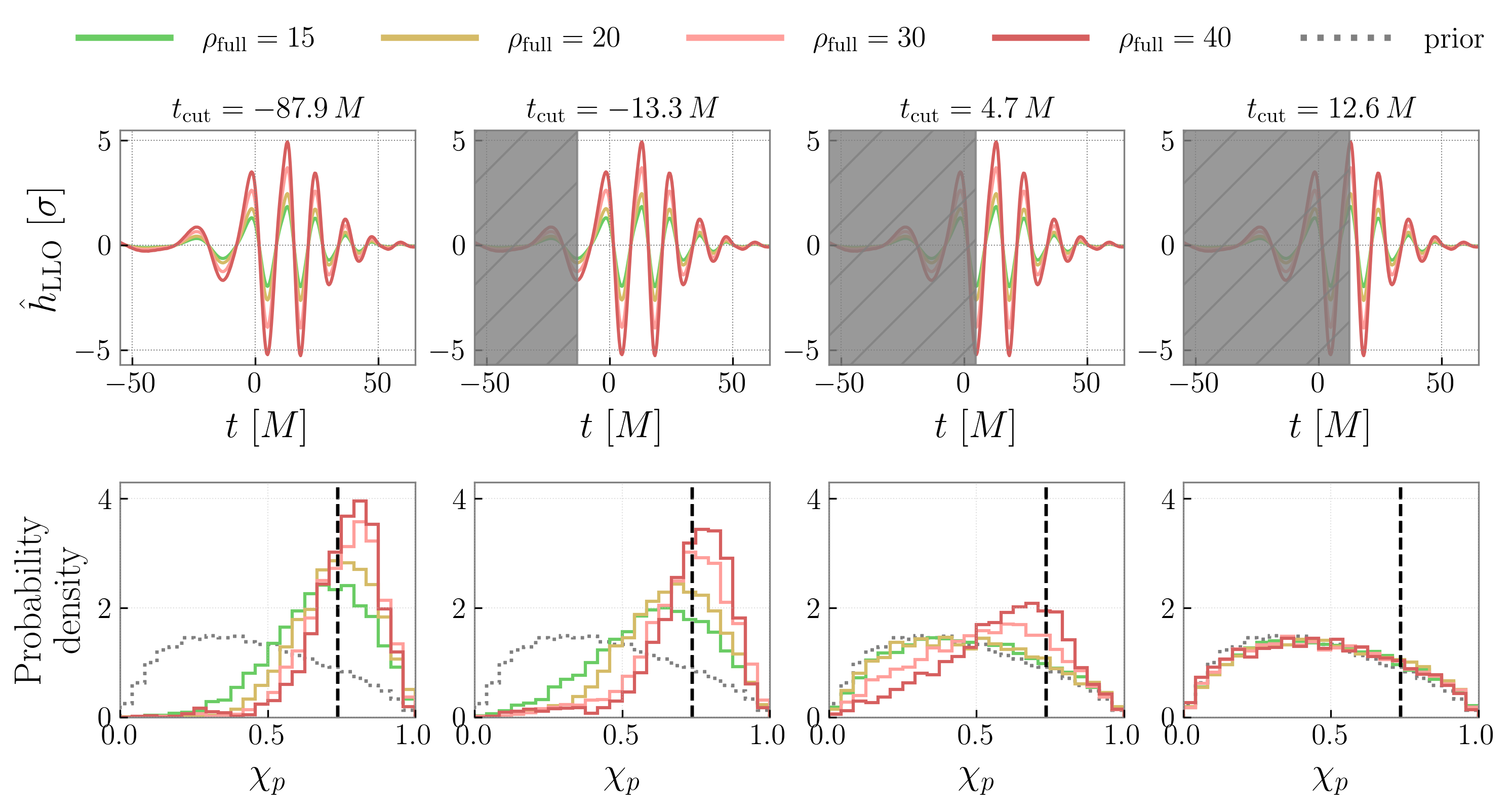}
    \caption{
    Evolution of the $\chip$ posterior for the GW190521 $\maxL$ waveform at different full-signal optimal network SNRs: $\rho_{\rm full}=$ 15 (green), 20 (yellow), 30 (pink), and 40 (red). 
    \textit{Top row:} Whitened strain $\hat h$ in LLO. The gray hatched shading hides data excluded from a given analysis. 
    \textit{Bottom row:} Posteriors for $\chip$ inferred from the non-shaded data in the corresponding column in the top row. 
    The prior is shown as gray dotted, while the black dashed vertical line is the true value. 
    Louder signals have more sharply peaked $\chip$ posteriors at all times.
    The $\rho_{\rm full}=30$ and $40$ waveforms retain some information about $\chip$ in the post-peak data alone (third column).
    See Ref.~\cite{animation_figure02} for an animation of this figure, including more cutoff times and SNRs.
    }
    \label{fig:different_SNRs}
\end{figure*}
\begin{figure}[]
    \centering
    \includegraphics[width=\columnwidth]{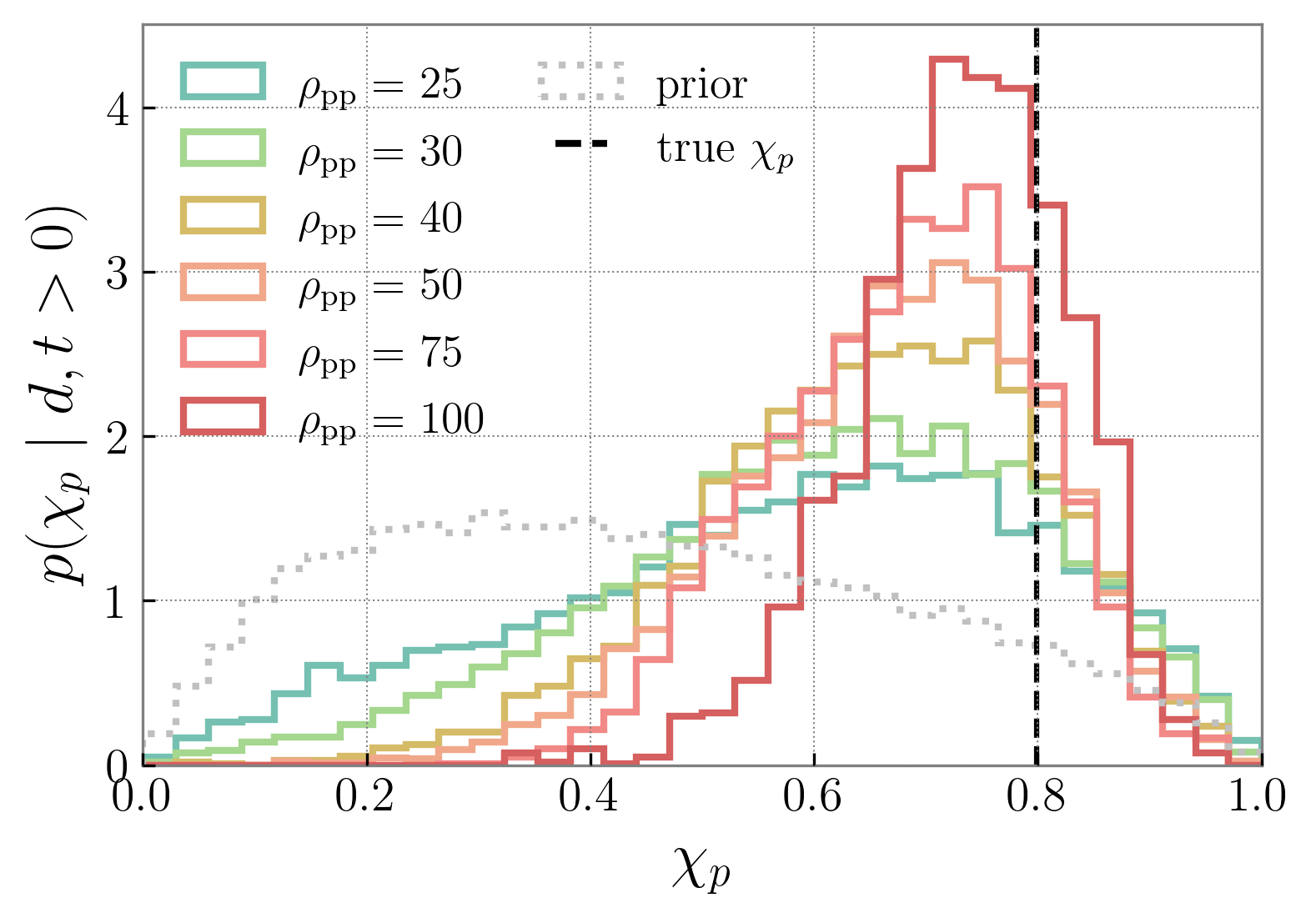}
    \includegraphics[width=\columnwidth]{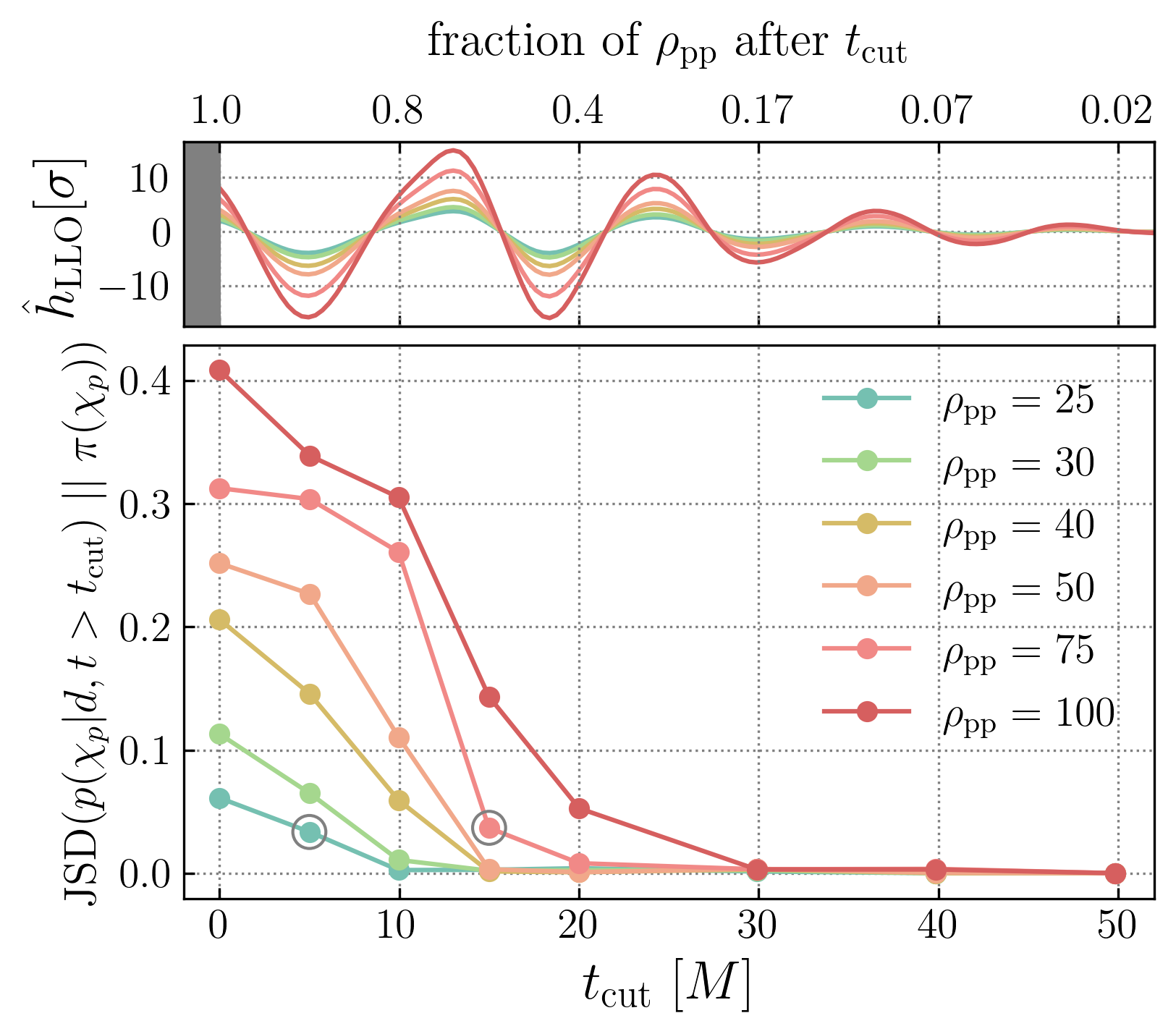}
    \caption{Information about $\chip$ in the ringdown of the $\maxL$ waveform at different SNRs. 
    These signals have $\rho_{\rm pp}=25,30,40,50,75,100$, which correspond to full-signal SNRs of $\rho = 30.5, 36.5, 48.7, 60.9, 91.4, 121.8$. For reference, the $\maxL$ waveform has $\rho_{\rm pp} = 11.4$.
    \textit{Top panel:} 
    Posteriors on $\chip$ from the post-peak data ($t > 0\,M$) for different post-peak SNRs (different colors).
    The ringdown alone has measurable information about $\chip$, as the posteriors are not consistent with the prior (gray dotted).
    \textit{Bottom panel:} 
    JSD between the $\chip$ posterior and prior from data after different cutoff times (horizontal axis) for different SNRs (different colors). 
    The whitened strain of each signal in LLO is plotted above, with the upper axis labeling the fraction of $\rho_{\rm pp}$ remaining after $\tcut$.
    The two points circled in gray are an example of posteriors with comparable JSDs originating from signals with different SNRs. 
    For loud enough signals, $\chip$ can be measured up to late times of $20\,M$ into the ringdown.
     }
    \label{fig:different_RD_SNRs}
\end{figure}

Having verified its similarity over cutoff time to the true event, we now use GW190521's $\maxL$ waveform as a starting point to explore the impact of various parameters on waveform morphology and parameter measurability. 
We simulate a suite of signals from BBHs with the same six intrinsic spin degrees of freedom and mass ratio.
This means each system has the \textit{same intrinsic spin evolution}. 
We also fix the time of coalescence and sky position (right ascension and declination).
One parameter at a time, we then vary the SNR (Sec.~\ref{subsec:SNR}), total mass (Sec.~\ref{subsec:total-mass}), and three extrinsic angles (phase, inclination, and polarization; Sec.~\ref{subsec:phase}), each of which changes the \textit{observed} waveform projected onto detectors in different ways.
The parameters we target roughly correspond to the three degrees of freedom of a
simple sine wave: SNR controls the signal amplitude, the total mass determines
the frequency and observable length, and the extrinsic angles (while fixing the SNR) determine the
phase.
Past work~\cite{Miller:2023ncs, Romero-Shaw:2022fbf} has shown that the number of observable cycles and their strain amplitude strongly matter for the determination of intrinsic BBH properties like precession.
Thus, we investigate how \textit{the way that we observe the same BBH system} impacts our ability to infer such properties. 
In doing so, we hope to illuminate whether GW190521's parameters were fine-tuned to measure precession, and if the morphological origin of its precession information is universal.

\subsection{Imprints of spin precession in the ringdown}
\label{subsec:SNR}

In Fig.~\ref{fig:real_vs_maxL} we showed that, similar to the real data, the last inspiral cycle is necessary to constrain $\chip$ in GW190521's $\maxL$ waveform.
Now, we explore whether this is a general characteristic of the signal—meaning the ringdown alone carries no information about precession—or if the ringdown does encode precession information, but it can only be extracted at a higher SNR, as would be expected from Refs.~\cite{Siegel:2023lxl,Zhu:2023fnf}.
To do this, we scale the luminosity distance of the $\maxL$ parameters (while fixing the detector-frame total mass) to give a desired network optimal SNR, leveraging the two's inverse proportionality~\cite{Hogg:1999ad}.
We then repeat the analysis of Sec.~\ref{sec:maxL_injection}.
In this section, we denote the full-signal SNR as as $\rho_{\rm full}$ and the post-peak (ringdown) SNR as $\rho_{\rm pp}$.

Results for various full-signal SNRs and time cuts are shown in Fig.~\ref{fig:different_SNRs}.
As expected, louder signals yield $\chip$ posteriors that are increasingly constrained away from the prior when the full signal is analyzed---seen, by proxy, in the first column.
Up to $\rho_{\rm full}=20$, the post-cutoff results behave analogously to GW190521, which has $\rho_{\rm full}=14$: seeing the final pre-merger cycle in comparison to the loud merger is crucial for constraining $\chip$ away from the prior, as seen in the transition from the second to the third columns for $\rho_{\rm full}=15$ (green) and $20$ (yellow). 
For the $\rho_{\rm full}=30$ and $40$ cases, however, the post-$t=0$ posteriors for $\chip$ are \textit{not} identical to the prior (third column), meaning that if a GW190521-like signal is loud enough, there is some evidence for precession in the post-peak signal alone.

We next investigate how far into the ringdown $\chip$ can be measured for increasingly loud GW190521-like signals.
We generate waveforms with increasing \textit{post-peak} SNRs ($\rho_{\rm pp}$), i.e., $\rho$ over the region where $t > 0$, with results shown in Fig.~\ref{fig:different_RD_SNRs}.
The top panel of Fig.~\ref{fig:different_RD_SNRs} shows the posteriors for $\chip$ when analyzing just the post-peak data (i.e., post-$t_{\rm cut} = 0$). 
We again use the \textsc{NRSur7dq4} waveform, \textit{not} ringdown-specific models like a sum of quasi-normal modes (QNMs), e.g.,~\cite{Teukolsky:1973,Chandrasekhar:1975,Kokkotas:1999bd,Detweiler:1980,Dreyer:2003bv,Berti:2005ys,Siegel:2023lxl,Zhu:2023fnf,Capano:2021etf,Hamilton:2021pkf,Finch:2021iip}.
For all cases (various colors), $\chip$ is constrained well away from the prior (gray dotted).
\textbf{\textit{Precession can be measured from the ringdown alone using an IMR waveform model, given sufficient SNR.}}
This is not surprising, as the inspiral dynamics are imprinted onto QNM complex amplitudes~\cite{Zhu:2023fnf,Capano:2021etf,Cheung:2023vki, Pacilio:2024tdl,MaganaZertuche:2024ajz,MaganaZertuche:2024ajz,Hamilton:2021pkf,Hamilton:2023znn,Finch:2021iip}, and these should be informative at sufficiently high SNR, as we discuss more in the conclusions.
While our results use {\sc NRSur7dq4}, other parametrized waveform models, such as {\sc IMRPhenomXO4a}~\cite{IMRPhenomXO4a} or {\sc SEOBNRv5PHM$_{\rm w/asym}$}~\cite{SEOBNRv5PHM_withAsym}, may yield similar results.

We next cut off increasingly more data and repeat the analysis.
To visualize the signal being excluded with each cut, the middle panel of Fig.~\ref{fig:different_RD_SNRs} shows the whitened waveforms in LLO.
The fraction of $\rho_{\rm pp}$ remaining in the signal after each $\tcut$ is labeled on the upper axis. 
For example, the data after $\tcut = 20\,M$ has $40\%$ of the SNR of the data after $\tcut = 0\,M$. 
This is independent of the total SNR.

In the bottom panel, we plot the JSD between the $\chip$ posterior $p(\chip | t > t_{\rm cut})$ and prior $\pi(\chip)$ as a function of cutoff time for each signal (different colors). 
Across all SNRs, as more data are excluded, i.e., $t_{\rm cut}$ increases, the JSDs decrease, indicating that the $\chip$ posterior becomes increasingly consistent with the prior.
For $\rho_{\rm pp} \leq 50$, the JSD between the $\chip$ posterior and prior is $\gtrsim 10^{-2}$ for up to $10{-}15\,M$ into the ringdown, indicating informative $\chip$ posteriors.
For the even higher SNR signals of $\rho_{\rm pp} = \{75,100\}$, there is information up to $t_{\rm cut} \sim 20\,M$.
\textbf{\textit{Pre-merger spin precession dynamics are imprinted very late into the ringdown.}}
A log-scaled version of this plot and for other parameters can be found in Fig.~\ref{fig:mass_in_RD} in Appendix~\ref{app:other_params_in_RD}.
The total mass, for example, can be measured up to $\tcut =40{-}50\,M$.

In the bottom panel of Fig.~\ref{fig:different_RD_SNRs}, the two points circled in gray are an example of posteriors with comparable JSDs originating from signal regions with different SNRs. 
The $\tcut=5\,M$, $\rho_\mathrm{pp} = 25$ case (teal) yields a JSD of $0.033$ over a region of data with SNR $0.89 \cdot 25 = 22.2$.
The $\tcut=15\,M$, $\rho_{\rm pp} = 75$ analysis (coral), on the other hand, has a comparable JSD of $0.037$ but an SNR that is a factor of two larger: $\rho = 0.62 \cdot 75 = 46.5$.
\textbf{\textit{It is thus not just the total SNR of the signal in a region that
matters, but the actual morphology contained in that region}.}

\subsection{Observing more inspiral bolsters spin precession inference}
\label{subsec:total-mass}

%
\begin{figure*}
    \centering
    \includegraphics[width=0.99\textwidth]{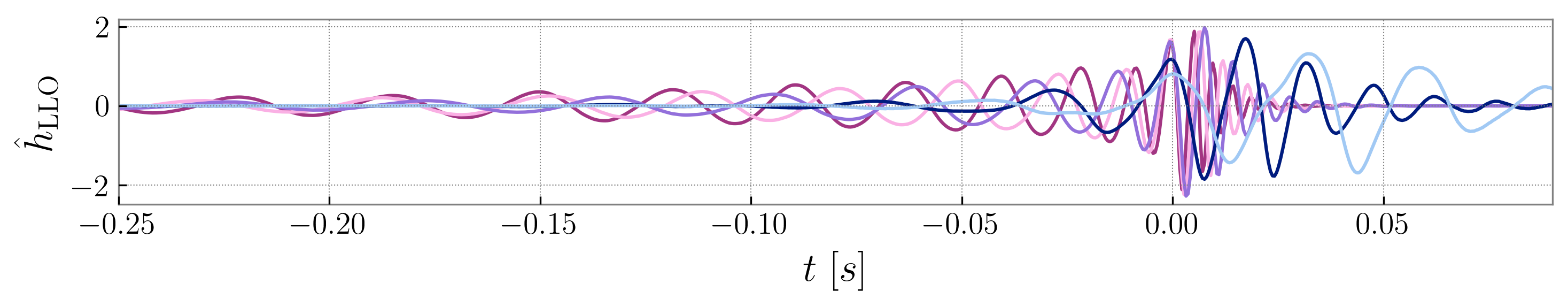}
    \includegraphics[width=0.99\textwidth]{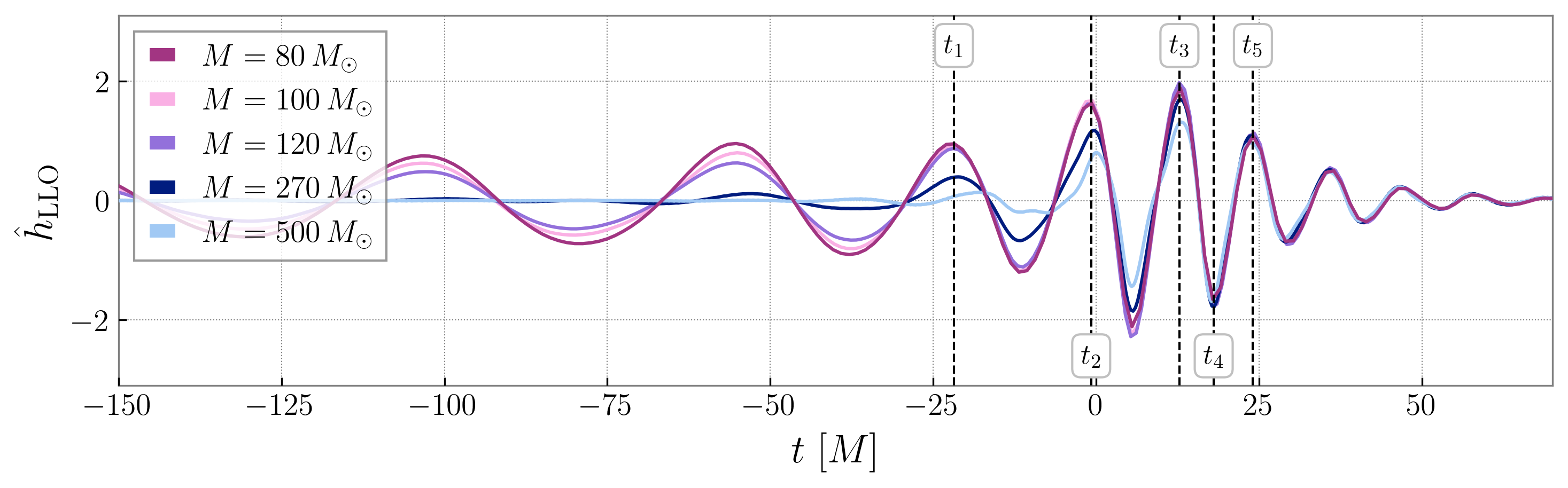}
    \includegraphics[width=\textwidth]{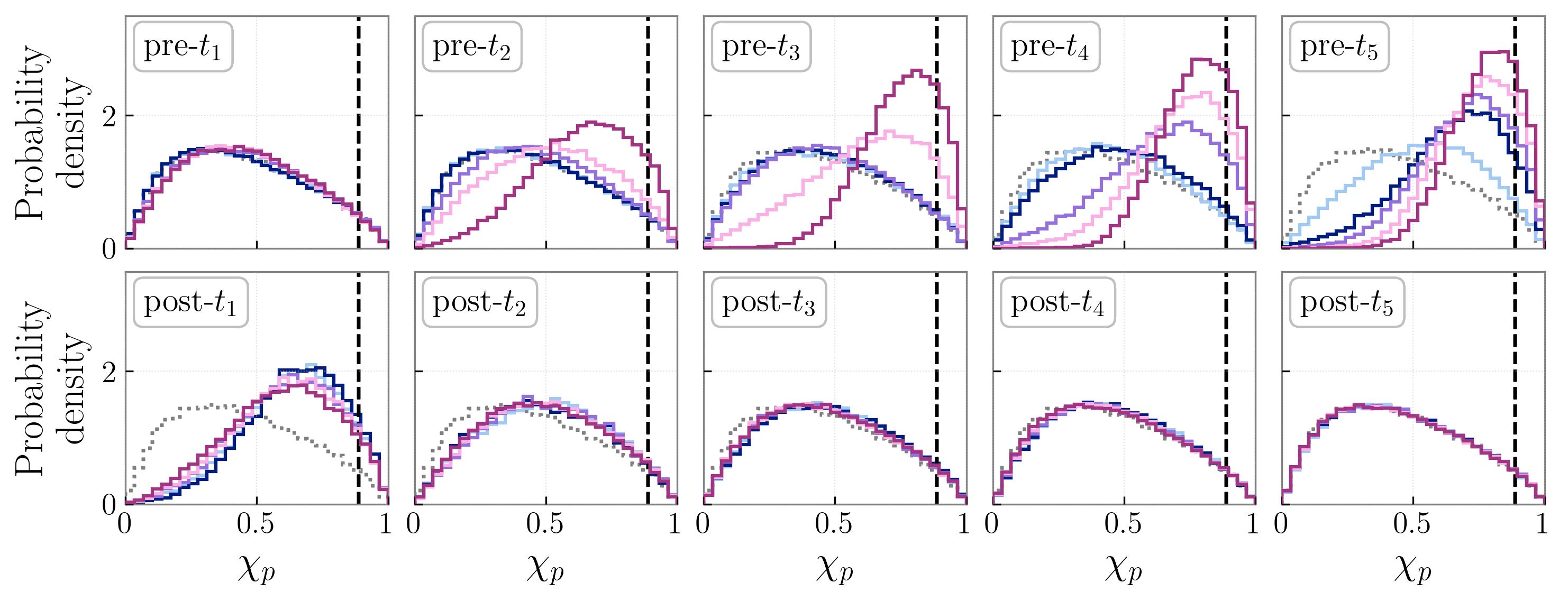}
    \includegraphics[width=\textwidth]{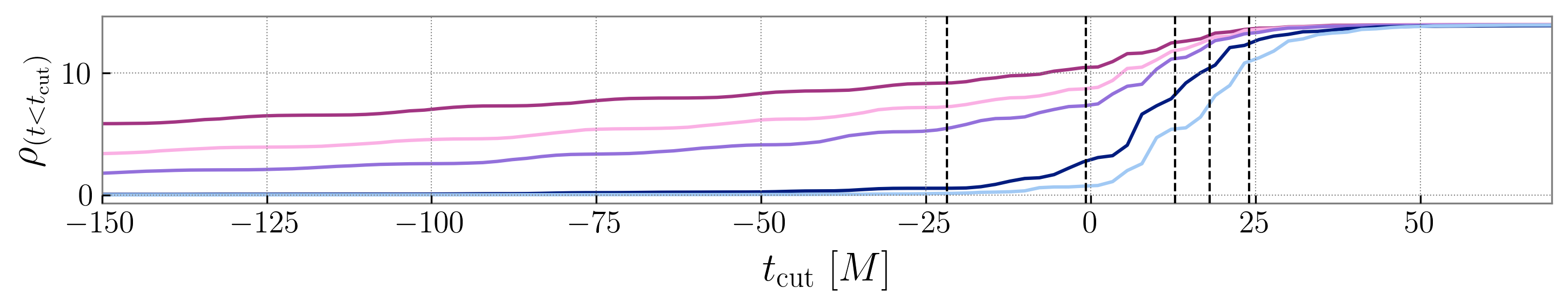}
    \caption{Waveforms and inference results for signals with different detector-frame total masses but the same SNR: $M=80\,\Msun$ (magenta), $100\,\Msun$ (pink), $120\,\Msun$ (purple), $270\,\Msun$ (dark blue), and $500\,\Msun$ (light blue). 
    \textit{First row:} Whitened waveforms in LLO, with the time axis plotted in units of seconds. 
    \textit{Second row:} Whitened waveforms in LLO, with the time axis plotted in units of detector-frame total mass. 
    Since all other parameters are the same, the waveform cycles line up when plotted in units of $M$.
    Cutoff times $t_i \in \{ -21.8,  -0.8, 12.8, 18.0, 24.1\}\,M$, corresponding to specific peaks/troughs of the signals, are labeled with vertical black dashed lines.
    \textit{Third and fourth rows:} $\chip$ posteriors for the pre- and post-cutoff analyses for each signal (corresponding colors to top two rows). 
    We exclude the full-signal analysis results, as they are indistinguishable from the pre-$t_5$ results.
    \textit{Fifth row:} SNR accumulation over time for the different signals.
    See Ref.~\cite{animation_figure04} for an animation of this figure, including more cutoff times and masses.
    }
    \label{fig:different_Mtotal}
\end{figure*}

Having explored the measurability of spin precession in the post-peak (ringdown) data, we now turn to the pre-peak (inspiral) data.
Observing more (less) of the inspiral maps to increasing (decreasing) the binary's merger frequency, which is achieved by decreasing (increasing) the system's detector-frame total mass~\cite{Maggiore:2007ulw}, due to the detector's low-frequency sensitivity cutoff.
For GW190521 and its $\maxL$ waveform, spin precession was not measurable in the inspiral alone---only one or two cycles are present for $t<0\,$M, as can be seen in the third row of Fig.~\ref{fig:real_vs_maxL}.

Figure \ref{fig:different_Mtotal} shows results of measuring spin precession in signals with various total masses, ranging from $M=80\,\Msun$ to $500\,\Msun$.
Corresponding results for the measured total mass and mass ratio for these signals are provided in Appendix~\ref{app:masses}.
Since all scales of the signal depend on the total mass, we here define waveform parameters for the simulated signals and during inference at $t=-100\,M$, ensuring that spin parameters are specified at the same point in the orbital evolution for all masses.%
\footnote{
This is different from other sections, where we map the spins to $t=-100\,M$ in post-processing. The $\maxL$ parameters for GW190521 analyzed at $t_{\rm ref}=-100\,M$ are given in the rightmost column of Table~\ref{tab:GW190521_maxL_params}.}
Defining parameters in this way guarantees that only $M$ (and thus the amount of visible inspiral) varies, while the intrinsic spin configuration remains unchanged~\cite{Biscoveanu:2021nvg, Varma:2019csw}.
We fix the total SNR of the system to that of the $\maxL$ signal ($\rho=14$).
The corresponding merger frequencies are listed in the second column of Table~\ref{tab:inspiral_and_rd_SNR_different_masses}. 
The top row of Fig.~\ref{fig:different_Mtotal} shows the whitened strain in LLO with time in units of seconds. 
Each signal has a different merger frequency (in Hz), and correspondingly a different fraction of SNR in the inspiral versus ringdown.
For example, the $M=80\,\Msun$ signal (magenta) has the longest visible inspiral, while the $500\,\Msun$ signal (light blue) has the longest visible ringdown.
In the second row we repeat this plot but with time in units of $M$.
Again we see that different signals have different amounts of observable inspiral, and additionally, all of the signals' peaks/troughs line up cycle-by-cycle when plotted in these time units.

We again truncate the data at different times and independently analyze before and after each cutoff. 
We use cutoff times of fixed $M$ that correspond to peaks/troughs of the signals (and thus to different numbers of seconds per signal), indicated by the vertical dashed lines in the second row of Fig.~\ref{fig:different_Mtotal} and labeled as $t_i$.
The posteriors for $\chip$ from the pre-cutoff analyses are shown in the third row of Fig.~\ref{fig:different_Mtotal}. 
For reference, pre-$t_2$ corresponds most closely to just the inspiral data being analyzed.
We exclude the full-signal analysis results from the figure, as they are indistinguishable from the pre-$t_5$ results. 
For all cutoff times, the lower the mass, the more informative the pre-cutoff signal is about spin precession.
More visible inspiral for a system with the same intrinsic spin configuration yields a stronger $\chip$ constraint, as \textbf{\textit{precession has a clear imprint on the inspiral}}~\cite[e.g.,][]{Arun:2008kb,Schmidt:2014iyl,Boyle:2014ioa}.
For the lowest-mass signal we analyze ($M=80\,\Msun$; magenta), the $\chip$ posterior is constrained away from the prior when \textit{just} the inspiral is analyzed, with hints also present in the $M=100\,\Msun$ system (pink).

The post-cutoff posteriors are shown in the fourth row of Fig.~\ref{fig:different_Mtotal}.
Although louder pre-peak data yield more informative $\chip$ constraints, the same is not true about the post-peak data at these masses and SNR: here, in contrast to Sec.~\ref{subsec:SNR}, increasing the total mass does \textit{not} make the ringdown notably more informative about spin precession for these test cases, as can be seen in the post-$t_2$ and onward $\chip$ posteriors. 
The signal with the loudest post-peak SNR is $500\,\Msun$ with $\rho_{\rm pp} = 12$, which is consistent with being uninformative with respect to the varying-SNR analyses presented in Sec.~\ref{subsec:SNR}. 
There, the $\chip$ information in the ringdown shows up somewhere between a $\rho_{\rm pp}$ of $17$ and $25$.
It is likely that for our test signals, the ringdown SNR is simply not large enough to measure precession.

Contrasting the pre- and post-cutoff results, the data between $t_1$ and $t_2$, i.e., the last pre-merger cycle, are necessary for measuring $\chip$ in these different-mass signals.
This is true regardless of how many cycles are observable before $t_1$.
For example, we consider the $M=80\,\Msun$ case where precession is the most strongly constrained.
The pre-$t_1$ and post-$t_2$ analyses are both uninformative about $\chip$, while the pre-$t_2$ and post-$t_1$ analyses \textit{are} informative.
The final pre-merger cycle again seems to inform $\chip$ measurements, no matter
the total mass and amount of observable inspiral.
In Sec.~\ref{subsec:phase}, we explore whether the cycle's importance holds true under different extrinsic angular configurations.
While we directly observe that the orbital angular momentum tilts away from the line of sight in GW190521---regardless of total mass---we do not necessarily expect this behavior to be universal.
Still, for GW190521, the importance of the final pre-merger cycle persists at different total masses.

\begin{figure}
    \centering
     \includegraphics[width=\columnwidth]{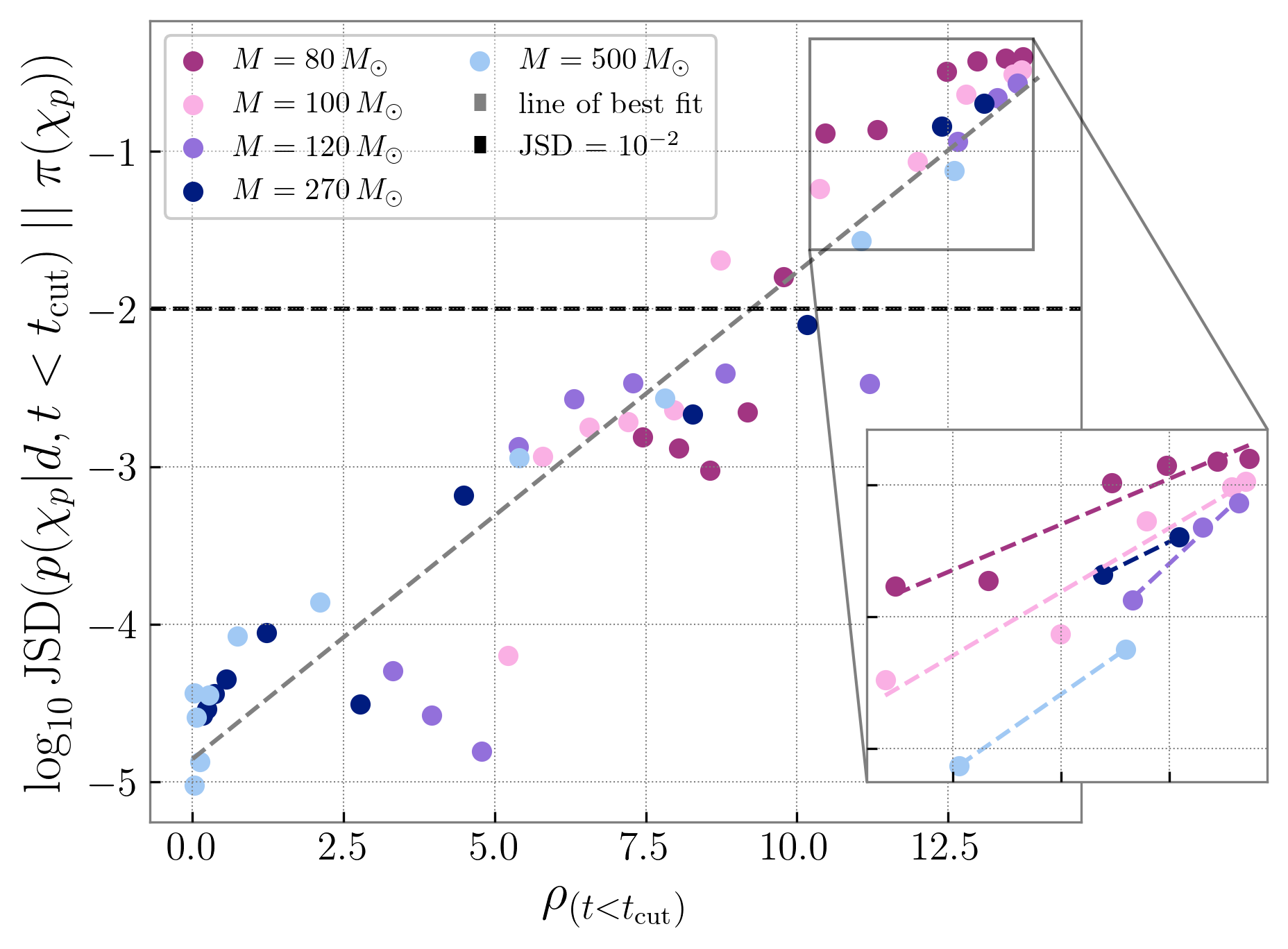}
    \caption{
    JSD between the pre-cutoff $\chip$ posterior and the prior versus the pre-cutoff SNR for different total masses (colors) and a range of cutoff times.
    The gray dashed line is the global line of best fit across all masses.
    The horizontal black dashed line marks the threshold we define for an informative posterior, $\mathrm{JSD} \gtrsim 10^{-2}$.
    The inset zooms in on the informative region, with lines of best fit for each mass individually (colored dashed lines).
    }
    \label{fig:JSD_vs_SNR}
\end{figure}

Even though all signals in Fig.~\ref{fig:different_Mtotal} have the same total SNR, this SNR \textit{accumulates} differently over time.
This is shown in the bottom row, where we plot $\rho_{(t<\tcut)}$: the SNR of the signal before the corresponding time $\tcut$ on the horizontal axis. 
The pre- and post-peak SNRs and their ratios are also given in Table~\ref{tab:inspiral_and_rd_SNR_different_masses}.
The relationship between total mass, pre-peak SNR, and $\chip$ measurement strength is strictly monotonic: the larger the mass, the lower the pre-peak SNR, the less informative the signal is about $\chip$.
To aid in understanding the nature of this correlation, Fig.~\ref{fig:JSD_vs_SNR} shows the JSD between each pre-cutoff $\chip$ posterior and prior versus the pre-cutoff SNR for many cutoff times.
These two quantities are linearly correlated,\footnote{The Pearson correlation coefficient~\cite{Pearson:1895} between the pre-cutoff SNR and the $\log_{10} \mathrm{JSD}$ is $0.95$ with a $p$-value of $10^{-29}$, strongly indicating a linear trend.} with a global line of best fit shown with a gray dashed line.

In other words, the SNR over a region of inspiral data has the \textit{leading-order} effect on the measurability of $\chip$ in that region, following common intuition. 
But as we found in Sec.~\ref{subsec:SNR}, the SNR \textit{alone} is not the full picture. 
The morphology contained within a given region bolsters or weakens precession measurability on top of the dominant effect of SNR, as we highlight in the inset axis of Fig.~\ref{fig:JSD_vs_SNR}.
In the regime where $\chip$ is measurable (above the black dashed horizontal line), the $M=80\,\Msun$ signal is the most informative about precession (highest JSDs) at a fixed pre-cutoff SNR, and the JSDs decrease as total mass increases. 
The fact that, across the board, spin precession constraints weaken with total mass makes the fact that we mostly measure precession in high-mass events in real data~\cite{GWTC3} all the more intriguing.

\begin{table}
    \centering
    \renewcommand{\arraystretch}{1.5}
    \begin{tabular}{ c | c | c | c | c }
          \hline\hline  
          $M$ & Merger $f$ & Pre-peak $\rho$ & Post-peak $\rho$  & $\rho$ ratio \\
          \hline 
          $80\,\Msun$  & 205\,Hz & 10.5 & 8.3  & 1.3\\ 
          $100\,\Msun$ & 164\,Hz & 8.7  & 9.5 &  0.9 \\
          $120\,\Msun$ & 137\,Hz & 7.4  & 10.4 & 0.7 \\ 
          $270\,\Msun$ & 59\,Hz  & 2.9  & 11.6 & 0.3 \\ 
          $500\,\Msun$ & 31\,Hz  & 0.8  & 12.2 & 0.1 \\
        \hline\hline  
        \end{tabular}
        \caption{Merger frequency and SNR information of the signals with different total masses shown in Fig.~\ref{fig:different_Mtotal}.
        For comparison, the $\maxL$ signal has a merger frequency of 56\,Hz.
        The method for extracting the merger frequencies from the data is given in Appendix~\ref{app:measuring_merger_freq}; see Fig~\ref{fig:merger_freq}.    
        The third and fourth columns list the optimal network SNRs for the pre-peak data (inspiral) and post-peak data (ringdown).
        Their ratio is given in the final column.
        The SNR is monotonically increasing/decreasing for the ringdown/inspiral as total mass increases.
        All systems have the same total SNR.
        }
    \label{tab:inspiral_and_rd_SNR_different_masses}
\end{table}
%

\subsection{The extrinsic angles of GW190521 were not finely tuned to measure spin precession}
\label{subsec:phase}

%
\begin{figure*}
    \centering
    \includegraphics[width=\textwidth]{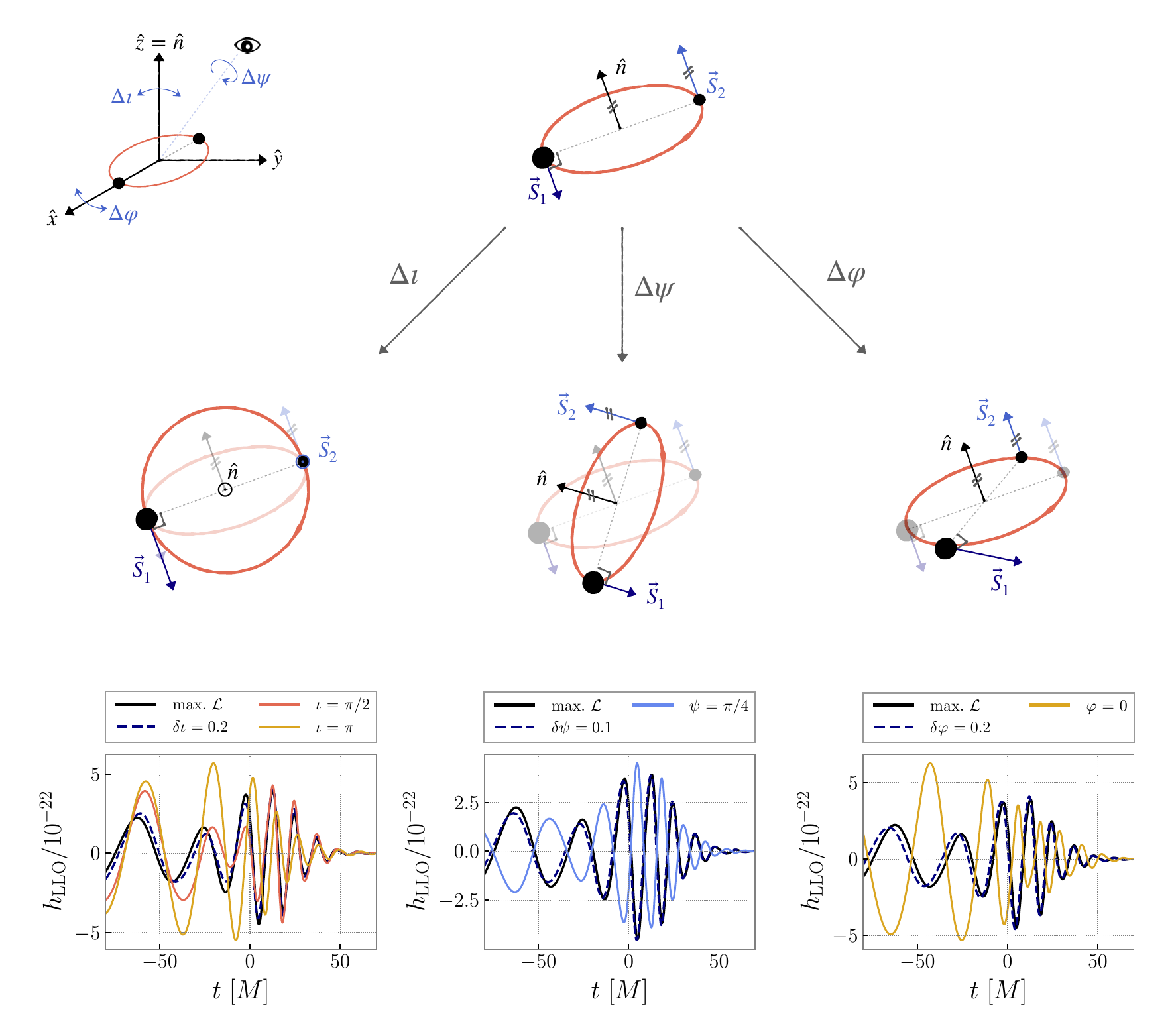}
    \caption{
    Schematic diagram explaining the effect of changing three extrinsic angle parameters on a binary's configuration and the resultant gravitational waveform: inclination $\iota$ (left), polarization $\psi$ (middle), and phase $\varphi$ (right). 
    Changing each angle is equivalent to a rigid rotation of one of the Euler angles, as illustrated in the smaller diagram in the upper left, showing the binary's frame with respect to a stationary observer (eyeball). 
    For the remaining diagrams, the binary is drawn on the plane of the sky, i.e., as viewed from an observer on Earth. 
    The quasicircular orbit of the BBH is drawn in red, the spins $\vec S_i$ of the primary and secondary BHs are in blue, and the normal vector to the orbital plane $\hat n$ is in black. 
    For demonstrative purposes, $\vec S_1$ lies in the orbital plane, perpendicular to the line connecting the BHs, while $\vec S_2$ is parallel to $\hat n$.
    Changing the inclination angle changes the degree to which the binary points toward/away from the viewer.
    This changes the projected ellipticity of the orbit onto the plane of the sky (the true binary is circular). 
    Varying the polarization angle rotates $\hat n$ around the line of sight, leaving the shape of the ellipse projected onto the sky unchanged but altering its orientation.
    Finally, changing the phase angle keeps $\hat n$ fixed, but rotates the orbit around $\hat n$, keeping the spin components rigidly fixed in the binary's source frame.
    Under each diagram, we also show several examples of waveforms where the corresponding angle differs from GW190521's $\maxL$ (black) while the rest are kept constant. 
    We show the effect of a small perturbation of each angle (dark blue dashed), and some extreme cases (yellow, red, and light blue).
    All signals have the same SNR; examples where just the angles are varied, leading to different SNRs, are shown in Fig.~\ref{fig:angles_appendix} in Appendix~\ref{app:angles}.}
    \label{fig:angles_diagram}
\end{figure*}

The measurability of spin precession in GW190521-like signals has been shown to depend sensitively on the exact intrinsic spin configuration of the system~\cite{Biscoveanu:2021nvg}, making GW190521's informative $\chip$ measurement seemingly fine tuned. 
We now explore the impact of the \textit{extrinsic} projection and the orientation of the binary, specifically looking at three angular parameters: \textit{inclination} $\iota$, \textit{polarization} $\psi$, and \textit{phase} $\varphi$.
We keep sky position (right ascension and declination) fixed for all simulated signals and in parameter estimation.
The extrinsic angular parameters do not control a binary's intrinsic dynamics but do impact the \textit{observed} GW signal---we provide equations for quantifying these dependencies in Appendix \ref{app:angles}.

The phase and polarization are not astrophysically significant, meaning that they are expected to be randomly distributed across the binaries in the universe: there is no known astrophysical process or selection effect~\cite{Malmquist:1922,Loredo:2004nn,Mandel:2018mve} that could tune the polarization or phase angle of a binary to specific values with respect to the Earth.
Inclination too is expected to be approximately isotropically distributed in the astrophysical population.%
\footnote{
The only BBH formation environment predicted to yield specific values of $\iota$ is the disks of active galactic nuclei (AGN), where BBHs' $\iota$ are preferentially aligned or anti-aligned with the orbital angular momentum of the disk~\cite[e.g.,][]{McKernan:2019beu,McKernan:2024kpr}.
This produces an astrophysical $\iota$ distribution with two sub-populations: one that is isotropic, and another clustered at specific inclinations of BBH-forming AGN-disks.
Since AGN disk orientations are themselves isotropic with respect to Earth, the positions of these over-densities are also randomly distributed, and have not yet been observed with GW data~\cite{Isi:2023dlk}.
}
However, due to selection effects, the inclination is not isotropic in the \textit{observed} population. At a fixed distance/SNR, we are most sensitive to inclinations of $\iota=0$ (face on) or $\pi$ (face off); see Appendix~\ref{app:angles}. 
The degeneracy between inclination and luminosity distance \cite{Maggiore:2007ulw,Schutz:2011tw,Usman:2018imj,Fairhurst:2017mvj,Fairhurst:2023idl} further impacts detectability. 
Thus, if the extrinsic angles of GW190521 are finely tuned (to non-face on/off configurations)---meaning that even a slight variation would render precession unobservable---that could be an indication that the waveform model is overfitting.
If the true signal lies outside the waveform model’s manifold, inference may yield sharply peaked posteriors in a narrow region of parameter space that best approximates the data~\cite{Okounkova:2022grv}, indicating model inefficiency.
Alternatively, we may have simply been lucky: perhaps only heavy BBHs with a finely tuned configuration lead to an observable precession cycle, and GW190521 happened to be one such case.

Figure~\ref{fig:angles_diagram} provides a schematic illustration of the effect of $\iota$, $\psi$, and $\varphi$ on the observed configuration of a BBH's (non-eccentric) orbit and resultant GW emission. 
The top row depicts an example BBH as it would appear on the plane of the sky to an observer on Earth, with the orbit drawn in red and the individual black hole spins, $\vec S_i$, in blue.
The normal vector $\hat n$ to the binary's orbital plane is shown in black, pointing at an angle out of the page.
In the second row, we sketch how varying $\iota$, $\psi$, and $\varphi$ (from left to right) alters the observed BBH's geometry. 
In the third row, we show the impact of changing the three angles away from GW190521's $\maxL$ configuration (black) on the emitted GW signal projected onto LLO, showing both small perturbations (dark blue dashed) and more extreme deviations (yellow, red, and light blue solid) for each parameter.
Within each column, all waveform parameters but two are kept constant: the varied angle, and luminosity distance (which is scaled such that all signals have the same SNR).
The specific combination of $\iota$, $\psi$, and $\varphi$ plays a key role in determining how much of a precession cycle is visible in a waveform. 

In the following bullet points, we detail how each extrinsic angle affects the BBH orbit and waveform. 
All angles are defined at the same time in the signal ($f=11\,\rm{Hz}$, corresponding to $t=-212\,M$, all systems have the same total mass). 
\begin{itemize}
    \item The \textit{inclination angle} $\iota$ determines the tilt of the binary's orbit relative to the observer's line of sight, i.e., the dot product between $\hat n$ and the line of sight. 
    While the binary's true orbit is circular, changing $\iota$ affects the  ellipticity of the orbit projected onto the plane of the sky. 
    If viewed face-on/off ($\iota=0$ or $\pi$), the projected orbit is a circle (left sketch in the second row of Fig.~\ref{fig:angles_diagram}); if viewed edge-on ($\iota = \pi/2$) it is a line.
    The inclination angle can significantly alter the observed signal
    morphology, as it changes what combination of GW radiation modes are
    observed; see Appendix~\ref{app:angles}.
    For instance, at $\iota = \pi$ (yellow waveform in lower-left subplot of Fig.~\ref{fig:angles_diagram}) the final pre-merger cycle has a large amplitude compared to the merger, whereas $\iota = \pi/2$ (red waveform) yields an extreme suppression of this same cycle.
    \item Changes in the \textit{polarization angle} $\psi$ rotate the projected orbit on the plane of the sky without altering its shape/ellipticity, shifting the waveform’s polarization state~\cite{Isi:2022mbx}.
    This is shown in the center diagram of the second row of Fig.~\ref{fig:angles_diagram}.   
    Crucially, this rotation does not make the $\hat n$ point more toward or away from the observer, but rather changes its azimuthal angle around the observer's line of sight. 
    In other words, changes in $\psi$ keep the dot product between $\hat n$ and the line of sight constant.
    Additionally, varying $\psi$ fully preserves the complex amplitude envelope of the waveform (Fig.~\ref{fig:angles_appendix}), but shifts the timing of the peaks and troughs of the strain, making it the most intuitive form of a ``phase shift" for precessing BBHs.
    \item Finally, variations in the so-called \textit{phase angle} $\varphi$ rotate the binary's orbit \textit{around} $\hat n$---equivalently, rotate the BHs along their fixed orbit---as shown in the right-hand diagram of the second row of Fig.~\ref{fig:angles_diagram}.
    Importantly, this rotation keeps the spin components rigidly fixed in the binary's source frame, i.e., the Cartesian components of the spins are the same in the frame of the source, where the $x$-axis is defined as the line from $m_2$ to $m_1$ at the reference time, and the $z$-axis is along $\hat n$~\cite{Varma:2019csw,Varma:2021csh}.
    In precessing BBHs---where higher-order modes are significant---$\varphi$ plays a more complicated role than the simple phase shift from which it derives its name, which only applies in the case of non-precessing, equal-mass systems; see Appendix~\ref{app:angles}.
    Like the inclination angle, $\varphi$ can significantly alter the signal morphology for precessing systems, as exemplified by the yellow waveform in the lower right-hand plot of Fig.~\ref{fig:angles_diagram}.
\end{itemize}
As mentioned, Appendix \ref{app:angles} provides further details about the role of inclination, polarization, and phase in precessing systems by giving equations. 
There we also contrast the impact of the phase angle on precessing systems versus with equal-mass non-spinning systems.

\begin{figure*}
    \centering
    \includegraphics[width=\textwidth]{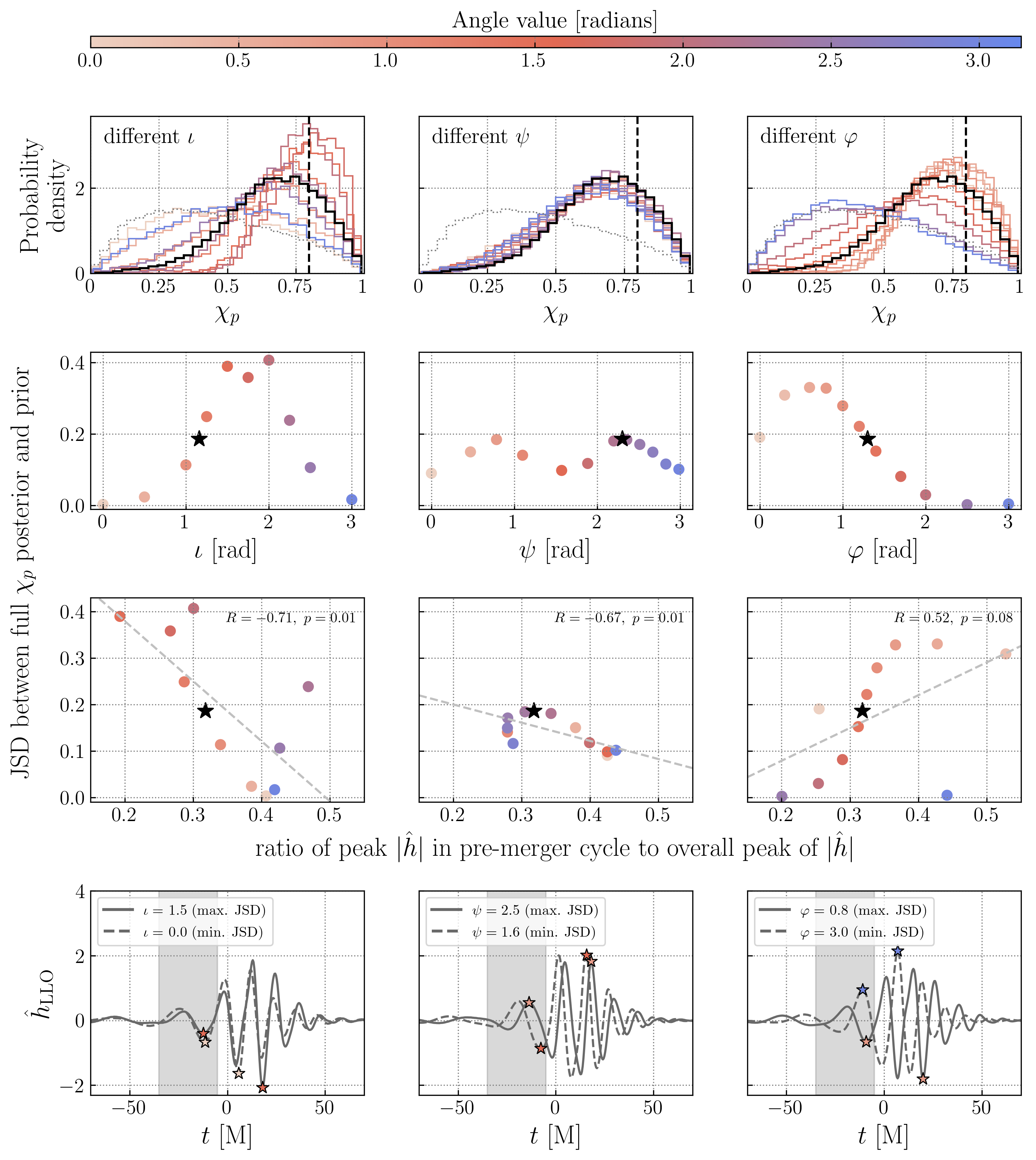}
    \caption{
    Results from full-duration signal analyses where one extrinsic angle at a time is changed from GW190521's $\maxL$ configuration: inclination angle $\iota$ (left column), polarization angle $\psi$ (middle column), and phase angle $\varphi$ (right column).
    (\textit{First row}) $\chip$ posteriors for all signals with varied angles (colorbar) compared to the $\maxL$ posterior (black) and prior (gray dotted).
    (\textit{Second row}) JSD between the marginalized $\chip$ posteriors and the prior as a function of each angle.
    GW190521's $\maxL$ signal is indicated with a black star.
    (\textit{Third row}) JSD versus the ratio of the peak whitened strain in the pre-merger cycle to the overall signal strain peak.
    In gray-dashed are lines of best fit, intended to visualize the net trend strength and direction. 
    Pearson $R$ correlation coefficients~\cite{Pearson:1895} and associated $p$-values are in each upper right corner.
    (\textit{Fourth row}) Comparisons of the whitened strain in LLO of signals which are the most informative (maximum JSD, solid) versus least informative (minimum JSD, dashed) about $\chip$.
    Stars show the values extracted to calculate the $x$-axis quantity in the third row.
    }
    \label{fig:JS_divs_different_angles}
\end{figure*}

Figure~\ref{fig:JS_divs_different_angles} shows results from  full signals with a range of inclinations (left column), polarizations (middle column), and phases (right column).
All signals are scaled to have the same full-signal SNR as GW190521's $\maxL$, and---aside from the one differing angle---share its other parameters. 
In inference, only the sky position, time of coalescence, and polarization are fixed to their injected values, as was true in the analyses of the previous sections.
The top row shows $\chip$ posteriors for each signal (various colors), compared to GW190521's $\maxL$ posterior (black) and the prior (gray dotted).
The second row plots the JSD between each posterior and the prior as a function of the varied angle.
With all other parameters fixed to GW190521's $\maxL$ configuration at $\rho=14$, we are most sensitive to precession at $\iota \sim 1.5-2$ radians, $\psi\sim0.8$ or $2.2$ radians, and $\varphi \sim 0.8$ radians.
\textbf{\textit{GW190521's $\maxL$ reconstruction is \textit{not} the most informative about spin precession over all the extrinsic angle configurations}}. 
GW190521's $\maxL$ signal is tied for most informative over the polarization angle (although the JSD variation between the different values of $\psi$ is minute), and falls squarely in the middle over inclination and phase angle. 
\textbf{\textit{This configuration---which is very similar to the real GW190521 data---is not exceptionally fine-tuned to measure spin precession}}.

Each angle affects the measurability of $\chip$ in different ways. 
In our test cases, the polarization angle $\psi$ does not significantly impact the $\chip$ constraints for the full signal,\footnote{While our inference was conducted with $\psi$ fixed, we have verified that the results remain robust when varying it: we perform inference for a selection of test cases while varying $\psi$ and find results unchanged.} whereas $\iota$ and $\varphi$ each have much more dramatic effect, yielding $\chip$ posteriors that range from uninformative ($\mathrm{JSD} \sim 10^{-3}$) to being very informative ($\mathrm{JSD}\sim0.4$), even though all signals have the same SNR.
The strong influence of $\iota$ is expected~\cite{Biscoveanu:2021nvg,Xu:2022zza}.
Sensitivity to $\varphi$ is also reasonable, given its impact on signal
morphology (Fig.~\ref{fig:angles_diagram}), stemming from its effect on the
superposition of higher-order
modes~\cite{Varma:2014jxa,CalderonBustillo:2015lrt,Varma:2016dnf} (Appendix \ref{app:angles}).
The polarization angle has been suggested to affect the measurability of precession for certain inclinations, based on its impact on the precessing SNR~\cite{Fairhurst:2019vut,Xu:2022zza,Green:2020ptm}.
\citet{Xu:2022zza} find that the precessing SNR is sensitive to $\psi$ (cf. their Fig.~8), although that study was based on a different GW190521-like signal with a different value of $\varphi$ and different intrinsic spin configuration. 
The individual effects of extrinsic angles are difficult to disentangle; varying them independently while holding others fixed, such as when calculating the precessing SNR, fails to fully capture their joint influence across the full parameter space.
As a result, polarization may impact $\chip$ measurability in some inclination-phase configurations (as in \citet{Xu:2022zza}), but not in others (as seen in our results). 
Ultimately, it is the position within the higher-dimensional space of extrinsic (and intrinsic) parameters that governs the influence on precession measurability~\cite{Biscoveanu:2021nvg}.

Next, we evaluate how precession measurability varies with signal strength at different times.
We focus on the relationship between the pre-merger versus merger strain amplitude.
\citet{Miller:2023ncs} found an anti-correlation between the pre-merger strain
amplitude and $\chip$ measurability, which we now test across the various extrinsic angle
configurations.
The third row of Fig.~\ref{fig:JS_divs_different_angles} shows the ratio of the signal's peak whitened strain amplitude in the final pre-merger cycle to that in the merger, plotted against the JSD between the full $\chip$ posterior and prior. 
The peak merger amplitude (denominator of ratio) is the maximal value of absolute value of the whitened strain $|\hat h|$ across all times.
Defining the final pre-merger is not precise nor universal, as it involves deciding the time at which the inspiral ends and merger begins.
In Fig.~\ref{fig:JS_divs_different_angles}, the peak final pre-merger strain amplitude (numerator of ratio) is calculated as the first local maxima of $|\hat h|$ under the conditions $t < -5\,M$ (shaded gray band) and $|\hat h| < 1$.  
We choose $-5M$ as a heuristic time for defining the end of the final pre-merger cycle because it is before peak emission but not so far into the inspiral that the signal becomes unobservable for GW190521's mass.
The $|\hat h|$ cap of $1$ removes ambiguity for several cases (e.g., $\varphi=0$) whose final-pre merger cycle seems, by eye, to be pushed even earlier.
In our accompanying data release~\cite{suppl_figs}, we repeat the analysis for two additional choices of both the transition time between inspiral/merger and the upper cap for $|\hat h|$ in the pre-merger.

The relationships between $\chip$ informativeness and the relative strength of the pre-merger cycle's amplitude are not linear.%
\footnote{
The strain ratio and JSD have Pearson $R$ correlation coefficients~\cite{Pearson:1895} of $-0.71$, $-0.67$, and $0.52$ when varying inclination, polarization, and phase respectively. The first two relationships have $p$-values of 0.01, indicating that the linear fit is good. For phase, the $p$-value is 0.08, indicating the linear relationship is insufficient to describe the data. 
This can all be confirmed by eye, looking at the third row of Fig.~\ref{fig:JS_divs_different_angles}.
}
The best-fit lines (gray dashed) are intended to illustrate first-order trends, but clearly do not capture the underlying correlational structure.
If the findings for GW190521 are generic, we would expect a consistent negative correlation, meaning smaller strain amplitude ratios correlate with more informative $\chip$ posteriors. 
However, \textbf{\textit{the relationship between $\chip$ and the ratio of pre-merger to overall strain is not generic}}.
The trends are in general weak, especially when varying $\psi$, and are not
consistently negative. 
The overall trend found when varying $\varphi$ is positive (although not well fit by a line), the opposite to what one might have expected.
The bottom row of Fig.~\ref{fig:JS_divs_different_angles} highlights the signals that are the most (solid lines) and least (dashed lines) informative about $\chip$.
The stars mark the pre-merger and overall strain values whose ratio is plotted in the third row. 
Interestingly, in these extreme cases, the trend from GW190521 \textit{does} hold: the most informative signals exhibit a more suppressed pre-merger cycle (gray shaded region) than the least.
This is particularly curious for $\varphi$ where the overall trend is not negative---the $\varphi = 3$ signal is an outlier in its uninformative $\chip$ measurement which we believe relates the fact that it has a cycle at $t\sim 5\,M$ with a peak amplitude in LLO right below our threshold of $|\hat h| < 1$~\cite[cf.][]{suppl_figs}.

\begin{figure*}
    \centering
    \includegraphics[width=0.9\textwidth]{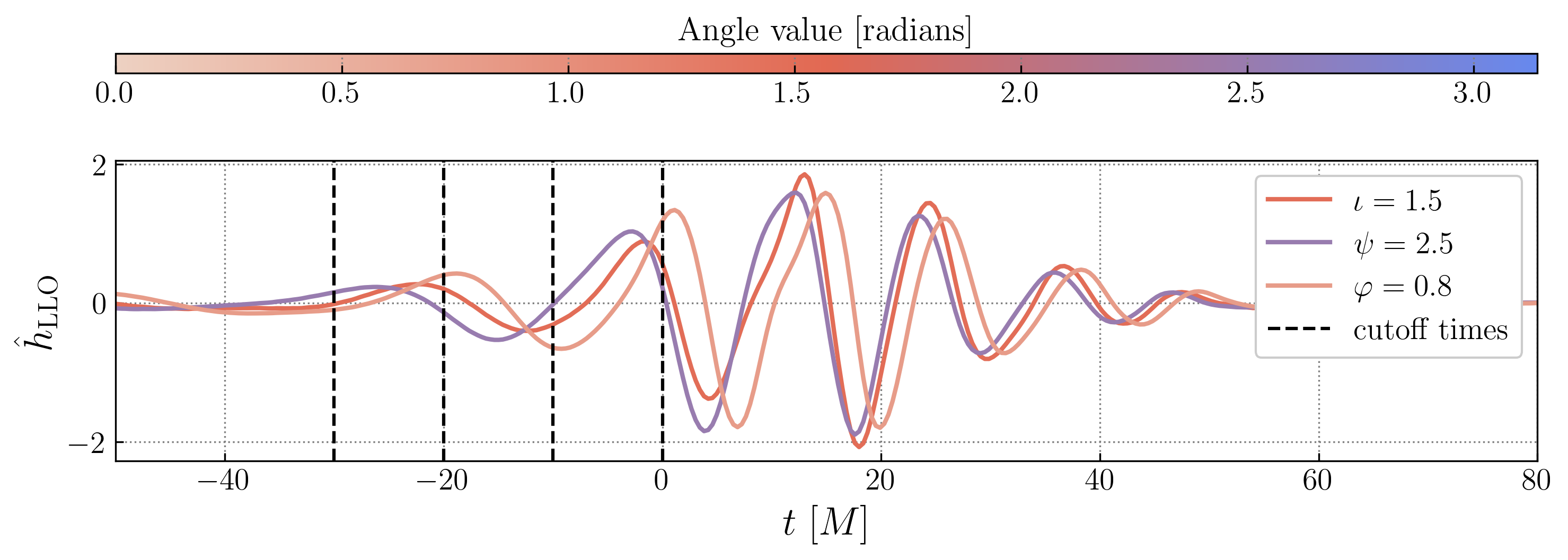}
    \includegraphics[width=0.9\textwidth]{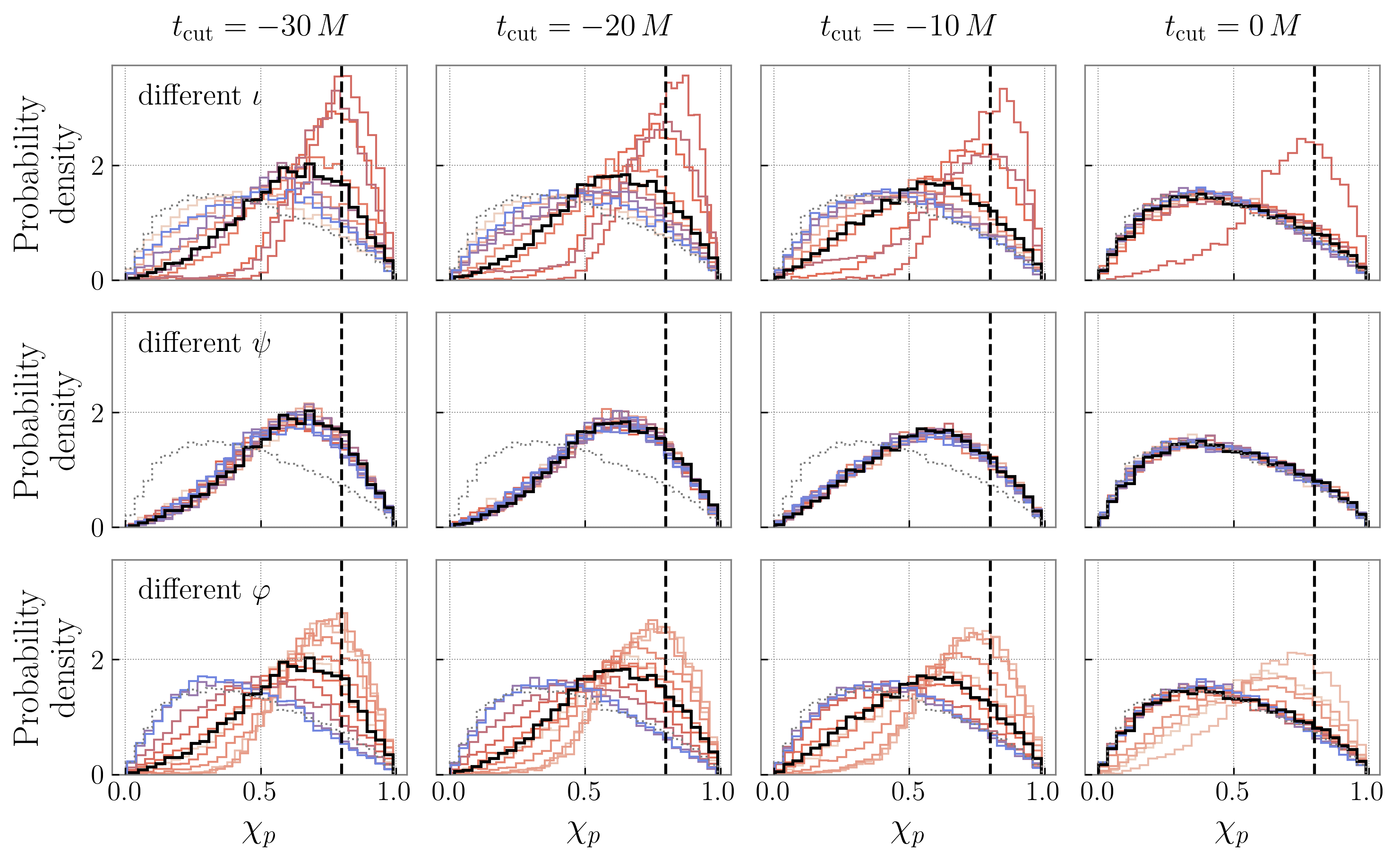}
    \caption{
    Results from post-cutoff analyses for signals with various extrinsic angles. 
    (\textit{First row}) Waveforms from the most informative variation of each angle, as identified in Fig.~\ref{fig:JS_divs_different_angles}, with cutoff times shown for reference (black dashed lines). 
    All signals have the same total mass.
    (\textit{Second through fourth rows})
    $\chip$ posteriors for the post-cutoff analyses where $\tcut = -30\,M$ (first column), $-20\,M$ (second column), $-10\,M$ (third column), and $0\,M$ (fourth column). 
    Each row shows posteriors from simulated signals where one extrinsic angle is changed from the $\maxL$ injection (black): inclination angle (top row), polarization angle (middle row), and phase angle (bottom row).
    The color bar indicates the value of the angle that is varied from the $\maxL$ configuration.
    The $\chip$ prior is shown in gray dotted lines. 
    We here only show post-cutoff analysis results, as the pre-cutoff $\chip$ posteriors for all signals at the selected $\tcut$ values are consistent with the prior.
    }
    \label{fig:different_angles_time_evolution}
\end{figure*}

Having studied how the extrinsic phases matter for spin precession measurability, we now look further into what part of the signal impacts these measurements by turning to the pre-/post-cutoff-time $\chip$ results.
Figure~\ref{fig:different_angles_time_evolution} shows post-cutoff results for $\tcut \in \{-30,-20,-10,0\}\,M$ (left to right columns). 
The pre-cutoff results at these times for all angular configurations are consistent with the prior and are, therefore, omitted.
The whitened waveforms in LLO for the most informative variation of each angle from the full-signal analyses are shown in the top row for reference, with the cutoff times marked with black dashed lines.
Results with varied $\iota$, $\psi$, and $\varphi$ are given in each of rows two through four.

Just like for the full signal, varying $\psi$ does not affect the post-cutoff posteriors (third row of Fig.~\ref{fig:different_angles_time_evolution}).
Against our intuition, all posteriors evolve identically despite differences in the timing of peaks and troughs of the strain.
Here we are cutting at fixed \textit{times} in the strain, which do not correspond to a fixed number of cycles from merger for different angular configurations.
The polarization angle does not change the signal's amplitude envelope (Appendix \ref{app:angles}). 
Thus, we posit that the \textit{amplitude envelope may matter more for constraining precession than the precise timing or values of strain maxima and zero-crossings}.
Furthermore, the polarization angle does not affect the evolution of a binary's inclination over time~\cite{suppl_figs}, meaning it does not change what part of a precessional cycle is visible in the data, complicating our interpretation from GW190521~\cite{Miller:2023ncs} of what underlying physics leads to precession being measurable.

On the other hand, $\iota$ and $\varphi$ matter substantially for $\chip$ evolution(second and fourth rows of Fig.~\ref{fig:different_angles_time_evolution}), and the evolution of inclination over time~\cite{suppl_figs}.
\textbf{\textit{A subset of signals with specific values of $\iota$ and $\varphi$ have ringdowns}} (post-peak data, fourth column of Fig.~\ref{fig:different_angles_time_evolution}) \textbf{\textit{that are informative about $\chip$ alone, even at the same SNR as GW190521.}}
At GW190521's SNR and intrinsic spin configuration, observable amplitude modulations in the inspiral due to changing the direction of GW emission are not prerequisite to measuring precession---it can instead be observed by its imprint via mode mixing on the ringdown alone~\cite{Zhu:2023fnf}.

\section{Conclusions}\label{sec:conclusion}

Ever since the detection of GW190521, there has been a need to understand how spin precession can be imprinted in the signals from heavy compact binaries, which are only observable in current ground-based gravitational-wave detectors for a few cycles.
\citet{Miller:2023ncs} used time-domain inference to localize the measurement of precession in GW190521: it originates from a subtle interaction between the pre- and post-peak signal, rather than from any isolated segment of the data, e.g., inspiral or ringdown alone.
In this paper, we built on \citet{Miller:2023ncs} and explored the morphological imprints of spin precession in simulated GW190521-like (high-mass, highly-precessing) GW signals using time-domain inference.
The goal was to understand whether the features observed in GW190521 can be replicated in simulated signals and whether its configuration was special in any regard, building on past studies \cite{Biscoveanu:2021nvg,Xu:2022zza}.
We focused on the effects of SNR, total mass, and extrinsic angular parameters on the measurability of spin precession over a range of time segments, with an emphasis on pre- and post-peak data.

We first showed that GW190521's behavior is indeed mimicked by its $\maxL$ injection, for both full and truncated data (Fig.~\ref{fig:real_vs_maxL}), and that as we increase the SNR of this signal precession becomes, as expected, better constrained (Fig.~\ref{fig:different_SNRs}).
When varying the total mass at the same SNR, we found that signals from lighter systems yield more informative $\chip$ posteriors. 
Precession is known to be measurable in the inspiral \cite[e.g.,][]{Arun:2008kb,Schmidt:2014iyl,Boyle:2014ioa}, so this trend is expected, as lower-mass systems have a longer visible inspiral. 
For a low enough total mass ($M\lesssim100\,\Msun$), precession is measurable when \textit{only} the inspiral data are visible (Fig.~\ref{fig:different_Mtotal}).
For higher-mass events, precession can become more or less apparent for certain angular configurations.
We find that the ability to measure spin precession in GW190521-like signals is sensitive (although \textit{not fine-tuned}) to the extrinsic angular configuration.
Moreover, the specific waveform cycles that inform the $\chip$ measurement differ as we change the system's angular configuration.
While there are more than a few combinations of inclination, polarization, and phase that lead to $\chip$ being constrained away from the prior, specific values of the inclination and phase angles do render precession unmeasurable (Fig.~\ref{fig:JS_divs_different_angles}).
At GW190521's $\maxL$ inclination and phase, the polarization angle does not strongly affect $\chip$ constraints for either the full or any pre-/post-cutoff data, which may be explained by their shared amplitude envelope. 

At high enough SNRs, we can measure precession in the ringdown alone using an IMR waveform model (Fig.~\ref{fig:different_SNRs}).
When the post-peak SNR is sufficiently large ($\rho_{\rm pp}\sim 100$), precession information remains in the signal up to ${\sim} 20\,M$ after the peak (Fig.~\ref{fig:different_RD_SNRs}).
Past work has explored this same idea with QNM ringdown models, for both for GW190521 itself~\cite{Siegel:2023lxl, Capano:2021etf,Capano:2022zqm,GW190521_astro} and various precessing numerical relativity simulations~\cite{Hamilton:2021pkf, Hamilton:2023znn, Finch:2021iip, OShaughnessy:2012iol, Zhu:2023fnf} rather than an IMR waveform model like {\sc NRSur7dq4}. 
The fact that the ringdown has imprints of precession is not unexpected for sufficiently high SNRs.
\citet{Zhu:2023fnf} found that, for near-equal-mass binaries, the ratio between the amplitudes of the 200:220 and 210:220 modes are correlated with the remnant spin misalignment angle---a proxy for spin-orbit misalignment and, therefore, precession.
In this vein, \citet{Siegel:2023lxl} found that the GW190521 ringdown data are consistent with a significant 210 QNM amplitude, which would be unexpected for a non-precessing BBH.

Even at GW90521's mass and SNR of $\sim 14$, certain combinations of inclination and phase render precession measurable in the ringdown (Fig.~\ref{fig:different_angles_time_evolution}).
This finding is in agreement to the complementary QNM studies discussed above~\cite[e.g.,][]{Zhu:2023fnf,Siegel:2023lxl}: the information about spins before the merger is expected to be encoded in not just the observed QNM complex frequency spectrum---which suffers from degeneracies inherent to the map from inspiral to remnant properties---but also these \textit{modes' relative amplitudes and phases}.
A direct consequence of the no-hair conjecture~\cite[e.g.,][]{Israel:1967} in General Relativity is that complex ringdown frequencies are uniquely determined by mass and spin; multiple sets of progenitor properties can yield the same BH remnant, and thus the same ringdown frequencies.
Even though these degeneracies may explain why, for example, we are able to infer precession from an SNR $\sim 10$ inspiral (Fig.~\ref{fig:different_Mtotal}) but not an SNR $\sim 10$ ringdown (Fig.~\ref{fig:different_RD_SNRs}) generated by a BBH with the same spin and angular configuration, they do not entirely govern the circumstances that render precession's imprint informative.
Which ringdown modes manifest and with what complex amplitudes are determined by the masses and spin magnitudes \textit{plus} the BBH's full spin/angular configuration~\cite{Mitman:2025hgy}.
Modeling the imprint of spin precession and other inspiral dynamics on the ringdown is an area of active research~\cite{Capano:2021etf,Cheung:2023vki, Pacilio:2024tdl,MaganaZertuche:2024ajz,MaganaZertuche:2024ajz,Hamilton:2021pkf,Hamilton:2023znn,Finch:2021iip,Mitman:2025hgy}, and a closer comparison with NR surrogate findings is warranted in a future study.

\citet{Biscoveanu:2021nvg} and \citet{Xu:2022zza} have both conducted complementary studies on measuring spin precession in high-mass BBHs, both using frequency-domain inference. 
\citet{Biscoveanu:2021nvg} explored various total mass and spin configurations at selected mass ratios, inclinations, and SNRs. 
They found that precession measurability is sensitive to the \textit{intrinsic} spin dynamic, such as spin tilt angles. 
Furthermore, they concluded that the uncertainty in spin measurements increases with the total mass, as do we. 
They simulated signals and measured spins at a fixed frequency in Hz rather than a fixed time in units of $M$, meaning their different mass signals \textit{also} had different intrinsic spin configurations. 
When comparing results with spins fixed at the same reference time, they found a stronger trend (cf. their Fig.~11), which is even more in agreement with ours.
Formulating our results in terms of the symmetric 90\% credible interval width ($\rm{CI_{90}}$) used in \citet{Biscoveanu:2021nvg} to assess a posterior's precision yields analogous results: between a total mass of $80$ and $250\,\Msun$, the $\rm{CI_{90}}$ for full-signal $\chip$ posteriors increases by $\sim 0.2$.
In terms of the extrinsic angles, they also observed improved $\chip$ measurement at higher inclinations (1.5 radians vs. 0.5 radians), consistent with our findings.

\citet{Xu:2022zza} examined the role of detector noise in GW190521-like signals, concluding that it impacts spin precession measurability in the expected way. 
They also analyzed variations in intrinsic spin configuration, SNR, and inclination/polarization.
They suggested that spin precession is not measurable in the full signal until SNR 45, whereas we found that at this SNR spin precession is measurable in the ringdown alone, and measurable in the full signal at SNR 14. 
In their test cases, they found that inclination matters less and polarization matters more than we do.
Together, these studies and ours highlight the dependence of spin precession measurability on multiple interdependent parameters.

More broadly, we found that not just the SNR in a region of a signal matters for constraining precession, but also the specific signal morphology contained within that region (Figs.~\ref{fig:different_RD_SNRs} and \ref{fig:JSD_vs_SNR}).
Furthermore, the morphological features needed to measure spin precession are not the same for all high-mass signals. 
We do not find that a loud merger and quiet pre-merger cycle consistently drive spin precession constraints in high-mass signals, as was the case for GW190521. 
While precession inference sometimes hinges on this morphology, the informativeness of a $\chip$ posterior is not universally anti-correlated with the ratio of the signal's pre-merger to overall amplitude peak (Fig.~\ref{fig:JS_divs_different_angles}).
As we vary SNR, total mass, and the extrinsic angles, different signals lose information about $\chip$ at different times in their evolution.

Sometimes, specific waveform features directly correlate with physical properties, e.g., the merger frequency and total mass, or the ratio of the 221:220 ringdown mode amplitudes and the angle between the remnant BH's spin and the progenitor's angular momentum vector~\cite{Zhu:2023fnf}.
We tested whether an analogous universal feature links $\chip$ and the late inspiral and merger strain amplitudes, but did not identify any such relationship.

Our results show that, while not fine-tuned, our ability to infer precession is \textit{sensitive} to a BBH's total mass and angular parameters at current detector sensitivity.
While this is not a selection effect in the traditional sense—as it does not impact the \textit{detectability} of events—it does analogously influence the \textit{measurability} of the associated astrophysical phenomena. 
This sensitivity necessitates hand selecting a small number of highly precessing events on which to perform follow-up studies such as this.
We may be mis-identifying other events as non-precessing due to imprecise measurements.
Perhaps there is an observed GW whose spin precession \textit{would} have confidently been observed if the system had a different extrinsic angular configuration. 
Looking to the future, ground-based GW detectors will become increasingly sensitive, alleviating these concerns.

In LIGO-Virgo-KAGRA's upcoming fifth observing run (``$A_+$ sensitivity") and future ``$A_\sharp$ sensitivity,"GW190521's SNR would be $\sim 40$ and $\sim 80$ respectively~\cite{Aplus_sensitivity, Asharp_sensitivity}.
In Cosmic Explorer, it would increase to $\sim 450$~\cite{CE_sensitivity}. 
This means that, had this event been observed at even $A_+$ sensitivity, precession would have been constrained in GW190521's post-peak data alone. 
Looking ahead, if precessing BBHs are abundant in the cosmos—as theoretical models suggest—we will be able to identify that they are indeed precessing more robustly, opening the door to a stronger understanding of how and where BBHs form.


\section*{Code and Data Availability}

The code {\tt tdinf~\cite{tdinf}} was written by the same authors as this manuscript and was used to perform all of the time-domain analyses presented in this work.
Notebooks to generate all the figures appearing in the text are available on Github at Ref.~\cite{github_release}.
The repository additionally contains supplemental figures~\cite{suppl_figs} and animations showing results from more cutoff times and simulated signals:
\begin{itemize}
    \item Ref.~\cite{animation_figure01}: Animation of $M$, $q$, and $\chip$ posteriors for GW190521 and its $\maxL$ injection, similar to Fig.~\ref{fig:real_vs_maxL}.
    \item Ref.~\cite{animation_figure02}:  Animation of $\chip$ posteriors for different SNR signals, similar to Fig.~\ref{fig:different_SNRs}.
    \item Ref.~\cite{animation_figure04}:  Animation of $\chip$ posteriors for different total mass signals, similar to Fig.~\ref{fig:different_Mtotal}.
\end{itemize}
Data necessary to reproduce figures and animations are available on Zenodo at Ref.~\cite{zenodo_release}. 
The data release includes posteriors for $M$, $q$, $\chieff$, and $\chip$ for each analysis. 
Additional data are available upon request.


\acknowledgements

We thank Harrison Siegel, Rhiannon Udall, Eliot Finch, and Will Farr for helpful discussions about ringdown analyses and parameter estimation, as well as Barry McKernan and K.~E.~Saavik Ford for insights into the expected astrophysical BBH inclination angle distribution.
SJM and KC were supported by NSF Grants PHY-2308770 and PHY-2409001.
The Flatiron Institute is a division of the Simons Foundation.
VV acknowledges support from NSF Grant No. PHY-2309301 and UMass Dartmouth's
Marine and Undersea Technology (MUST) Research Program funded by the Office of
Naval Research (ONR) under Grant No. N00014-23-1-2141.
SH was supported by the National Science Foundation Graduate Research Fellowship under Grant DGE-1745301.
This material is based upon work supported by NSF's LIGO Laboratory which is a major facility fully funded by the National Science Foundation.
The authors are thankful for LIGO Laboratory computing resources, funded by the National Science Foundation Grants PHY-0757058 and PHY-0823459, as well as the Hawk computing cluster provided by Cardiff University and supported by STFC Grants ST/I006285/1 and ST/V005618/1.
This research has made use of data or software obtained from the Gravitational Wave Open Science Center (gwosc.org)~\cite{GW190521_gwosc,opendata_O3}, a service of the LIGO Scientific Collaboration, the Virgo Collaboration, and KAGRA. 
This material is based upon work supported by NSF's LIGO Laboratory which is a major facility fully funded by the National Science Foundation, as well as the Science and Technology Facilities Council (STFC) of the United Kingdom, the Max-Planck-Society (MPS), and the State of Niedersachsen/Germany for support of the construction of Advanced LIGO and construction and operation of the GEO600 detector. Additional support for Advanced LIGO was provided by the Australian Research Council. Virgo is funded, through the European Gravitational Observatory (EGO), by the French Centre National de Recherche Scientifique (CNRS), the Italian Istituto Nazionale di Fisica Nucleare (INFN) and the Dutch Nikhef, with contributions by institutions from Belgium, Germany, Greece, Hungary, Ireland, Japan, Monaco, Poland, Portugal, Spain. KAGRA is supported by Ministry of Education, Culture, Sports, Science and Technology (MEXT), Japan Society for the Promotion of Science (JSPS) in Japan; National Research Foundation (NRF) and Ministry of Science and ICT (MSIT) in Korea; Academia Sinica (AS) and National Science and Technology Council (NSTC) in Taiwan.

Software: \texttt{tdinf}~\cite{tdinf}, \textsc{emcee}~\cite{emcee}, \textsc{LALSuite}~\cite{lalsuite}, 
\textsc{PyCBC}~\cite{pycbc}, 
\textsc{gwpy}~\cite{gwpy},
\textsc{numpy}~\cite{numpy}, \textsc{scipy}~\cite{scipy},
\textsc{h5py}~\cite{h5py}, \textsc{h5ify}~\cite{h5ify}, \textsc{matplotlib}~\cite{matplotlib}, \textsc{seaborn}~\cite{seaborn}, \textsc{ringdown}~\cite{Isi:2019aib, Isi:2021iql}, \textsc{gwtools}~\cite{gwtools}, \textsc{pandas}~\cite{pandas1,pandas2}.

\appendix

\section{Measuring more mass and spin parameters in the post-peak data and exploring conditional priors}
\label{app:other_params_in_RD}

\begin{figure*}
    \centering
    \includegraphics[width=0.85\textwidth]{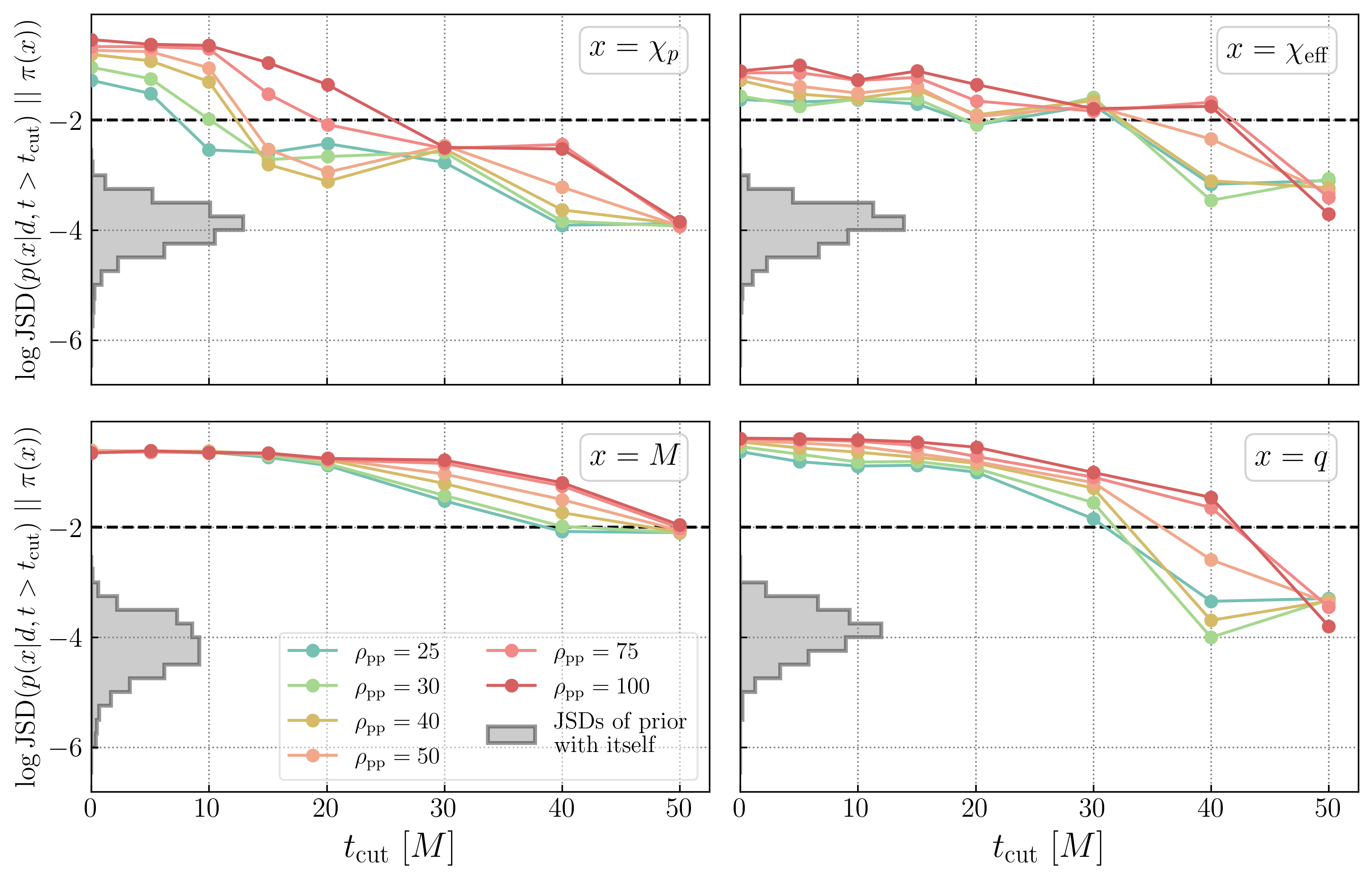}
    \caption{Jensen Shannon Divergence ($\log_{10}$ JSD) between the posterior $p(x|d,t>\tcut)$ and prior $\pi(x)$ for $x\in \{\chip, \chieff, M,q\}$ from data after different cutoff times $\tcut$ (horizontal axis) for different signals with different post-peak SNRs ($\rho_{\rm pp}$; different colors). 
    Distributions of the JSD between two sets of $N=3000$ samples from the marginal prior for each parameter are shown in gray histograms; our conservative benchmark for measurability of $\log_{10} \rm JSD =-2$ is indicated by the black dashed line.
    See Fig.~\ref{fig:different_RD_SNRs} and the corresponding text for more details.
    The top left panel is the same as the bottom panel of Fig.~\ref{fig:different_RD_SNRs} except in log-space.  
    }
    \label{fig:mass_in_RD}
\end{figure*}

Section~\ref{subsec:SNR} explores how late into the post-peak signal $\chip$ can be measured at different SNRs by computing the JSD between the post-cutoff posterior and prior as the cutoff time $\tcut$ increases.
Figure~\ref{fig:mass_in_RD} presents an analogous version to the bottom panel of Fig.~\ref{fig:different_RD_SNRs}, here showing the $\log_{10}{\rm JSD}$ evolution of more parameters: $\chieff$, $M$, and $q$, along with $\chip$ for comparison. 
The signals lose information about each parameter at different times: first $\chip$, then $\chieff$, then $q$, and finally $M$. 
The effective spin $\chieff$ remains the least informative throughout the whole signal, never yielding a JSD greater than $0.01$, even in the highest-SNR case. 
The total mass $M$ has the most information at late times in the ringdown, yielding JSDs $\gtrsim 10^{-2}$ (our approximate threshold for measurability) up to $\tcut = 40{-}50\,M$ after the signal's peak. 
This is expected, since $M$ is directly proportional to the frequency of the ringdown modes, which is a primary observable~\cite{Isi:2019aib,Teukolsky:1973}.

After the point where posteriors become uninformative, their actual $\log_{10} \rm JSD$ values become less meaningful, due to the increase in fractional uncertainty intrinsic to finite random sampling.
Posteriors for $\chip$ with, e.g., $\mathrm{JSD} = 10^{-3}$ and $\mathrm{JSD} = 10^{-4}$, are indistinguishable by eye. 
For reference, in Fig.~\ref{fig:mass_in_RD} we plot JSDs between distributions approximated by sets of $N=3000$ random samples from each marginal prior (gray histograms), which is effectively a null distribution for what the JSD can look like when a parameter is fully uninformed by the data.
Our conservative threshold for measurability, $\log_{10} \rm JSD = -2$, is shown with a black line, slightly above these histograms.
At $t=30\,M$, both the $\chieff$ and $\chip$ JSD curves cluster at a value just above or below this threshold. 
There is no apparent unusual feature in the waveforms or posteriors at this time.
The clustering likely reflects random variation as posteriors become poorly constrained. 
From this point onward, the $\chip$ posteriors at all SNRs are uninformative. 
The correlation between $\chip$ and the (uninformative) $\chieff$ explains its clustering at the same cutoff time.

Correlated priors between two (or more) parameters can lead to spuriously strong one-dimensional JSDs, as discussed in \citet{Gangardt:2022ltd}.
We use the data in Fig.~\ref{fig:mass_in_RD} to explore the degree to which conditional priors are affecting our results.
As we are primarily concerned with measuring spin precession, we investigate correlations between $\chip$ and the other parameters. 
The inference prior is uniform in $M$, $q$, and spin magnitudes, and is isotropic in spin angles. 
The prior between $M$ and $\chip$ is uncorrelated, which we validate by calculating a negligible Pearson correlation coefficient~\cite{Pearson:1895} between samples drawn from the prior.
The test yields a $p$-value of $0.81$, indicating non-correlation.
On the other hand, the priors on $q$, $\chieff$, and $\chip$ are all correlated~\cite{Callister:2021gxf,Iwaya:2024zzq}.

To assess how conditional priors from measuring $\chieff$ and $q$ influence the $\chip$ results, we compute the JSD between the full $\chip$ prior and the $\chip$ prior conditioned on $\chieff$ and/or $q$ being close to the true value.
We focus on conditioning using the $\rho_{\rm pp}=100$ post-$\tcut=0$ posterior because it provides the most informative measurements, making potential effects the most pronounced.
We define “close” for $\chieff$ as within one standard deviation of the posterior relative to the injected value. 
For $q$, since all posteriors peak near the true value at $q=1$, we define “close” as greater than the $5^{\rm th}$ percentile of the posterior.
Conditioning on $\chieff$ adds a negligible $2\times10^{-3}$ nats of information to $\chip$, while conditioning on $q$ or both parameters increases it by $2\times10^{-2}$ nats. 
This suggests that $\chieff$ has little impact on $\chip$ constraints, which aligns with Fig.~\ref{fig:mass_in_RD} where $\chieff$ is shown to contribute less information than the mass parameters.
However, we must evaluate whether and how $q$ influences our $\chip$ measurements.

\begin{figure}
    \centering
    \includegraphics[width=\columnwidth]{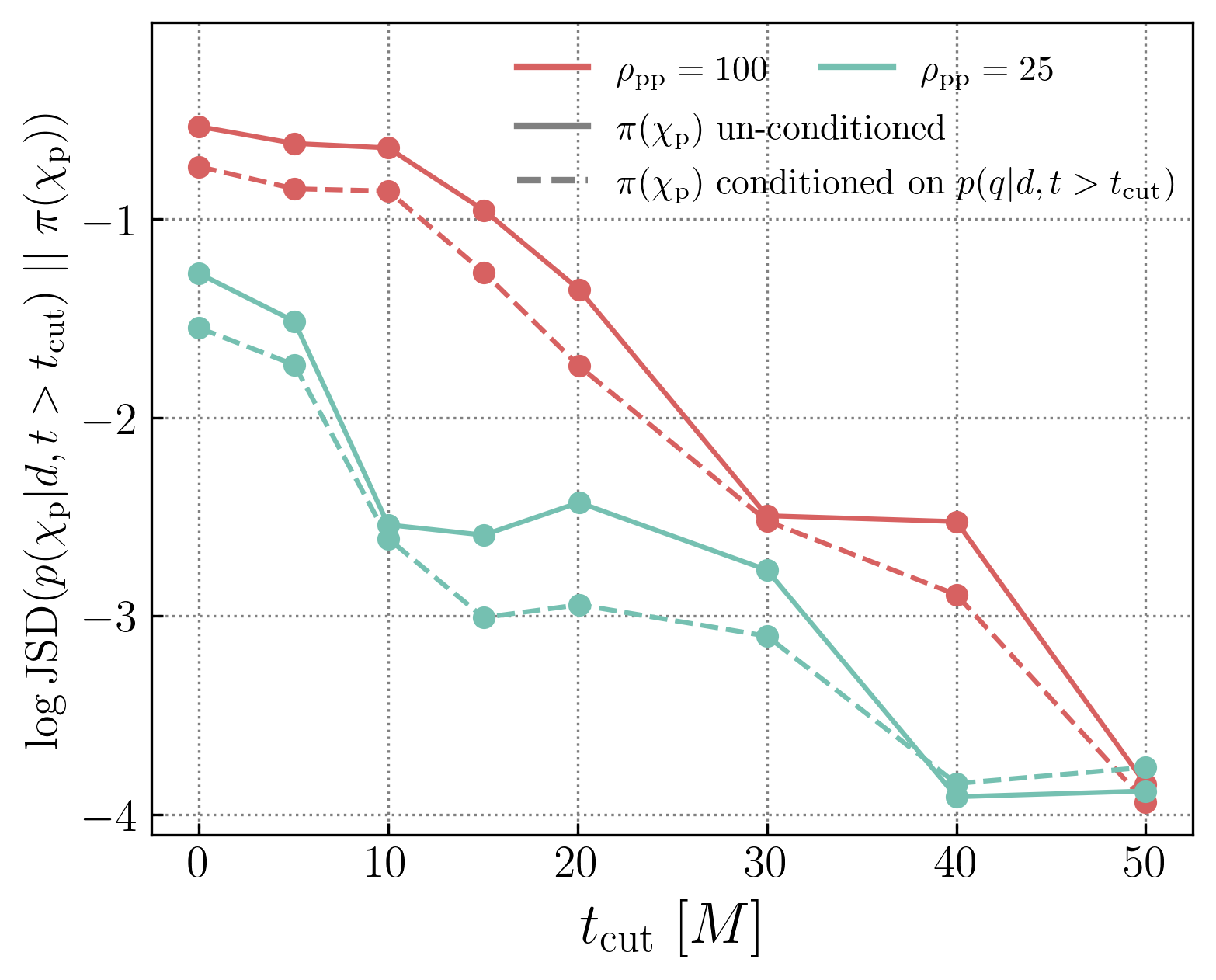}
    \caption{$\log_{10}{\rm JSD}$ between the $\chip$ posterior and full prior $\pi(\chip)$ (solid lines) compared to that computed with $\pi(\chip)$ conditioned on the $q$ posterior (dashed lines), as a function of cutoff time $\tcut$ for post-peak SNRs of $\rho_{\rm pp}=$ 100 (red) and 25 (teal). }
    \label{fig:conditioning_chip_prior}
\end{figure}

Instead of defining the measurability of precession by comparing a given $\chip$ posterior to its \textit{full} prior, we instead calculate the JSD between the posterior and the \textit{conditional} prior~\cite{Gangardt:2022ltd}. 
Figure~\ref{fig:conditioning_chip_prior} shows how the JSD is affected by this correction as a function of cutoff time, for the highest (red) and lowest (teal) SNR case of Fig.~\ref{fig:mass_in_RD}. 
For each cutoff time, we compute the conditional $\chip$ prior by selecting only $\chip$ prior samples where the corresponding $q$ exceeds the $5^{\rm th}$ percentile of the post-cutoff posterior. 
The JSDs from the unconditioned $\chip$ priors (solid lines) are always larger than those from the priors conditioned on $q$ (dashed lines), although the effect is small. 
Crucially, the conditioning does not alter the cutoff time at which the $\chip$ posteriors become uninformative (i.e., JSD $\lesssim 10^{-2}$), nor the general trend. 
In the $\rho_{\rm pp}=25$ case, this transition remains between $5{-}10\,M$ after the signal peak, while for $\rho_{\rm pp}=100$, it remains at $20{-}30\,M$.

\section{Extracting the merger frequency}
\label{app:measuring_merger_freq}

\begin{figure}
    \centering
    \includegraphics[width=\columnwidth]{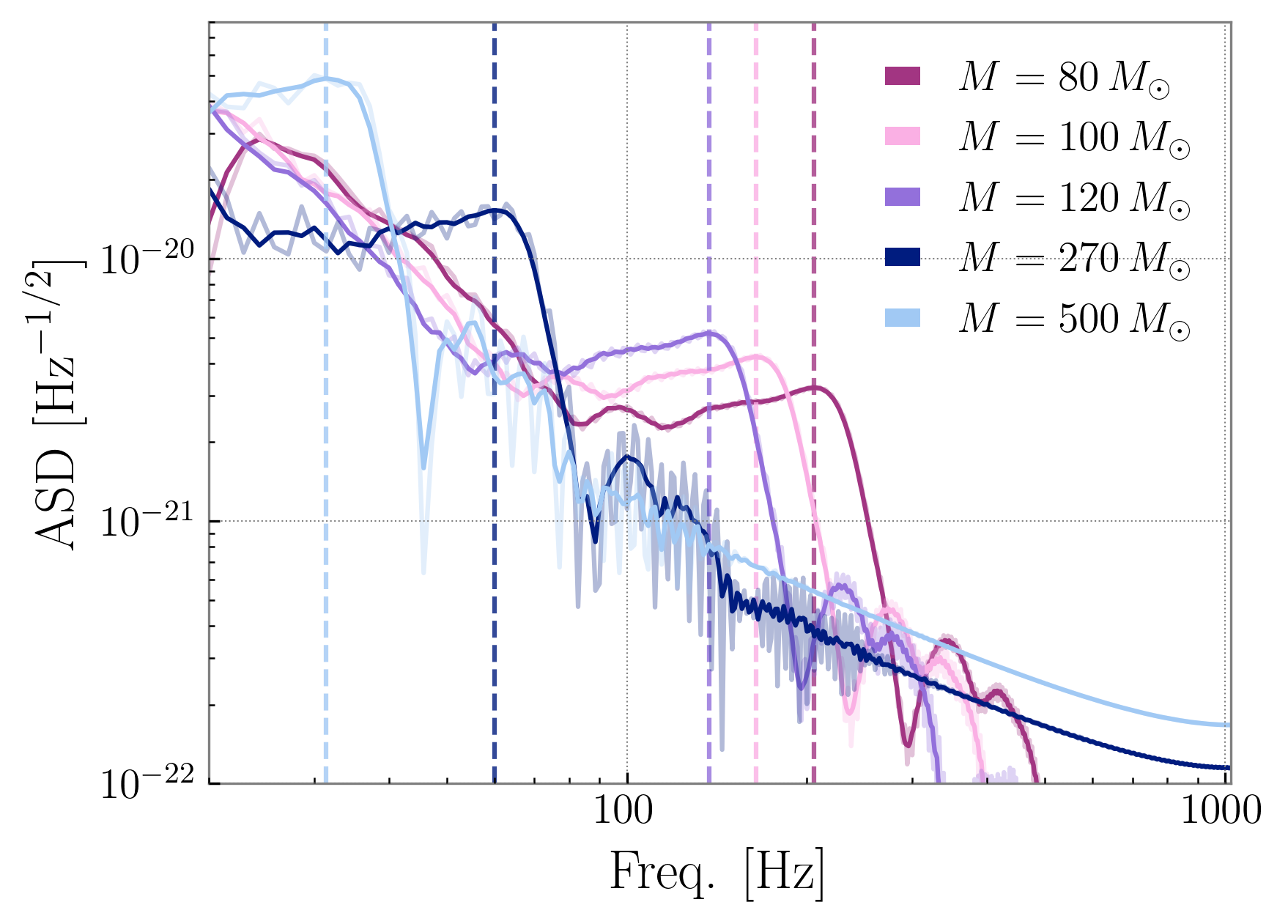}
    \caption{The ASD in LLO versus frequency for five signals with different total masses (different colors). 
    The lighter shade is the true signal, while the overlaid darker shade is the signal with a smoothing filter applied. 
    The dashed vertical lines indicate the merger frequencies, i.e., the frequencies at which each smoothed ASD has a local maximum corresponding to merger.
    These frequencies are listed in Table~\ref{tab:inspiral_and_rd_SNR_different_masses}.}
    \label{fig:merger_freq}
\end{figure}

We define the merger frequency of a GW signal as the frequency at which the signal reaches its maximum amplitude. 
We extract it in the frequency domain; cf. Fig.~\ref{fig:merger_freq} for the five different total mass cases presented in Sec.~\ref{subsec:total-mass}. 
To average out fluctuations in the signal (lighter lines), we apply a smoothing Savitzky-Golay filter~\cite{Savitzky:1964} (yields the darker lines).
We then find the region in frequency space over which the smoothed signal has a net-increasing amplitude spectral density (ASD), and calculate the frequency at which the ASD is the largest over that region.
We identify these values as the merger frequencies and mark them with dashed vertical lines in Fig.~\ref{fig:merger_freq}.
The figure only shows data in the LIGO Livingston detector; we extract consistent merger frequencies from LIGO Hanford and Virgo.
While we opt for a data-driven approach, merger frequencies can also be calculated using the post-Newtonian formalism~\cite{Blanchet:2013haa,Nagar:2018zoe} or numerical relativity surrogate dynamics~\cite{Varma:2019csw}.

\section{Measuring the mass parameters for systems with different $M$}
\label{app:masses}

\begin{figure*}
    \centering
    \includegraphics[width=0.85\textwidth]{figure_04b.png}
    \includegraphics[width=0.85\textwidth]{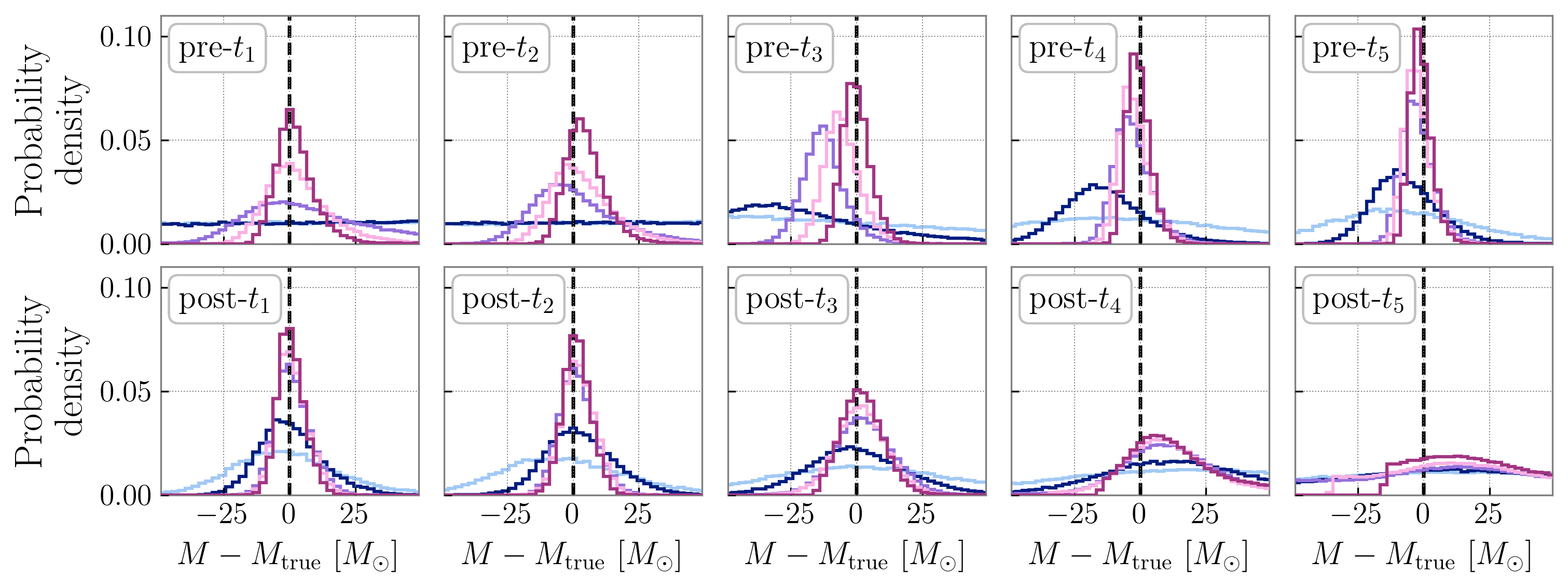}
    \includegraphics[width=0.85\textwidth]{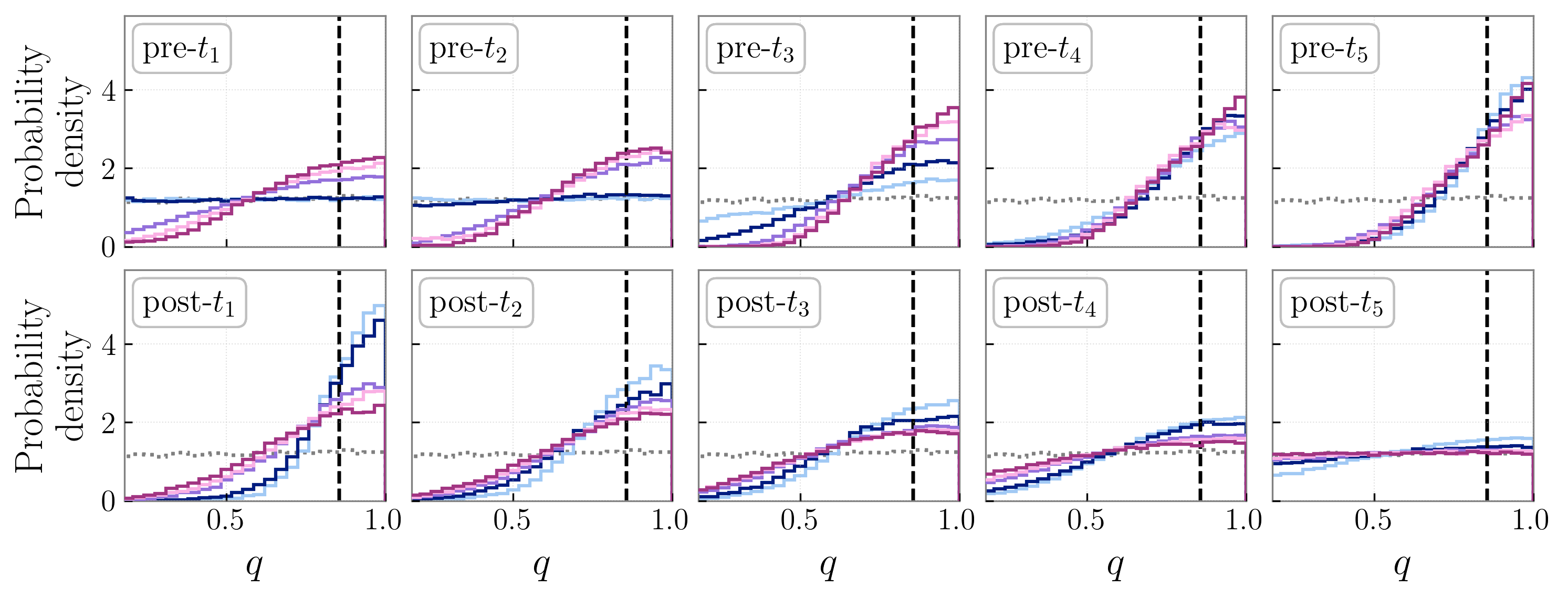}
    \caption{Evolution of the total mass $M$ and mass ratio $q$ as a function of cutoff time for signals with different total masses, presented in Sec.~\ref{subsec:total-mass}.
    Since all signals shown here have different total masses, we here plot posteriors on the inferred total mass minus the true total mass for a more direct comparison.
    The top panel is copied from Fig.~\ref{fig:different_Mtotal} in the main text for reference.
    See the caption of Fig.~\ref{fig:different_Mtotal} for more details.}
    \label{fig:other_params_diff_mass_runs}
\end{figure*}

Section~\ref{subsec:total-mass} looks at signals with different total masses and thus different amounts of visible inspiral. 
In the main text we presented the evolution of the $\chip$ posteriors as various data are excluded; see Fig.~\ref{fig:different_Mtotal}.
Analogous plots for the $M$ and $q$ posteriors are provided in Fig.~\ref{fig:other_params_diff_mass_runs}. 
For the three lowest mass signals, $M$ and $q$ can be measured in the inspiral alone (pre-$t_2$ and earlier).
For all masses, $M$ and $q$ can be measured in the ringdown alone (post-$t_2$), albeit weakly for the highest mass case.
The total mass is always measured better for the lower-mass systems. 
On the other hand, the mass ratio is better measured in post-cutoff data for the higher-mass systems.
Only the $M\leq 120\,\Msun$ signals contain information about $q$ before $t_1$ and $t_2$, while all signals have similar $q$ information before $t_4$ and $t_5$. 
These mass parameters behave differently than $\chip$, which hinges specifically on including the data between $t_1$ and $t_2$ (final pre-merger cycle).

%

\section{Impact of extrinsic angles: Equations and waveforms}
\label{app:angles}

\begin{figure}
    \centering
    \includegraphics[width=0.98\columnwidth]{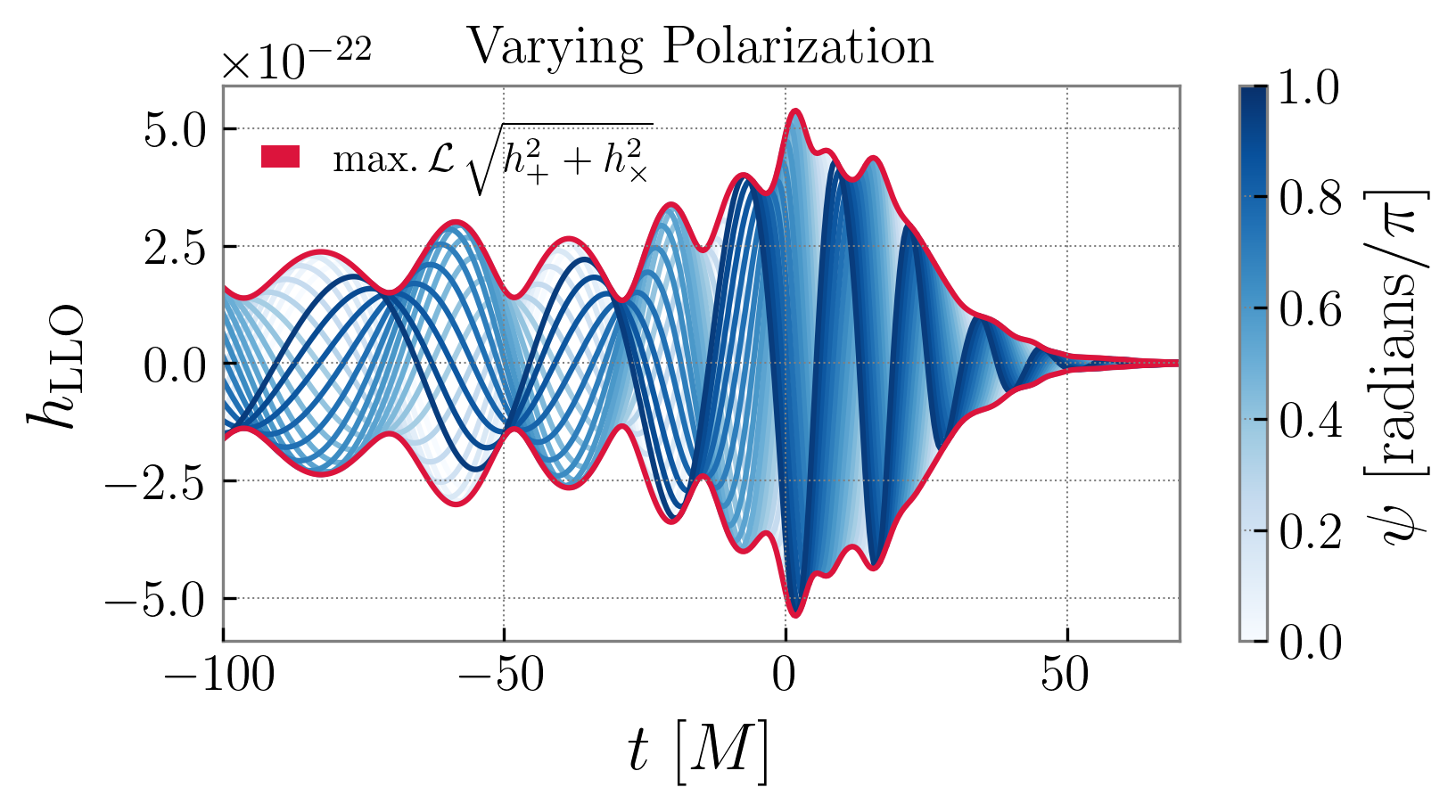}
    \includegraphics[width=0.98\columnwidth]{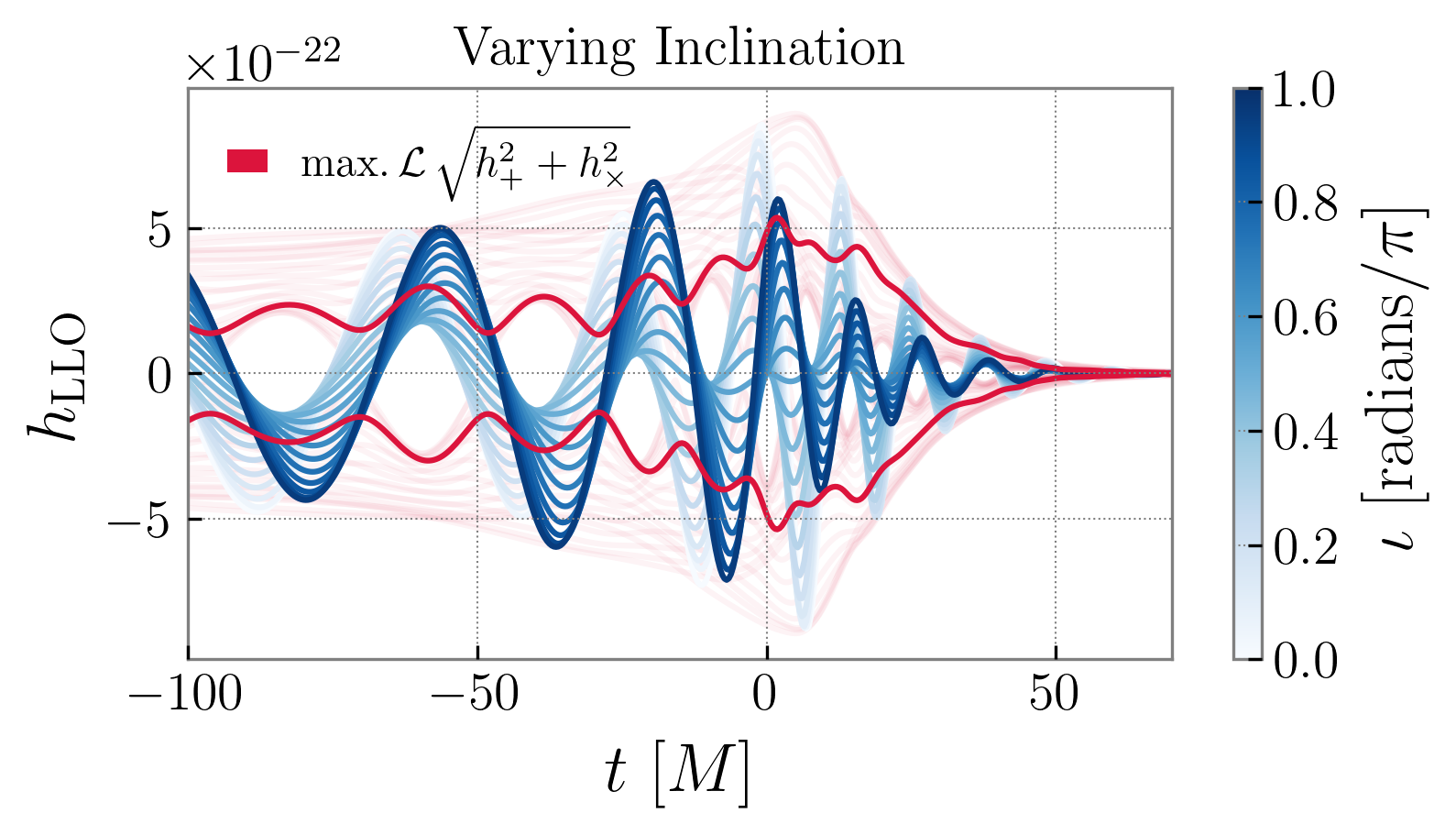}
    \includegraphics[width=0.98\columnwidth]{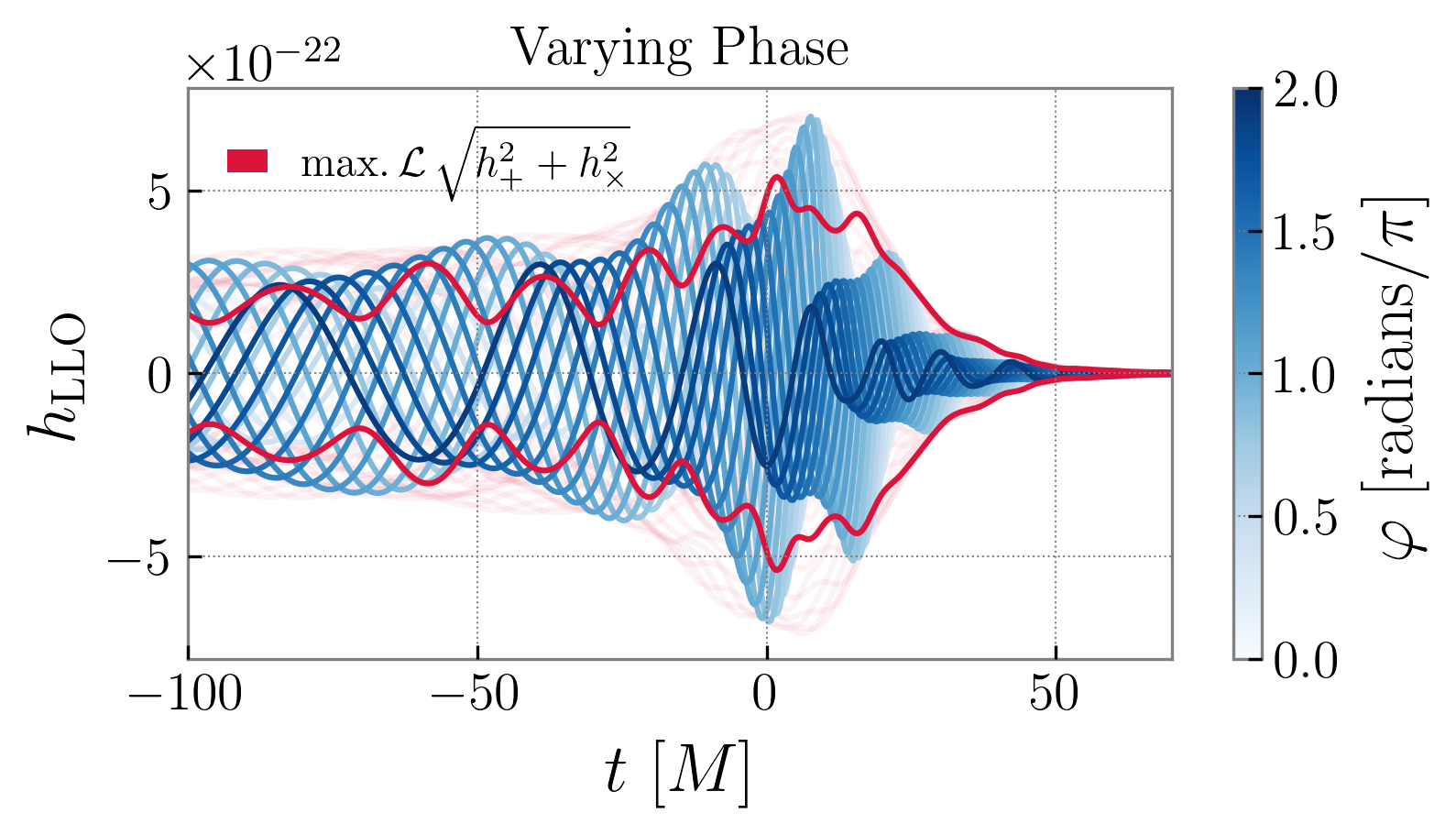}
    \includegraphics[width=0.98\columnwidth]{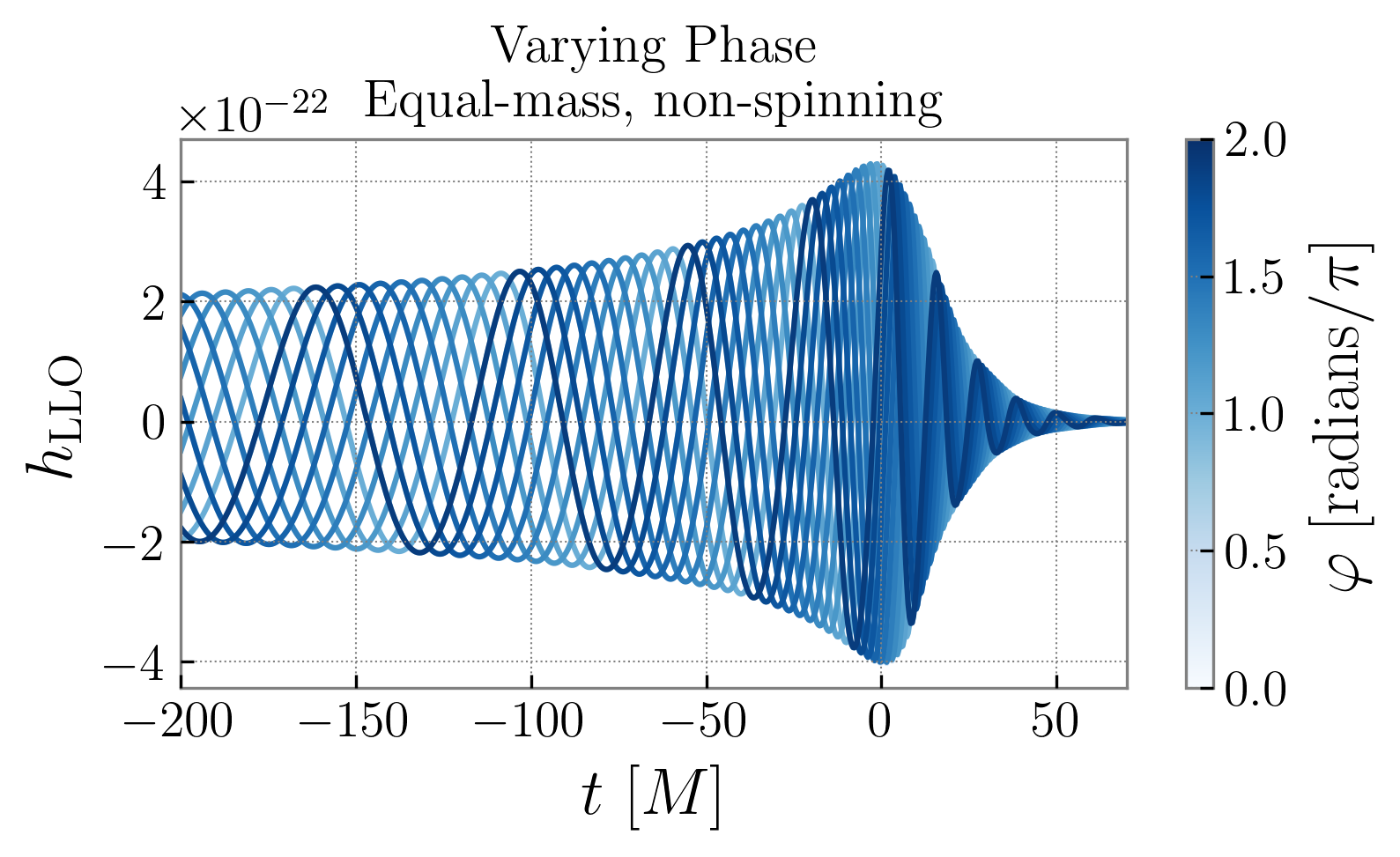}
    \caption{Effect of varying the polarization (top), inclination (second), and phase angle (third) over a range of angles (blues, colorbar) from the $\maxL$ configuration on a projected waveform in LLO, at a fixed SNR. 
    The $\maxL$ waveform amplitude envelope (red) is invariant under changes in polarization, but not inclination or phase angle. 
    The pink traces show the amplitude envelope for each signal; for $\psi$ they all fall directly under the red.
    For comparison, varying the phase angle of an equal-mass, non-spinning BBH is shown in the bottom panel.}
    \label{fig:angles_appendix}
\end{figure}

Section~\ref{subsec:phase} explores the impact of three extrinsic angles---inclination $\iota$, polarization $\psi$, and phase $\varphi$---on GW morphology and resultant measurability of spin precession. 
Figure~\ref{fig:angles_diagram} illustrates how variations in each angle influence both the observed binary configuration on the plane of the sky and the corresponding GW signal.
Here, we additionally provide analytic expressions that capture how each angle affects a GW signal.
Changing the polarization angle by $\Delta\psi$ causes the two GW polarizations  $h_+$ and $h_\times$ to mix, cf. Eq.~(36) of \citet{Isi:2022mbx},
\begin{equation}
\begin{split}
  h_+ \rightarrow h_+' = h_+ \cos 2 \Delta\psi - h_\times \sin 2 \Delta\psi\,, \\
  h_\times \rightarrow h_\times' = h_+ \sin 2 \Delta\psi + h_\times \cos 2 \Delta\psi\,.
\end{split}
\end{equation}
With some simple trigonometry, we can see that this transformation does not change the strain amplitude~$h = \sqrt{h_+^2 + h_\times^2}$. 
In the top panel of Fig.~\ref{fig:angles_appendix}, we confirm this visually: the amplitude envelope for GW190521's $\maxL$ waveform (red) is conserved as $\psi$ is varied.
The amplitude envelope traces for all the other signals shown fall directly under that from the $\maxL$.
For reference, these signals are plotted in LLO, with the amplitude envelopes scaled by the antenna pattern amplitude.
This is not the case when $\iota$ or $\varphi$ are varied, as shown via the amplitude envelope traces (pink) in the second and third panels of Fig.~\ref{fig:angles_appendix}.

In the frame of the source~\cite{Kidder:2007rt}, the two polarizations can be represented by a complex time series decomposed into the sum of spin $-2$ weighted spherical harmonics, dependent on the inclination angle $\iota$ and phase $\varphi$ \cite{Varma:2016dnf,CalderonBustillo:2015lrt,Mills:2020thr,Blanchet:2013haa}:
\begin{equation}
    h_+ - i h_\times = \sum^{\infty}_{\ell=2} \sum^{\ell}_{m=-\ell} Y^{(-2)}_{\ell m}(\iota, \varphi) h_{\ell m}(\lambda, t)\,,
\end{equation}
assuming we are infinitely away from the source. 
The modes $h_{\ell m}$ are each a function of time $t$ and intrinsic parameters $\lambda$ (masses and spins), and can be expressed in terms of their amplitude $A_{\ell m}$ and phase $\Phi_{\ell m}$ evolution 
\begin{equation}
    h_{\ell m}(\lambda, t) = A_{\ell m}(\lambda, t) e^{-i \Phi_{\ell m}(\lambda, t)}\,,
\end{equation}
which are in turn approximated by post-Newtonian theory, cf. Eqs.~(218, 327-329) of \citet{Blanchet:2013haa} or Eq.~(4.17) of \citet{Arun:2008kb}, only for the inspiral phase.
The spherical harmonics are separable for $\iota$ and $\varphi$: 
\begin{equation}
    Y^{(-2)}_{\ell m}(\iota, \varphi) = \sqrt{\frac{2\ell + 1}{4\pi}} d^{\ell m}_{(-2)}(\iota) e^{i m \varphi}\,,
\end{equation}
where $d^{\ell m}_{(-2)}(\iota)$ are the spin -2 Wigner $d$ functions~\cite{Wigner:1959}.
The dominant $\ell = 2$, $m=\pm2$ modes are~\cite{Mills:2020thr}
\begin{equation}
\begin{split}
  Y^{(-2)}_{2,2}(\iota, \varphi) & = \sqrt{\frac{5}{4\pi}}\cos^4\Big(\frac{\iota}{2}\Big) e^{2 i\varphi}\,, \\
  Y^{(-2)}_{2,-2}(\iota, \varphi) & = \sqrt{\frac{5}{4\pi}}\sin^4\Big(\frac{\iota}{2}\Big) e^{-2 i\varphi}\,,
\end{split}
\end{equation}
from which it can be shown that~\cite{Isi:2022mbx}
\begin{equation}
\begin{split}
  h_+ & = \sqrt{\frac{5}{4\pi}} \Bigg( \frac{1 + \cos^2 \iota}{2}\Bigg) \,A_{22} \, \cos \big(\Phi_{22} - 2\varphi\big)\,, \\
  h_\times & = \sqrt{\frac{5}{4\pi}} \cos\iota \,A_{22} \, \sin \big(\Phi_{22} - 2\varphi\big)\,,
\end{split}
\end{equation}
assuming planar symmetry (e.g. non-precessing systems).
Written in this way, the $\iota$ and $\varphi$ dependencies of the 22 mode are apparent~\cite{Maggiore:2007ulw}: 
\begin{equation}
  h_+ \propto \frac{1 + \cos^2 \iota}{2}~,~~ h_\times \propto \cos\iota\,,
\end{equation}
while $\varphi$ is a simple phase shift. 
For non-precessing, equal-mass binaries, the $22$ mode closely approximates the signal~\cite{CalderonBustillo:2015lrt,Mishra:2016whh}.
The bottom panel of Fig.~\ref{fig:angles_appendix} shows waveforms from such a system over a range of $\varphi$ values, illustrating that $\varphi$ just adds a phase shift to the signal while leaving the strain amplitude evolution unchanged. 
The non-spinning, equal-mass case underlies the general intuition of $\varphi$ being the signal’s ``phase."
However, in systems that are high-mass, precessing, and/or have unequal masses, the 22 mode alone is an insufficient approximation, and excluding higher-order modes leads to biased parameter estimates~\cite{Shaik:2019dym}. 
In these cases, interference between multiple $\ell m$ modes makes the role of $\varphi$ more complicated than a trivial phase shift, as shown in the third row of Fig.~\ref{fig:angles_appendix}.

\bibliographystyle{apsrev4-2} 
\bibliography{OurRefs.bib}

@ARTICLE{aLIGO,
    author = {{Aasi}, J. and {Abbott}, B.~P. and
    {Abbott}, R. and {Abbott}, T. and {Abernathy}, M.~R. and others},
    title = "{Advanced LIGO}",
    journal = {Classical and Quantum Gravity},
    year = "2015",
    month = "Apr",
    volume = {32},
    number = {7},
    eid = {074001},
    pages = {074001},
    doi = {10.1088/0264-9381/32/7/074001},
    archivePrefix = {arXiv},
    eprint = {1411.4547},
    primaryClass = {gr-qc}
}

@ARTICLE{aVirgo,
    author = {{Acernese}, F. and {Agathos}, M. and {Agatsuma}, K. and {Aisa}, D. and
    {Allemandou}, N. and others},
    title = "{Advanced Virgo: a second-generation interferometric gravitational wave detector}",
    journal = {Classical and Quantum Gravity},
    year = "2015",
    month = "Jan",
    volume = {32},
    number = {2},
    eid = {024001},
    pages = {024001},
    doi = {10.1088/0264-9381/32/2/024001},
    archivePrefix = {arXiv},
    eprint = {1408.3978},
    primaryClass = {gr-qc},
    adsurl = {https://ui.adsabs.harvard.edu/abs/2015CQGra..32b4001A}
}

@article{KAGRA,
    author = "Akutsu, T. and others",
    collaboration = "KAGRA",
    title = "{Overview of KAGRA: Detector design and construction history}",
    eprint = "2005.05574",
    archivePrefix = "arXiv",
    primaryClass = "physics.ins-det",
    doi = "10.1093/ptep/ptaa125",
    journal = "PTEP",
    volume = "2021",
    number = "5",
    pages = "05A101",
    year = "2021"
}

@article{LIGOScientific:2019hgc,
    author = "Abbott, Benjamin P and others",
    collaboration = "LIGO Scientific, Virgo",
    title = "{A guide to LIGO\textendash{}Virgo detector noise and extraction of transient gravitational-wave signals}",
    eprint = "1908.11170",
    archivePrefix = "arXiv",
    primaryClass = "gr-qc",
    doi = "10.1088/1361-6382/ab685e",
    journal = "Class. Quant. Grav.",
    volume = "37",
    number = "5",
    pages = "055002",
    year = "2020"
}

@ARTICLE{O3-sensitivity,
       author = "Buikema, A. and others",
      journal = {\prd},
         year = 2020,
        month = sep,
       volume = {102},
       number = {6},
          eid = {062003},
        pages = {062003},
          doi = {10.1103/PhysRevD.102.062003},
archivePrefix = {arXiv},
       eprint = {2008.01301},
 primaryClass = {astro-ph.IM},
       adsurl = {https://ui.adsabs.harvard.edu/abs/2020PhRvD.102f2003B}
}

@article{LIGOScientific:2016gtq,
    author = "Abbott, B. P. and others",
    collaboration = "LIGO Scientific, Virgo",
    title = "{Characterization of transient noise in Advanced LIGO relevant to gravitational wave signal GW150914}",
    eprint = "1602.03844",
    archivePrefix = "arXiv",
    primaryClass = "gr-qc",
    doi = "10.1088/0264-9381/33/13/134001",
    journal = "Class. Quant. Grav.",
    volume = "33",
    number = "13",
    pages = "134001",
    year = "2016"
}

@article{LIGO:2021ppb,
    author = "Davis, Derek and others",
    collaboration = "LIGO",
    title = "{LIGO detector characterization in the second and third observing runs}",
    eprint = "2101.11673",
    archivePrefix = "arXiv",
    primaryClass = "astro-ph.IM",
    reportNumber = "P2000495",
    doi = "10.1088/1361-6382/abfd85",
    journal = "Class. Quant. Grav.",
    volume = "38",
    number = "13",
    pages = "135014",
    year = "2021"
}

@article{GWTC2,
    author = "Abbott, R. and others",
    collaboration = "LIGO Scientific, Virgo",
    title = "{GWTC-2: Compact Binary Coalescences Observed by LIGO and Virgo During the First Half of the Third Observing Run}",
    eprint = "2010.14527",
    archivePrefix = "arXiv",
    primaryClass = "gr-qc",
    reportNumber = "P2000061",
    doi = "10.1103/PhysRevX.11.021053",
    journal = "Phys. Rev. X",
    volume = "11",
    pages = "021053",
    year = "2021"
}

@article{GWTC2.1,
    author = "Abbott, R. and others",
    collaboration = "LIGO Scientific, VIRGO",
    title = "{GWTC-2.1: Deep extended catalog of compact binary coalescences observed by LIGO and Virgo during the first half of the third observing run}",
    eprint = "2108.01045",
    archivePrefix = "arXiv",
    primaryClass = "gr-qc",
    reportNumber = "LIGO-P2100063",
    doi = "10.1103/PhysRevD.109.022001",
    journal = "Phys. Rev. D",
    volume = "109",
    number = "2",
    pages = "022001",
    year = "2024"
}

@article{GWTC3,
    author = "Abbott, R. and others",
    collaboration = "KAGRA, VIRGO, LIGO Scientific",
    title = "{GWTC-3: Compact Binary Coalescences Observed by LIGO and Virgo during the Second Part of the Third Observing Run}",
    eprint = "2111.03606",
    archivePrefix = "arXiv",
    primaryClass = "gr-qc",
    reportNumber = "LIGO-P2000318",
    doi = "10.1103/PhysRevX.13.041039",
    journal = "Phys. Rev. X",
    volume = "13",
    number = "4",
    pages = "041039",
    year = "2023"
}

@misc{GW190521_gwosc,
  author = "Abbott, R. and others",
  collaboration = "LIGO Scientific, Virgo",
  title        = {{GW190521}},
  month        = aug,
  year         = 2023,
  publisher    = {GWOSC},
  doi          = {10.7935/1502-wj52},
  url          = {https://gwosc.org/eventapi/html/O3_Discovery_Papers/GW190521/v2/}
}

@misc{GW190521_PE,
  author = "Abbott, R. and others",
  collaboration = "LIGO Scientific, Virgo",
  title        = {{GW190521 posterior samples}},
  month        = nov,
  year         = 2021,
  publisher    = {LIGO Document Control Center},
  url          = {https://dcc.ligo.org/public/0168/P2000158/004/GW190521_posterior_samples.html}
}

@article{opendata_O1O2,
    author = "Abbott, Rich and others",
    collaboration = "LIGO Scientific, Virgo",
    title = "{Open data from the first and second observing runs of Advanced LIGO and Advanced Virgo}",
    eprint = "1912.11716",
    archivePrefix = "arXiv",
    primaryClass = "gr-qc",
    reportNumber = "LIGO-P1900206",
    doi = "10.1016/j.softx.2021.100658",
    journal = "SoftwareX",
    volume = "13",
    pages = "100658",
    year = "2021"
}

@article{opendata_O3,
    author = "Abbott, R. and others",
    collaboration = "KAGRA, VIRGO, LIGO Scientific",
    title = "{Open Data from the Third Observing Run of LIGO, Virgo, KAGRA, and GEO}",
    eprint = "2302.03676",
    archivePrefix = "arXiv",
    primaryClass = "gr-qc",
    reportNumber = "LIGO-P2200316",
    doi = "10.3847/1538-4365/acdc9f",
    journal = "Astrophys. J. Suppl.",
    volume = "267",
    number = "2",
    pages = "29",
    year = "2023"
}

@article{LIGOScientific:2020kqk,
    author = "Abbott, R. and others",
    collaboration = "LIGO Scientific, Virgo",
    title = "{Population Properties of Compact Objects from the Second LIGO-Virgo Gravitational-Wave Transient Catalog}",
    eprint = "2010.14533",
    archivePrefix = "arXiv",
    primaryClass = "astro-ph.HE",
    reportNumber = "LIGO-P2000077",
    doi = "10.3847/2041-8213/abe949",
    journal = "Astrophys. J. Lett.",
    volume = "913",
    number = "1",
    pages = "L7",
    year = "2021"
}

@misc{Aplus_sensitivity,
  author = "Barsotti, Lisa and McCuller, Lee and Evans, Matthew and Fritschel, Peter",
  title        = {{The $A+$ design curve}},
  month        = mar,
  year         = 2018,
  publisher    = {LIGO Document Control Center},
  url          = {https://dcc.ligo.org/LIGO-T1800042/public}
}

@misc{Asharp_sensitivity,
  author = "Kuns, Kevin and Fritschel, Peter",
  title        = {{$A\sharp$ Strain Sensitivity}},
  month        = sept,
  year         = 2023,
  publisher    = {LIGO Document Control Center},
  url          = {https://dcc.ligo.org/LIGO-T2300041/public}
}

@misc{CE_sensitivity,
  author = "Kuns, Kevin and Fulda, Paul and Barsotti, Lisa and Evans, Matthew",
  title        = {{Cosmic Explorer Strain Sensitivity}},
  month        = oct,
  year         = 2023,
  publisher    = {Cosmic Explorer Document Control Center},
  url          = {https://dcc.cosmicexplorer.org/CE-T2000017/public}
}

@article{GW190521_detection,
    author = "Abbott, R. and others",
    collaboration = "LIGO Scientific, Virgo",
    title = "{GW190521: A Binary Black Hole Merger with a Total Mass of $150  M_{\odot}$}",
    eprint = "2009.01075",
    archivePrefix = "arXiv",
    primaryClass = "gr-qc",
    doi = "10.1103/PhysRevLett.125.101102",
    journal = "Phys. Rev. Lett.",
    volume = "125",
    number = "10",
    pages = "101102",
    year = "2020"
}

@article{GW190521_astro,
    author = "Abbott, R. and others",
    collaboration = "LIGO Scientific, Virgo",
    title = "{Properties and Astrophysical Implications of the 150 M$_\odot$ Binary Black Hole Merger GW190521}",
    eprint = "2009.01190",
    archivePrefix = "arXiv",
    primaryClass = "astro-ph.HE",
    reportNumber = "LIGO-P2000021",
    doi = "10.3847/2041-8213/aba493",
    journal = "Astrophys. J. Lett.",
    volume = "900",
    number = "1",
    pages = "L13",
    year = "2020"
}

@unpublished{GW231123,
    author = "Abac, A. G. and others",
    collaboration = "LIGO Scientific, VIRGO, KAGRA",
    title = "{GW231123: a Binary Black Hole Merger with Total Mass 190-265 $M_{\odot}$}",
    eprint = "2507.08219",
    archivePrefix = "arXiv",
    primaryClass = "astro-ph.HE",
    reportNumber = "DCC: P2500026-v6",
    month = "7",
    year = "2025"
}

@article{LVK_IMBH,
    author = "Abbott, Rich and others",
    collaboration = "LIGO Scientific, VIRGO, KAGRA",
    title = "{Search for intermediate-mass black hole binaries in the third observing run of Advanced LIGO and Advanced Virgo}",
    eprint = "2105.15120",
    archivePrefix = "arXiv",
    primaryClass = "astro-ph.HE",
    reportNumber = "LIGO-P2100025",
    doi = "10.1051/0004-6361/202141452",
    journal = "Astron. Astrophys.",
    volume = "659",
    pages = "A84",
    year = "2022"
}

@article{GWTC3_TGR,
    author = "Abbott, R. and others",
    collaboration = "LIGO Scientific, VIRGO, KAGRA",
    title = "{Tests of General Relativity with GWTC-3}",
    eprint = "2112.06861",
    archivePrefix = "arXiv",
    primaryClass = "gr-qc",
    reportNumber = "LIGO-P2100275",
    doi = "10.1103/PhysRevD.112.084080",
    journal = "Phys. Rev. D",
    volume = "112",
    number = "8",
    pages = "084080",
    year = "2025"
}

@misc{lalsuite,
       author         = "{LIGO Scientific Collaboration}",
       title          = "{LIGO} {A}lgorithm {L}ibrary - {LALS}uite",
       howpublished   = "Free software (GPL)",
       doi            = "10.7935/GT1W-FZ16",
       publisher = "git.ligo.org",
       year           = "2019"
 }

@ARTICLE{emcee,
       author = {{Foreman-Mackey}, Daniel and {Hogg}, David W. and {Lang}, Dustin and {Goodman}, Jonathan},
        title = "{emcee: The MCMC Hammer}",
      journal = {Publications of the Astronomical Society of the Pacific},
         year = 2013,
        month = mar,
       volume = {125},
       number = {925},
        pages = {306},
          doi = {10.1086/670067},
archivePrefix = {arXiv},
       eprint = {1202.3665},
 primaryClass = {astro-ph.IM},
       adsurl = {https://ui.adsabs.harvard.edu/abs/2013PASP..125..306F}
}

@Article{numpy,
 title         = {Array programming with {NumPy}},
 author        = {Charles R. Harris and K. Jarrod Millman and St{\'{e}}fan J.
                 van der Walt and Ralf Gommers and Pauli Virtanen and David
                 Cournapeau and Eric Wieser and Julian Taylor and Sebastian
                 Berg and Nathaniel J. Smith and Robert Kern and Matti Picus
                 and Stephan Hoyer and Marten H. van Kerkwijk and Matthew
                 Brett and Allan Haldane and Jaime Fern{\'{a}}ndez del
                 R{\'{i}}o and Mark Wiebe and Pearu Peterson and Pierre
                 G{\'{e}}rard-Marchant and Kevin Sheppard and Tyler Reddy and
                 Warren Weckesser and Hameer Abbasi and Christoph Gohlke and
                 Travis E. Oliphant},
 year          = {2020},
 month         = sep,
 journal       = {Nature},
 volume        = {585},
 number        = {7825},
 pages         = {357--362},
 doi           = {10.1038/s41586-020-2649-2},
 publisher     = {Springer Science and Business Media {LLC}},
 url           = {http://doi.org/10.1038/s41586-020-2649-2}
}

@ARTICLE{scipy,
  author  = {Virtanen, Pauli and Gommers, Ralf and Oliphant, Travis E. and
            Haberland, Matt and Reddy, Tyler and others},
  title   = {{{SciPy} 1.0: Fundamental Algorithms for Scientific
            Computing in Python}},
  journal = {Nature Methods},
  year    = {2020},
  volume  = {17},
  pages   = {261--272},
  adsurl  = {https://rdcu.be/b08Wh},
  doi     = {10.1038/s41592-019-0686-2},
	url = {http://doi.org/10.1038/s41592-019-0686-2}
}

@article{matplotlib,
	Author = {Hunter, J. D.},
	Doi = {10.1109/MCSE.2007.55},
	url = {http://doi.org/10.1109/MCSE.2007.55},
	Journal = {Computing In Science \& Engineering},
	Number = {3},
	Pages = {90--95},
	Publisher = {IEEE COMPUTER SOC},
	Title = {Matplotlib: A 2D graphics environment},
	Volume = {9},
	Year = 2007}

@article{seaborn,
    doi = {10.21105/joss.03021},
    url = {https://doi.org/10.21105/joss.03021},
    year = {2021},
    publisher = {The Open Journal},
    volume = {6},
    number = {60},
    pages = {3021},
    author = {Michael L. Waskom},
    title = {seaborn: statistical data visualization},
    journal = {Journal of Open Source Software}
 }

@misc{h5py,
  author       = {Andrew Collette and
                  Thomas Kluyver and
                  Thomas A Caswell and
                  James Tocknell and
                  Jerome Kieffer and others},
  title        = {h5py: 3.2.1},
  month        = mar,
  year         = 2021,
  publisher    = {Zenodo},
  version      = {3.2.1},
  url          = {https://doi.org/10.5281/zenodo.4584676}}

@misc{h5ify,
  author = {Matthew Mould},
  title = {h5ify: Save Python dictionaries into HDF5 files; load HDF5 files into Python dictionaries.},
  year = 2023,
  url = {https://github.com/mdmould/h5ify}
}

@misc{gwtools,
  author = {Chad Galley},
  title = {\texttt{GWTools}: A collection of useful gravitational wave tools},
  year = 2022,
  url = {https://bitbucket.org/chadgalley/gwtools/src/master/}
}

@software{pandas1,
    author       = {The pandas development team},
    title        = {pandas-dev/pandas: Pandas},
    month        = feb,
    year         = 2020,
    publisher    = {Zenodo},
    version      = {latest},
    doi          = {10.5281/zenodo.3509134},
    url          = {https://doi.org/10.5281/zenodo.3509134}
}

@InProceedings{pandas2,
  author    = { {W}es {M}c{K}inney },
  title     = { {D}ata {S}tructures for {S}tatistical {C}omputing in {P}ython },
  booktitle = { {P}roceedings of the 9th {P}ython in {S}cience {C}onference },
  pages     = { 56 - 61 },
  year      = { 2010 },
  editor    = { {S}t\'efan van der {W}alt and {J}arrod {M}illman },
  doi       = { 10.25080/Majora-92bf1922-00a }
}

@software{pycbc,
	author = {Alex Nitz and Ian Harry and Duncan Brown and Christopher M. Biwer and Josh Willis and Tito Dal Canton and Collin Capano and Thomas Dent and Larne Pekowsky and Gareth S Cabourn Davies and Soumi De and Miriam Cabero and Shichao Wu and Andrew R. Williamson and Bernd Machenschalk and Duncan Macleod and Francesco Pannarale and Prayush Kumar and Steven Reyes and dfinstad and Sumit Kumar and M{\'a}rton T{\'a}pai and Leo Singer and Praveen Kumar and veronica-villa and maxtrevor and Bhooshan Uday Varsha Gadre and Sebastian Khan and Stephen Fairhurst and Arthur Tolley},
	doi = {10.5281/zenodo.10473621},
	month = jan,
	publisher = {Zenodo},
	title = {gwastro/pycbc: v2.3.3 release of PyCBC},
	url = {https://doi.org/10.5281/zenodo.10473621},
	version = {v2.3.3},
	year = 2024}

@article{gwpy,
    title = "{GWpy: A Python package for gravitational-wave astrophysics}",
   author = {{Macleod}, D.~M. and {Areeda}, J.~S. and {Coughlin}, S.~B. and {Massinger}, T.~J. and {Urban}, A.~L.},
  journal = {SoftwareX},
   volume = 13,
    pages = 100657,
     year = 2021,
     issn = {2352-7110},
      doi = {10.1016/j.softx.2021.100657},
      url = {https://www.sciencedirect.com/science/article/pii/S2352711021000029},
}

@software{tdinf,
    author = {Miller, Simona J. and Hourihane, Sophie and Isi, Maximiliano and Udall, Rhiannon and Chatziioannou, Katerina},
    doi = {10.5281/zenodo.16865525},
    month = aug,
    year = 2025,
    publisher = {Zenodo},
    title = {{\tt tdinf}: time domain parameter estimation for gravitational-wave signals},
    url = {https://doi.org/10.5281/zenodo.16865525},
    version = {v1.0.0}
}

@article{Varma:2019csw,
    author = "Varma, Vijay and Field, Scott E. and Scheel, Mark A. and Blackman, Jonathan and Gerosa, Davide and Stein, Leo C. and Kidder, Lawrence E. and Pfeiffer, Harald P.",
    title = "{Surrogate models for precessing binary black hole simulations with unequal masses}",
    eprint = "1905.09300",
    archivePrefix = "arXiv",
    primaryClass = "gr-qc",
    doi = "10.1103/PhysRevResearch.1.033015",
    journal = "Phys. Rev. Research.",
    volume = "1",
    pages = "033015",
    year = "2019"
}

@article{IMRPhenomXO4a,
    author = "Thompson, Jonathan E. and Hamilton, Eleanor and London, Lionel and Ghosh, Shrobana and Kolitsidou, Panagiota and Hoy, Charlie and Hannam, Mark",
    title = "{PhenomXO4a: a phenomenological gravitational-wave model for precessing black-hole binaries with higher multipoles and asymmetries}",
    eprint = "2312.10025",
    archivePrefix = "arXiv",
    primaryClass = "gr-qc",
    reportNumber = "LIGO-P2300437",
    doi = "10.1103/PhysRevD.109.063012",
    journal = "Phys. Rev. D",
    volume = "109",
    number = "6",
    pages = "063012",
    year = "2024"
}

@unpublished{SEOBNRv5PHM_withAsym,
    author = "Estell{\'e}s, H{\'e}ctor and Buonanno, Alessandra and Enficiaud, Raffi and Foo, Cheng and Pompili, Lorenzo",
    title = "{Adding equatorial-asymmetric effects for spin-precessing binaries into the SEOBNRv5PHM waveform model}",
    eprint = "2506.19911",
    archivePrefix = "arXiv",
    primaryClass = "gr-qc",
    month = "6",
    year = "2025"
}

@article{Ajith:2009bn,
    author = "Ajith, P. and others",
    title = "{Inspiral-merger-ringdown waveforms for black-hole binaries with non-precessing spins}",
    eprint = "0909.2867",
    archivePrefix = "arXiv",
    primaryClass = "gr-qc",
    doi = "10.1103/PhysRevLett.106.241101",
    journal = "Phys. Rev. Lett.",
    volume = "106",
    pages = "241101",
    year = "2011"
}

@article{Allen:2004gu,
    author = "Allen, Bruce",
    title = "{${\chi}^{2}$ time-frequency discriminator for gravitational wave detection}",
    eprint = "gr-qc/0405045",
    archivePrefix = "arXiv",
    doi = "10.1103/PhysRevD.71.062001",
    journal = "Phys. Rev. D",
    volume = "71",
    pages = "062001",
    year = "2005"
}

@article{Apostolatos:1994,
  title = {Spin-induced orbital precession and its modulation of the gravitational waveforms from merging binaries},
  author = {Apostolatos, Theocharis A. and Cutler, Curt and Sussman, Gerald J. and Thorne, Kip S.},
  journal = {Phys. Rev. D},
  volume = {49},
  issue = {12},
  pages = {6274--6297},
  numpages = {0},
  year = {1994},
  month = {Jun},
  publisher = {American Physical Society},
  doi = {10.1103/PhysRevD.49.6274},
  url = {https://link.aps.org/doi/10.1103/PhysRevD.49.6274}
}

@article{Arun:2008kb,
    author = "Arun, K. G. and Buonanno, Alessandra and Faye, Guillaume and Ochsner, Evan",
    title = "{Higher-order spin effects in the amplitude and phase of gravitational waveforms emitted by inspiraling compact binaries: Ready-to-use gravitational waveforms}",
    eprint = "0810.5336",
    archivePrefix = "arXiv",
    primaryClass = "gr-qc",
    doi = "10.1103/PhysRevD.79.104023",
    journal = "Phys. Rev. D",
    volume = "79",
    pages = "104023",
    year = "2009",
    note = "[Erratum: Phys.Rev.D 84, 049901 (2011)]"
}

@article{Baibhav:2022qxm,
    author = "Baibhav, Vishal and Doctor, Zoheyr and Kalogera, Vicky",
    title = "{Dropping Anchor: Understanding the Populations of Binary Black Holes with Random and Aligned-spin Orientations}",
    eprint = "2212.12113",
    archivePrefix = "arXiv",
    primaryClass = "astro-ph.HE",
    doi = "10.3847/1538-4357/acbf4c",
    journal = "Astrophys. J.",
    volume = "946",
    number = "1",
    pages = "50",
    year = "2023"
}

@article{Baird:2012cu,
    author = "Baird, Emily and Fairhurst, Stephen and sHannam, Mark and Murphy, Patricia",
    title = "{Degeneracy between mass and spin in black-hole-binary waveforms}",
    eprint = "1211.0546",
    archivePrefix = "arXiv",
    primaryClass = "gr-qc",
    doi = "10.1103/PhysRevD.87.024035",
    journal = "Phys. Rev. D",
    volume = "87",
    number = "2",
    pages = "024035",
    year = "2013"
}

@article{Berti:2005ys,
    author = "Berti, Emanuele and Cardoso, Vitor and Will, Clifford M.",
    title = "{On gravitational-wave spectroscopy of massive black holes with the space interferometer LISA}",
    eprint = "gr-qc/0512160",
    archivePrefix = "arXiv",
    doi = "10.1103/PhysRevD.73.064030",
    journal = "Phys. Rev. D",
    volume = "73",
    pages = "064030",
    year = "2006"
}

@article{Biscoveanu:2021nvg,
    author = "Biscoveanu, Sylvia and Isi, Maximiliano and Varma, Vijay and Vitale, Salvatore",
    title = "{Measuring the spins of heavy binary black holes}",
    eprint = "2106.06492",
    archivePrefix = "arXiv",
    primaryClass = "gr-qc",
    reportNumber = "LIGO document number P2100204",
    doi = "10.1103/PhysRevD.104.103018",
    journal = "Phys. Rev. D",
    volume = "104",
    number = "10",
    pages = "103018",
    year = "2021"
}

@article{Blackman:2017pcm,
    author = "Blackman, Jonathan and Field, Scott E. and Scheel, Mark A. and Galley, Chad R. and Ott, Christian D. and Boyle, Michael and Kidder, Lawrence E. and Pfeiffer, Harald P. and Szil\'agyi, B\'ela",
    title = "{Numerical relativity waveform surrogate model for generically precessing binary black hole mergers}",
    eprint = "1705.07089",
    archivePrefix = "arXiv",
    primaryClass = "gr-qc",
    reportNumber = "YITP-17-44",
    doi = "10.1103/PhysRevD.96.024058",
    journal = "Phys. Rev. D",
    volume = "96",
    number = "2",
    pages = "024058",
    year = "2017"
}

@article{Blanchet:2013haa,
    author = "Blanchet, Luc",
    title = "{Post-Newtonian Theory for Gravitational Waves}",
    eprint = "1310.1528",
    archivePrefix = "arXiv",
    primaryClass = "gr-qc",
    doi = "10.12942/lrr-2014-2",
    journal = "Living Rev. Rel.",
    volume = "17",
    pages = "2",
    year = "2014"
}

@published{Boyle:2014ioa,
    author = "Boyle, Michael and Kidder, Lawrence E. and Ossokine, Serguei and Pfeiffer, Harald P.",
    title = "{Gravitational-wave modes from precessing black-hole binaries}",
    eprint = "1409.4431",
    archivePrefix = "arXiv",
    primaryClass = "gr-qc",
    month = "9",
    year = "2014"
}

@article{Breschi:2019wki,
    author = "Breschi, Matteo and O'Shaughnessy, Richard and Lange, Jacob and Birnholtz, Ofek",
    title = "{IMR consistency tests with higher modes on gravitational signals from the second observing run of LIGO and Virgo}",
    eprint = "1903.05982",
    archivePrefix = "arXiv",
    primaryClass = "gr-qc",
    reportNumber = "LIGO-P1800365",
    doi = "10.1088/1361-6382/ab5629",
    journal = "Class. Quant. Grav.",
    volume = "36",
    number = "24",
    pages = "245019",
    year = "2019"
}

@article{CalderonBustillo:2015lrt,
    author = {Calder\'on Bustillo, Juan and Husa, Sascha and Sintes, Alicia M. and P\"urrer, Michael},
    title = "{Impact of gravitational radiation higher order modes on single aligned-spin gravitational wave searches for binary black holes}",
    eprint = "1511.02060",
    archivePrefix = "arXiv",
    primaryClass = "gr-qc",
    reportNumber = "LIGO-P1500184",
    doi = "10.1103/PhysRevD.93.084019",
    journal = "Phys. Rev. D",
    volume = "93",
    number = "8",
    pages = "084019",
    year = "2016"
}

@article{Callister:2020vyz,
    author = "Callister, Thomas A. and Farr, Will M. and Renzo, Mathieu",
    title = "{State of the Field: Binary Black Hole Natal Kicks and Prospects for Isolated Field Formation after GWTC-2}",
    eprint = "2011.09570",
    archivePrefix = "arXiv",
    primaryClass = "astro-ph.HE",
    doi = "10.3847/1538-4357/ac1347",
    journal = "Astrophys. J.",
    volume = "920",
    number = "2",
    pages = "157",
    year = "2021"
}

@unpublished{Callister:2021gxf,
    author = "Callister, T.",
    title = "{A Thesaurus for Common Priors in Gravitational-Wave Astronomy}",
    eprint = "2104.09508",
    archivePrefix = "arXiv",
    primaryClass = "gr-qc",
    month = "4",
    year = "2021"
}

@article{Capano:2021etf,
    author = "Capano, Collin D. and Cabero, Miriam and Westerweck, Julian and Abedi, Jahed and Kastha, Shilpa and Nitz, Alexander H. and Wang, Yi-Fan and Nielsen, Alex B. and Krishnan, Badri",
    title = "{Multimode Quasinormal Spectrum from a Perturbed Black Hole}",
    eprint = "2105.05238",
    archivePrefix = "arXiv",
    primaryClass = "gr-qc",
    doi = "10.1103/PhysRevLett.131.221402",
    journal = "Phys. Rev. Lett.",
    volume = "131",
    number = "22",
    pages = "221402",
    year = "2023"
}

@article{Capano:2022zqm,
    author = "Capano, Collin D. and Abedi, Jahed and Kastha, Shilpa and Nitz, Alexander H. and Westerweck, Julian and Wang, Yi-Fan and Cabero, Miriam and Nielsen, Alex B. and Krishnan, Badri",
    title = "{Estimating false alarm rates of sub-dominant quasi-normal modes in GW190521}",
    eprint = "2209.00640",
    archivePrefix = "arXiv",
    primaryClass = "gr-qc",
    doi = "10.1088/1361-6382/ad84ae",
    journal = "Class. Quant. Grav.",
    volume = "41",
    number = "24",
    pages = "245009",
    year = "2024"
}

@article{Carullo:2019flw,
    author = "Carullo, Gregorio and Del Pozzo, Walter and Veitch, John",
    title = "{Observational Black Hole Spectroscopy: A time-domain multimode analysis of GW150914}",
    eprint = "1902.07527",
    archivePrefix = "arXiv",
    primaryClass = "gr-qc",
    doi = "10.1103/PhysRevD.99.123029",
    journal = "Phys. Rev. D",
    volume = "99",
    number = "12",
    pages = "123029",
    year = "2019",
    note = "[Erratum: Phys.Rev.D 100, 089903 (2019)]"
}

@ARTICLE{Chandrasekhar:1975,
       author = {{Chandrasekhar}, S. and {Detweiler}, S.},
        title = "{The Quasi-Normal Modes of the Schwarzschild Black Hole}",
      journal = {Proceedings of the Royal Society of London Series A},
         year = 1975,
        month = aug,
       volume = {344},
       number = {1639},
        pages = {441-452},
          doi = {10.1098/rspa.1975.0112},
       adsurl = {https://ui.adsabs.harvard.edu/abs/1975RSPSA.344..441C}
}

@article{Cheung:2023vki,
    author = "Cheung, Mark Ho-Yeuk and Berti, Emanuele and Baibhav, Vishal and Cotesta, Roberto",
    title = "{Extracting linear and nonlinear quasinormal modes from black hole merger simulations}",
    eprint = "2310.04489",
    archivePrefix = "arXiv",
    primaryClass = "gr-qc",
    doi = "10.1103/PhysRevD.109.044069",
    journal = "Phys. Rev. D",
    volume = "109",
    number = "4",
    pages = "044069",
    year = "2024",
    note = "[Erratum: Phys.Rev.D 110, 049902 (2024)]"
}

@article{Correia:2023ipz,
    author = "Correia, Alex and Capano, Collin D.",
    title = "{Sky marginalization in black hole spectroscopy and tests of the area theorem}",
    eprint = "2312.15146",
    archivePrefix = "arXiv",
    primaryClass = "gr-qc",
    doi = "10.1103/PhysRevD.110.044018",
    journal = "Phys. Rev. D",
    volume = "110",
    number = "4",
    pages = "044018",
    year = "2024"
}

@article{Cutler:1994ys,
    author = "Cutler, Curt and Flanagan, Eanna E.",
    title = "{Gravitational waves from merging compact binaries: How accurately can one extract the binary's parameters from the inspiral wave form?}",
    eprint = "gr-qc/9402014",
    archivePrefix = "arXiv",
    reportNumber = "GRP-369",
    doi = "10.1103/PhysRevD.49.2658",
    journal = "Phys. Rev. D",
    volume = "49",
    pages = "2658--2697",
    year = "1994"
}

@article{Davis:2020nyf,
    author = "Davis, Derek and White, Laurel V. and Saulson, Peter R.",
    title = "{Utilizing aLIGO Glitch Classifications to Validate Gravitational-Wave Candidates}",
    eprint = "2002.09429",
    archivePrefix = "arXiv",
    primaryClass = "gr-qc",
    doi = "10.1088/1361-6382/ab91e6",
    journal = "Class. Quant. Grav.",
    volume = "37",
    number = "14",
    pages = "145001",
    year = "2020"
}

@article{Davis:2022ird,
    author = "Davis, D. and Littenberg, T. B. and Romero-Shaw, I. M. and Millhouse, M. and McIver, J. and Di Renzo, F. and Ashton, G.",
    title = "{Subtracting glitches from gravitational-wave detector data during the third LIGO-Virgo observing run}",
    eprint = "2207.03429",
    archivePrefix = "arXiv",
    primaryClass = "astro-ph.IM",
    reportNumber = "P2200192",
    doi = "10.1088/1361-6382/aca238",
    journal = "Class. Quant. Grav.",
    volume = "39",
    number = "24",
    pages = "245013",
    year = "2022"
}

@ARTICLE{Detweiler:1980,
       author = {{Detweiler}, S.},
        title = "{Black holes and gravitational waves. III - The resonant frequencies of rotating holes}",
      journal = {\apj},
         year = 1980,
        month = jul,
       volume = {239},
        pages = {292-295},
          doi = {10.1086/158109},
       adsurl = {https://ui.adsabs.harvard.edu/abs/1980ApJ...239..292D}
}

@article{Doctor:2019ruh,
       author = {{Doctor}, Z. and {Wysocki}, D. and {O'Shaughnessy}, R. and {Holz}, D.~E. and {Farr}, B.},
        title = "{Black Hole Coagulation: Modeling Hierarchical Mergers in Black Hole Populations}",
      journal = {\apj},
         year = 2020,
        month = apr,
       volume = {893},
       number = {1},
          eid = {35},
        pages = {35},
          doi = {10.3847/1538-4357/ab7fac},
archivePrefix = {arXiv},
       eprint = {1911.04424},
 primaryClass = {astro-ph.HE},
       adsurl = {https://ui.adsabs.harvard.edu/abs/2020ApJ...893...35D}
}

@article{Dreyer:2003bv,
    author = "Dreyer, Olaf and Kelly, Bernard J. and Krishnan, Badri and Finn, Lee Samuel and Garrison, David and Lopez-Aleman, Ramon",
    title = "{Black hole spectroscopy: Testing general relativity through gravitational wave observations}",
    eprint = "gr-qc/0309007",
    archivePrefix = "arXiv",
    doi = "10.1088/0264-9381/21/4/003",
    journal = "Class. Quant. Grav.",
    volume = "21",
    pages = "787--804",
    year = "2004"
}

@article{Endres:2003,
  author={Endres, D.M. and Schindelin, J.E.},
  journal={IEEE Transactions on Information Theory}, 
  title={A new metric for probability distributions}, 
  year={2003},
  volume={49},
  number={7},
  pages={1858-1860},
  doi={10.1109/TIT.2003.813506}
}

@article{Fairhurst:2017mvj,
    author = "Fairhurst, Stephen",
    title = "{Localization of transient gravitational wave sources: beyond triangulation}",
    eprint = "1712.04724",
    archivePrefix = "arXiv",
    primaryClass = "gr-qc",
    reportNumber = "LIGO-P1300174",
    doi = "10.1088/1361-6382/aab675",
    journal = "Class. Quant. Grav.",
    volume = "35",
    number = "10",
    pages = "105002",
    year = "2018"
}

@article{Fairhurst:2019srr,
    author = "Fairhurst, Stephen and Green, Rhys and Hannam, Mark and Hoy, Charlie",
    title = "{When will we observe binary black holes precessing?}",
    eprint = "1908.00555",
    archivePrefix = "arXiv",
    primaryClass = "gr-qc",
    reportNumber = "LIGO-P1900224",
    doi = "10.1103/PhysRevD.102.041302",
    journal = "Phys. Rev. D",
    volume = "102",
    number = "4",
    pages = "041302",
    year = "2020"
}

@article{Fairhurst:2019vut,
    author = "Fairhurst, Stephen and Green, Rhys and Hoy, Charlie and Hannam, Mark and Muir, Alistair",
    title = "{Two-harmonic approximation for gravitational waveforms from precessing binaries}",
    eprint = "1908.05707",
    archivePrefix = "arXiv",
    primaryClass = "gr-qc",
    reportNumber = "LIGO-P1900225",
    doi = "10.1103/PhysRevD.102.024055",
    journal = "Phys. Rev. D",
    volume = "102",
    number = "2",
    pages = "024055",
    year = "2020"
}

@article{Fairhurst:2023idl,
    author = "Fairhurst, Stephen and Hoy, Charlie and Green, Rhys and Mills, Cameron and Usman, Samantha A.",
    title = "{Simple parameter estimation using observable features of gravitational-wave signals}",
    eprint = "2304.03731",
    archivePrefix = "arXiv",
    primaryClass = "gr-qc",
    doi = "10.1103/PhysRevD.108.082006",
    journal = "Phys. Rev. D",
    volume = "108",
    number = "8",
    pages = "082006",
    year = "2023"
}

@article{Farr:2017uvj,
    author = "Farr, Will M. and Stevenson, Simon and Coleman Miller, M. and Mandel, Ilya and Farr, Ben and Vecchio, Alberto",
    title = "{Distinguishing Spin-Aligned and Isotropic Black Hole Populations With Gravitational Waves}",
    eprint = "1706.01385",
    archivePrefix = "arXiv",
    primaryClass = "astro-ph.HE",
    reportNumber = "LIGO-P1700067",
    doi = "10.1038/nature23453",
    journal = "Nature",
    volume = "548",
    pages = "426",
    year = "2017"
}

@article{Finch:2021iip,
    author = "Finch, Eliot and Moore, Christopher J.",
    title = "{Modeling the ringdown from precessing black hole binaries}",
    eprint = "2102.07794",
    archivePrefix = "arXiv",
    primaryClass = "gr-qc",
    reportNumber = "LIGO document number P2100036-v2",
    doi = "10.1103/PhysRevD.103.084048",
    journal = "Phys. Rev. D",
    volume = "103",
    number = "8",
    pages = "084048",
    year = "2021"
}

@article{Fuller:2019sxi,
    author = "Fuller, Jim and Ma, Linhao",
    title = "{Most Black Holes are Born Very Slowly Rotating}",
    eprint = "1907.03714",
    archivePrefix = "arXiv",
    primaryClass = "astro-ph.SR",
    doi = "10.3847/2041-8213/ab339b",
    journal = "Astrophys. J. Lett.",
    volume = "881",
    number = "1",
    pages = "L1",
    year = "2019"
}

@article{Gangardt:2022ltd,
    author = "Gangardt, Daria and Gerosa, Davide and Kesden, Michael and De Renzis, Viola and Steinle, Nathan",
    title = "{Constraining black-hole binary spin precession and nutation with sequential prior conditioning}",
    eprint = "2204.00026",
    archivePrefix = "arXiv",
    primaryClass = "gr-qc",
    doi = "10.1103/PhysRevD.106.024019",
    journal = "Phys. Rev. D",
    volume = "106",
    number = "2",
    pages = "024019",
    year = "2022",
    note = "[Erratum: Phys.Rev.D 107, 109901 (2023)]"
}

@article{Gerosa:2015tea,
    author = "Gerosa, Davide and Kesden, Michael and Sperhake, Ulrich and Berti, Emanuele and O'Shaughnessy, Richard",
    title = "{Multi-timescale analysis of phase transitions in precessing black-hole binaries}",
    eprint = "1506.03492",
    archivePrefix = "arXiv",
    primaryClass = "gr-qc",
    doi = "10.1103/PhysRevD.92.064016",
    journal = "Phys. Rev. D",
    volume = "92",
    pages = "064016",
    year = "2015"
}

@article{Gerosa:2017kvu,
    author = "Gerosa, Davide and Berti, Emanuele",
    title = "{Are merging black holes born from stellar collapse or previous mergers?}",
    eprint = "1703.06223",
    archivePrefix = "arXiv",
    primaryClass = "gr-qc",
    doi = "10.1103/PhysRevD.95.124046",
    journal = "Phys. Rev. D",
    volume = "95",
    number = "12",
    pages = "124046",
    year = "2017"
}

@article{Gerosa:2018wbw,
    author = "Gerosa, Davide and Berti, Emanuele and O'Shaughnessy, Richard and Belczynski, Krzysztof and Kesden, Michael and Wysocki, Daniel and Gladysz, Wojciech",
    title = "{Spin orientations of merging black holes formed from the evolution of stellar binaries}",
    eprint = "1808.02491",
    archivePrefix = "arXiv",
    primaryClass = "astro-ph.HE",
    doi = "10.1103/PhysRevD.98.084036",
    journal = "Phys. Rev. D",
    volume = "98",
    number = "8",
    pages = "084036",
    year = "2018"
}

@article{Gerosa:2020aiw,
    author = "Gerosa, Davide and Mould, Matthew and Gangardt, Daria and Schmidt, Patricia and Pratten, Geraint and Thomas, Lucy M.",
    title = "{A generalized precession parameter $\chi_\mathrm{p}$ to interpret gravitational-wave data}",
    eprint = "2011.11948",
    archivePrefix = "arXiv",
    primaryClass = "gr-qc",
    doi = "10.1103/PhysRevD.103.064067",
    journal = "Phys. Rev. D",
    volume = "103",
    number = "6",
    pages = "064067",
    year = "2021"
}

@article{Ghosh:2016qgn,
    author = "Ghosh, Abhirup and others",
    title = "{Testing general relativity using golden black-hole binaries}",
    eprint = "1602.02453",
    archivePrefix = "arXiv",
    primaryClass = "gr-qc",
    reportNumber = "LIGO-P1500185-V10, ICTS-2016-1, LIGO-P1500185-V11",
    doi = "10.1103/PhysRevD.94.021101",
    journal = "Phys. Rev. D",
    volume = "94",
    number = "2",
    pages = "021101",
    year = "2016"
}

@article{Ghosh:2017gfp,
    author = "Ghosh, Abhirup and Johnson-Mcdaniel, Nathan K. and Ghosh, Archisman and Mishra, Chandra Kant and Ajith, Parameswaran and Del Pozzo, Walter and Berry, Christopher P. L. and Nielsen, Alex B. and London, Lionel",
    title = "{Testing general relativity using gravitational wave signals from the inspiral, merger and ringdown of binary black holes}",
    eprint = "1704.06784",
    archivePrefix = "arXiv",
    primaryClass = "gr-qc",
    reportNumber = "LIGO-P1700006, ICTS-2017-3",
    doi = "10.1088/1361-6382/aa972e",
    journal = "Class. Quant. Grav.",
    volume = "35",
    number = "1",
    pages = "014002",
    year = "2018"
}

@article{Green:2020ptm,
    author = "Green, Rhys and Hoy, Charlie and Fairhurst, Stephen and Hannam, Mark and Pannarale, Francesco and Thomas, Cory",
    title = "{Identifying when Precession can be Measured in Gravitational Waveforms}",
    eprint = "2010.04131",
    archivePrefix = "arXiv",
    primaryClass = "gr-qc",
    reportNumber = "LIGO-P2000376",
    doi = "10.1103/PhysRevD.103.124023",
    journal = "Phys. Rev. D",
    volume = "103",
    number = "12",
    pages = "124023",
    year = "2021"
}

@article{Hamilton:2021pkf,
    author = "Hamilton, Eleanor and London, Lionel and Thompson, Jonathan E. and Fauchon-Jones, Edward and Hannam, Mark and Kalaghatgi, Chinmay and Khan, Sebastian and Pannarale, Francesco and Vano-Vinuales, Alex",
    title = "{Model of gravitational waves from precessing black-hole binaries through merger and ringdown}",
    eprint = "2107.08876",
    archivePrefix = "arXiv",
    primaryClass = "gr-qc",
    doi = "10.1103/PhysRevD.104.124027",
    journal = "Phys. Rev. D",
    volume = "104",
    number = "12",
    pages = "124027",
    year = "2021"
}

@article{Hamilton:2023znn,
    author = "Hamilton, Eleanor and London, Lionel and Hannam, Mark",
    title = "{Ringdown frequencies in black holes formed from precessing black-hole binaries}",
    eprint = "2301.06558",
    archivePrefix = "arXiv",
    primaryClass = "gr-qc",
    doi = "10.1103/PhysRevD.107.104035",
    journal = "Phys. Rev. D",
    volume = "107",
    number = "10",
    pages = "104035",
    year = "2023"
}

@unpublished{Hogg:1999ad,
    author = "Hogg, David W.",
    title = "{Distance measures in cosmology}",
    eprint = "astro-ph/9905116",
    archivePrefix = "arXiv",
    month = "5",
    year = "1999"
}

@article{Hughes:2019zmt,
    author = "Hughes, Scott A. and Apte, Anuj and Khanna, Gaurav and Lim, Halston",
    title = "{Learning about black hole binaries from their ringdown spectra}",
    eprint = "1901.05900",
    archivePrefix = "arXiv",
    primaryClass = "gr-qc",
    doi = "10.1103/PhysRevLett.123.161101",
    journal = "Phys. Rev. Lett.",
    volume = "123",
    number = "16",
    pages = "161101",
    year = "2019"
}

@article{Isi:2019aib,
    author = "Isi, Maximiliano and Giesler, Matthew and Farr, Will M. and Scheel, Mark A. and Teukolsky, Saul A.",
    title = "{Testing the no-hair theorem with GW150914}",
    eprint = "1905.00869",
    archivePrefix = "arXiv",
    primaryClass = "gr-qc",
    reportNumber = "LIGO-P1900135",
    doi = "10.1103/PhysRevLett.123.111102",
    journal = "Phys. Rev. Lett.",
    volume = "123",
    number = "11",
    pages = "111102",
    year = "2019"
}

@article{Isi:2020tac,
    author = "Isi, Maximiliano and Farr, Will M. and Giesler, Matthew and Scheel, Mark A. and Teukolsky, Saul A.",
    title = "{Testing the Black-Hole Area Law with GW150914}",
    eprint = "2012.04486",
    archivePrefix = "arXiv",
    primaryClass = "gr-qc",
    reportNumber = "LIGO-P2000507",
    doi = "10.1103/PhysRevLett.127.011103",
    journal = "Phys. Rev. Lett.",
    volume = "127",
    number = "1",
    pages = "011103",
    year = "2021"
}

@unpublished{Isi:2021iql,
    author = "Isi, Maximiliano and Farr, Will M.",
    title = "{Analyzing black-hole ringdowns}",
    eprint = "2107.05609",
    archivePrefix = "arXiv",
    primaryClass = "gr-qc",
    reportNumber = "LIGO-P2100227",
    month = "7",
    year = "2021"
}

@article{Isi:2022mbx,
    author = "Isi, Maximiliano",
    title = "{Parametrizing gravitational-wave polarizations}",
    eprint = "2208.03372",
    archivePrefix = "arXiv",
    primaryClass = "gr-qc",
    reportNumber = "LIGO-P2200221",
    doi = "10.1088/1361-6382/acf28c",
    journal = "Class. Quant. Grav.",
    volume = "40",
    number = "20",
    pages = "203001",
    year = "2023"
}

@article{Isi:2023dlk,
    author = "Isi, Maximiliano and Farr, Will M. and Varma, Vijay",
    title = "{The Directional Isotropy of LIGO\textendash{}Virgo Binaries}",
    eprint = "2304.13254",
    archivePrefix = "arXiv",
    primaryClass = "gr-qc",
    reportNumber = "LIGO-P2300088",
    doi = "10.3847/1538-4357/ad0ec9",
    journal = "Astrophys. J.",
    volume = "962",
    number = "1",
    pages = "19",
    year = "2024"
}

@article{Israel:1967,
    author = "Israel, Werner",
    title = "{Event Horizons in Static Vacuum Space-Times}",
    journal = "Phys. Rev.",
    volume = "164",
    issue = "5",
    pages = "1776--1779",
    year = "1967",
    doi = "10.1103/PhysRev.164.1776"
}

@article{Iwaya:2024zzq,
    author = "Iwaya, Masaki and Kobayashi, Kazuya and Morisaki, Soichiro and Hotokezaka, Kenta and Kinugawa, Tomoya",
    title = "{Analytic joint priors of effective spin parameters and inference of the spin distribution of binary black holes}",
    eprint = "2412.14551",
    archivePrefix = "arXiv",
    primaryClass = "gr-qc",
    doi = "10.1103/PhysRevD.111.103046",
    journal = "Phys. Rev. D",
    volume = "111",
    number = "10",
    pages = "103046",
    year = "2025"
}

@article{Kalogera:1999tq,
    author = "Kalogera, Vassiliki",
    title = "{Spin orbit misalignment in close binaries with two compact objects}",
    eprint = "astro-ph/9911417",
    archivePrefix = "arXiv",
    doi = "10.1086/309400",
    journal = "Astrophys. J.",
    volume = "541",
    pages = "319--328",
    year = "2000"
}

@article{Kamaretsos:2012bs,
    author = "Kamaretsos, Ioannis and Hannam, Mark and Sathyaprakash, B.",
    title = "{Is black-hole ringdown a memory of its progenitor?}",
    eprint = "1207.0399",
    archivePrefix = "arXiv",
    primaryClass = "gr-qc",
    doi = "10.1103/PhysRevLett.109.141102",
    journal = "Phys. Rev. Lett.",
    volume = "109",
    pages = "141102",
    year = "2012"
}

@article{Kidder:1993,
  title = {Spin effects in the inspiral of coalescing compact binaries},
  author = {Kidder, Lawrence E. and Will, Clifford M. and Wiseman, Alan G.},
  journal = {Phys. Rev. D},
  volume = {47},
  issue = {10},
  pages = {R4183--R4187},
  numpages = {0},
  year = {1993},
  month = {May},
  publisher = {American Physical Society},
  doi = {10.1103/PhysRevD.47.R4183},
  url = {https://link.aps.org/doi/10.1103/PhysRevD.47.R4183}
}

@article{Kidder:2007rt,
    author = "Kidder, Lawrence E.",
    title = "{Using full information when computing modes of post-Newtonian waveforms from inspiralling compact binaries in circular orbit}",
    eprint = "0710.0614",
    archivePrefix = "arXiv",
    primaryClass = "gr-qc",
    doi = "10.1103/PhysRevD.77.044016",
    journal = "Phys. Rev. D",
    volume = "77",
    pages = "044016",
    year = "2008"
}

@article{Kimball:2020opk,
    author = "Kimball, Chase and Talbot, Colm and L. Berry, Christopher P. and Carney, Matthew and Zevin, Michael and Thrane, Eric and Kalogera, Vicky",
    title = "{Black Hole Genealogy: Identifying Hierarchical Mergers with Gravitational Waves}",
    eprint = "2005.00023",
    archivePrefix = "arXiv",
    primaryClass = "astro-ph.HE",
    reportNumber = "DCC P2000131",
    doi = "10.3847/1538-4357/aba518",
    journal = "Astrophys. J.",
    volume = "900",
    number = "2",
    pages = "177",
    year = "2020"
}

@article{Kokkotas:1999bd,
    author = "Kokkotas, Kostas D. and Schmidt, Bernd G.",
    title = "{Quasinormal modes of stars and black holes}",
    eprint = "gr-qc/9909058",
    archivePrefix = "arXiv",
    doi = "10.12942/lrr-1999-2",
    journal = "Living Rev. Rel.",
    volume = "2",
    pages = "2",
    year = "1999"
}

@article{Kullback:1951,
	author = {S. Kullback and R. A. Leibler},
	doi = {10.1214/aoms/1177729694},
	journal = {The Annals of Mathematical Statistics},
	number = {1},
	pages = {79 -- 86},
	publisher = {Institute of Mathematical Statistics},
	title = {{On Information and Sufficiency}},
	url = {https://doi.org/10.1214/aoms/1177729694},
	volume = {22},
	year = {1951}
}

@book{Kullback:1959,
  address = {New York},
  author = {Kullback, Solomon},
  booktitle = {{Information Theory and Statistics}},
  publisher = {Wiley},
  title = {{Information Theory and Statistics}},
  x-fetchedfrom = {Bibsonomy},
  year = 1959
}

@ARTICLE{Lin:1991,
  author={Lin, J.},
  journal={IEEE Transactions on Information Theory}, 
  title={Divergence measures based on the Shannon entropy}, 
  year={1991},
  volume={37},
  number={1},
  pages={145-151},
  doi={10.1109/18.61115}}

@article{Loredo:2004nn,
    author = "Loredo, Thomas J.",
    editor = "Fischer, Rainer and Preuss, Roland and von Toussaint, Udo",
    title = "{Accounting for source uncertainties in analyses of astronomical survey data}",
    eprint = "astro-ph/0409387",
    archivePrefix = "arXiv",
    doi = "10.1063/1.1835214",
    journal = "AIP Conf. Proc.",
    volume = "735",
    number = "1",
    pages = "195--206",
    year = "2004"
}

@article{MaganaZertuche:2024ajz,
    author = "Maga{\~n}a Zertuche, Lorena and others",
    title = "{High-precision ringdown surrogate model for nonprecessing binary black holes}",
    eprint = "2408.05300",
    archivePrefix = "arXiv",
    primaryClass = "gr-qc",
    doi = "10.1103/q7sy-g3kl",
    journal = "Phys. Rev. D",
    volume = "112",
    number = "2",
    pages = "024077",
    year = "2025"
}

@book{Maggiore:2007ulw,
    author = "Maggiore, Michele",
    title = "{Gravitational Waves. Vol. 1: Theory and Experiments}",
    doi = "10.1093/acprof:oso/9780198570745.001.0001",
    isbn = "978-0-19-171766-6, 978-0-19-852074-0",
    publisher = "Oxford University Press",
    year = "2007"
}

@ARTICLE{Malmquist:1922,
       author = {{Malmquist}, K.~G.},
        title = "{On some relations in stellar statistics}",
      journal = {Meddelanden fran Lunds Astronomiska Observatorium Serie I},
         year = 1922,
        month = mar,
       volume = {100},
        pages = {1-52},
       adsurl = {https://ui.adsabs.harvard.edu/abs/1922MeLuF.100....1M}
}

@article{Mandel:2018mve,
    author = "Mandel, Ilya and Farr, Will M. and Gair, Jonathan R.",
    title = "{Extracting distribution parameters from multiple uncertain observations with selection biases}",
    eprint = "1809.02063",
    archivePrefix = "arXiv",
    primaryClass = "physics.data-an",
    doi = "10.1093/mnras/stz896",
    journal = "Mon. Not. Roy. Astron. Soc.",
    volume = "486",
    number = "1",
    pages = "1086--1093",
    year = "2019"
}

@article{Mandel:2018hfr,
    author = "Mandel, Ilya and Farmer, Alison",
    title = "{Merging stellar-mass binary black holes}",
    eprint = "1806.05820",
    archivePrefix = "arXiv",
    primaryClass = "astro-ph.HE",
    doi = "10.1016/j.physrep.2022.01.003",
    journal = "Phys. Rept.",
    volume = "955",
    pages = "1--24",
    year = "2022"
}

@article{Mapelli:2018uds,
    author = "Mapelli, Michela",
    editor = "Coccia, E. and Silk, J. and Vittorio, N.",
    title = "{Astrophysics of stellar black holes}",
    eprint = "1809.09130",
    archivePrefix = "arXiv",
    primaryClass = "astro-ph.HE",
    doi = "10.3254/ENFI200005",
    journal = "Proc. Int. Sch. Phys. Fermi",
    volume = "200",
    pages = "87--121",
    year = "2020"
}

@article{McKernan:2019beu,
    author = "McKernan, B. and Ford, K. E. S. and O'Shaughnessy, R. and Wysocki, D.",
    title = "{Monte Carlo simulations of black hole mergers in AGN discs: Low $\chi_{\rm eff}$ mergers and predictions for LIGO}",
    eprint = "1907.04356",
    archivePrefix = "arXiv",
    primaryClass = "astro-ph.HE",
    doi = "10.1093/mnras/staa740",
    journal = "Mon. Not. Roy. Astron. Soc.",
    volume = "494",
    number = "1",
    pages = "1203--1216",
    year = "2020"
}

@article{McKernan:2024kpr,
    author = "McKernan, Barry and Ford, K. E. Saavik and Cook, Harrison E. and Delfavero, Vera and McPike, Emily and Nathaniel, Kaila and Postiglione, Jake and Ray, Shawn and O'Shaughnessy, Richard",
    title = "{McFACTS I: Testing the LVK AGN Channel with Monte Carlo for AGN Channel Testing and Simulation (McFACTS)}",
    eprint = "2410.16515",
    archivePrefix = "arXiv",
    primaryClass = "astro-ph.HE",
    doi = "10.3847/1538-4357/adf114",
    journal = "Astrophys. J.",
    volume = "990",
    number = "2",
    pages = "217",
    year = "2025"
}

@article{Menendez:1997,
	author = {M.L. Men{\'e}ndez and J.A. Pardo and L. Pardo and M.C. Pardo},
	doi = {https://doi.org/10.1016/S0016-0032(96)00063-4},
	issn = {0016-0032},
	journal = {Journal of the Franklin Institute},
	number = {2},
	pages = {307-318},
	title = {The Jensen-Shannon divergence},
	url = {https://www.sciencedirect.com/science/article/pii/S0016003296000634},
	volume = {334},
	year = {1997}
}

@article{Miller:2023ncs,
    author = "Miller, Simona J. and Isi, Maximiliano and Chatziioannou, Katerina and Varma, Vijay and Mandel, Ilya",
    title = "{GW190521: Tracing imprints of spin-precession on the most massive black hole binary}",
    eprint = "2310.01544",
    archivePrefix = "arXiv",
    primaryClass = "astro-ph.HE",
    reportNumber = "LIGO-P2300329",
    doi = "10.1103/PhysRevD.109.024024",
    journal = "Phys. Rev. D",
    volume = "109",
    number = "2",
    pages = "024024",
    year = "2024"
}

@article{Mills:2020thr,
    author = "Mills, Cameron and Fairhurst, Stephen",
    title = "{Measuring gravitational-wave higher-order multipoles}",
    eprint = "2007.04313",
    archivePrefix = "arXiv",
    primaryClass = "gr-qc",
    doi = "10.1103/PhysRevD.103.024042",
    journal = "Phys. Rev. D",
    volume = "103",
    number = "2",
    pages = "024042",
    year = "2021"
}

@article{Mishra:2016whh,
    author = "Mishra, Chandra Kant and Kela, Aditya and Arun, K. G. and Faye, Guillaume",
    title = "{Ready-to-use post-Newtonian gravitational waveforms for binary black holes with nonprecessing spins: An update}",
    eprint = "1601.05588",
    archivePrefix = "arXiv",
    primaryClass = "gr-qc",
    doi = "10.1103/PhysRevD.93.084054",
    journal = "Phys. Rev. D",
    volume = "93",
    number = "8",
    pages = "084054",
    year = "2016"
}

@article{Mitman:2025hgy,
    author = "Mitman, Keefe and others",
    title = "{Probing the ringdown perturbation in binary black hole coalescences with an improved quasinormal mode extraction algorithm}",
    eprint = "2503.09678",
    archivePrefix = "arXiv",
    primaryClass = "gr-qc",
    doi = "10.1103/qq1g-jlnw",
    journal = "Phys. Rev. D",
    volume = "112",
    number = "6",
    pages = "064016",
    year = "2025"
}

@article{Nagar:2018zoe,
	archiveprefix = {arXiv},
	author = {Nagar, Alessandro and others},
	doi = {10.1103/PhysRevD.98.104052},
	eprint = {1806.01772},
	journal = {Phys. Rev. D},
	number = {10},
	pages = {104052},
	primaryclass = {gr-qc},
	title = {{Time-domain effective-one-body gravitational waveforms for coalescing compact binaries with nonprecessing spins, tides and self-spin effects}},
	volume = {98},
	year = {2018}
}

@article{Nissanke:2009kt,
    author = "Nissanke, Samaya and Holz, Daniel E. and Hughes, Scott A. and Dalal, Neal and Sievers, Jonathan L.",
    title = "{Exploring short gamma-ray bursts as gravitational-wave standard sirens}",
    eprint = "0904.1017",
    archivePrefix = "arXiv",
    primaryClass = "astro-ph.CO",
    doi = "10.1088/0004-637X/725/1/496",
    journal = "Astrophys. J.",
    volume = "725",
    pages = "496--514",
    year = "2010"
}

@article{Nitz:2020mga,
    author = "Nitz, Alexander H. and Capano, Collin D.",
    title = "{GW190521 may be an intermediate mass ratio inspiral}",
    eprint = "2010.12558",
    archivePrefix = "arXiv",
    primaryClass = "astro-ph.HE",
    doi = "10.3847/2041-8213/abccc5",
    journal = "Astrophys. J. Lett.",
    volume = "907",
    number = "1",
    pages = "L9",
    year = "2021"
}

@article{Okounkova:2022grv,
    author = "Okounkova, Maria and Isi, Maximiliano and Chatziioannou, Katerina and Farr, Will M.",
    title = "{Gravitational wave inference on a numerical-relativity simulation of a black hole merger beyond general relativity}",
    eprint = "2208.02805",
    archivePrefix = "arXiv",
    primaryClass = "gr-qc",
    doi = "10.1103/PhysRevD.107.024046",
    journal = "Phys. Rev. D",
    volume = "107",
    number = "2",
    pages = "024046",
    year = "2023"
}

@article{OShaughnessy:2012iol,
    author = "O'Shaughnessy, R. and London, L. and Healy, J. and Shoemaker, D.",
    title = "{Precession during merger: Strong polarization changes are observationally accessible features of strong-field gravity during binary black hole merger}",
    eprint = "1209.3712",
    archivePrefix = "arXiv",
    primaryClass = "gr-qc",
    reportNumber = "LIGO-DCC-P1200015",
    doi = "10.1103/PhysRevD.87.044038",
    journal = "Phys. Rev. D",
    volume = "87",
    number = "4",
    pages = "044038",
    year = "2013"
}

@article{OShaughnessy:2017eks,
    author = "O'Shaughnessy, Richard and Gerosa, Davide and Wysocki, Daniel",
    title = "{Inferences about supernova physics from gravitational-wave measurements: GW151226 spin misalignment as an indicator of strong black-hole natal kicks}",
    eprint = "1704.03879",
    archivePrefix = "arXiv",
    primaryClass = "astro-ph.HE",
    reportNumber = "LIGO-P1700066",
    doi = "10.1103/PhysRevLett.119.011101",
    journal = "Phys. Rev. Lett.",
    volume = "119",
    number = "1",
    pages = "011101",
    year = "2017"
}

@article{Pacilio:2024tdl,
    author = "Pacilio, Costantino and Bhagwat, Swetha and Nobili, Francesco and Gerosa, Davide",
    title = "{Flexible mapping of ringdown amplitudes for nonprecessing binary black holes}",
    eprint = "2408.05276",
    archivePrefix = "arXiv",
    primaryClass = "gr-qc",
    doi = "10.1103/PhysRevD.110.103037",
    journal = "Phys. Rev. D",
    volume = "110",
    number = "10",
    pages = "103037",
    year = "2024"
}

@article{Payne:2022spz,
    author = "Payne, Ethan and Hourihane, Sophie and Golomb, Jacob and Udall, Rhiannon and Udall, Richard and Davis, Derek and Chatziioannou, Katerina",
    title = "{Curious case of GW200129: Interplay between spin-precession inference and data-quality issues}",
    eprint = "2206.11932",
    archivePrefix = "arXiv",
    primaryClass = "gr-qc",
    reportNumber = "LIGO DCC: P2200185",
    doi = "10.1103/PhysRevD.106.104017",
    journal = "Phys. Rev. D",
    volume = "106",
    number = "10",
    pages = "104017",
    year = "2022"
}

@article{Payne:2024ywe,
    author = "Payne, Ethan and Kremer, Kyle and Zevin, Michael",
    title = "{Spin Doctors: How to Diagnose a Hierarchical Merger Origin}",
    eprint = "2402.15066",
    archivePrefix = "arXiv",
    primaryClass = "gr-qc",
    reportNumber = "LIGO DCC P2400050",
    doi = "10.3847/2041-8213/ad3e82",
    journal = "Astrophys. J. Lett.",
    volume = "966",
    number = "1",
    pages = "L16",
    year = "2024"
}

@ARTICLE{Pearson:1895,
       author = {{Pearson}, Karl},
        title = "{Note on Regression and Inheritance in the Case of Two Parents}",
      journal = {Proceedings of the Royal Society of London Series I},
         year = 1895,
        month = jan,
       volume = {58},
        pages = {240-242},
       adsurl = {https://ui.adsabs.harvard.edu/abs/1895RSPS...58..240P}
}

@article{Racine:2008qv,
    author = "Racine, Etienne",
    title = "{Analysis of spin precession in binary black hole systems including quadrupole-monopole interaction}",
    eprint = "0803.1820",
    archivePrefix = "arXiv",
    primaryClass = "gr-qc",
    doi = "10.1103/PhysRevD.78.044021",
    journal = "Phys. Rev. D",
    volume = "78",
    pages = "044021",
    year = "2008"
}

@article{Rodriguez:2016vmx,
    author = "Rodriguez, Carl L. and Zevin, Michael and Pankow, Chris and Kalogera, Vasilliki and Rasio, Frederic A.",
    title = "{Illuminating Black Hole Binary Formation Channels with Spins in Advanced LIGO}",
    eprint = "1609.05916",
    archivePrefix = "arXiv",
    primaryClass = "astro-ph.HE",
    doi = "10.3847/2041-8205/832/1/L2",
    journal = "Astrophys. J. Lett.",
    volume = "832",
    number = "1",
    pages = "L2",
    year = "2016"
}

@article{Rodriguez:2019huv,
    author = "Rodriguez, Carl L. and Zevin, Michael and Amaro-Seoane, Pau and Chatterjee, Sourav and Kremer, Kyle and Rasio, Frederic A. and Ye, Claire S.",
    title = "{Black holes: The next generation\textemdash{}repeated mergers in dense star clusters and their gravitational-wave properties}",
    eprint = "1906.10260",
    archivePrefix = "arXiv",
    primaryClass = "astro-ph.HE",
    doi = "10.1103/PhysRevD.100.043027",
    journal = "Phys. Rev. D",
    volume = "100",
    number = "4",
    pages = "043027",
    year = "2019"
}

@article{Romero-Shaw:2020owr,
    author = "Romero-Shaw, I. M. and others",
    title = "{Bayesian inference for compact binary coalescences with bilby: validation and application to the first LIGO\textendash{}Virgo gravitational-wave transient catalogue}",
    eprint = "2006.00714",
    archivePrefix = "arXiv",
    primaryClass = "astro-ph.IM",
    doi = "10.1093/mnras/staa2850",
    journal = "Mon. Not. Roy. Astron. Soc.",
    volume = "499",
    number = "3",
    pages = "3295--3319",
    year = "2020"
}

@article{Romero-Shaw:2020thy,
    author = "Romero-Shaw, Isobel M. and Lasky, Paul D. and Thrane, Eric and Bustillo, Juan Calderon",
    title = "{GW190521: orbital eccentricity and signatures of dynamical formation in a binary black hole merger signal}",
    eprint = "2009.04771",
    archivePrefix = "arXiv",
    primaryClass = "astro-ph.HE",
    doi = "10.3847/2041-8213/abbe26",
    journal = "Astrophys. J. Lett.",
    volume = "903",
    number = "1",
    pages = "L5",
    year = "2020"
}

@article{Romero-Shaw:2022fbf,
    author = "Romero-Shaw, Isobel M. and Gerosa, Davide and Loutrel, Nicholas",
    title = "{Eccentricity or spin precession? Distinguishing subdominant effects in gravitational-wave data}",
    eprint = "2211.07528",
    archivePrefix = "arXiv",
    primaryClass = "astro-ph.HE",
    doi = "10.1093/mnras/stad031",
    journal = "Mon. Not. Roy. Astron. Soc.",
    volume = "519",
    number = "4",
    pages = "5352--5357",
    year = "2023"
}

@article{Roulet:2022kot,
    author = "Roulet, Javier and Olsen, Seth and Mushkin, Jonathan and Islam, Tousif and Venumadhav, Tejaswi and Zackay, Barak and Zaldarriaga, Matias",
    title = "{Removing degeneracy and multimodality in gravitational wave source parameters}",
    eprint = "2207.03508",
    archivePrefix = "arXiv",
    primaryClass = "gr-qc",
    doi = "10.1103/PhysRevD.106.123015",
    journal = "Phys. Rev. D",
    volume = "106",
    number = "12",
    pages = "123015",
    year = "2022"
}

@article{Savitzky:1964,
	author = {Savitzky, Abraham. and Golay, M. J. E.},
	doi = {10.1021/ac60214a047},
	eprint = {https://doi.org/10.1021/ac60214a047},
	journal = {Analytical Chemistry},
	number = {8},
	pages = {1627-1639},
	title = {Smoothing and Differentiation of Data by Simplified Least Squares Procedures.},
	url = {https://doi.org/10.1021/ac60214a047},
	volume = {36},
	year = {1964}}

@article{Schmidt:2010it,
    author = "Schmidt, Patricia and Hannam, Mark and Husa, Sascha and Ajith, P.",
    title = "{Tracking the precession of compact binaries from their gravitational-wave signal}",
    eprint = "1012.2879",
    archivePrefix = "arXiv",
    primaryClass = "gr-qc",
    doi = "10.1103/PhysRevD.84.024046",
    journal = "Phys. Rev. D",
    volume = "84",
    pages = "024046",
    year = "2011"
}

@article{Schmidt:2012rh,
    author = "Schmidt, Patricia and Hannam, Mark and Husa, Sascha",
    title = "{Towards models of gravitational waveforms from generic binaries: A simple approximate mapping between precessing and non-precessing inspiral signals}",
    eprint = "1207.3088",
    archivePrefix = "arXiv",
    primaryClass = "gr-qc",
    doi = "10.1103/PhysRevD.86.104063",
    journal = "Phys. Rev. D",
    volume = "86",
    pages = "104063",
    year = "2012"
}

@article{Schmidt:2014iyl,
    author = "Schmidt, Patricia and Ohme, Frank and Hannam, Mark",
    title = "{Towards models of gravitational waveforms from generic binaries II: Modelling precession effects with a single effective precession parameter}",
    eprint = "1408.1810",
    archivePrefix = "arXiv",
    primaryClass = "gr-qc",
    doi = "10.1103/PhysRevD.91.024043",
    journal = "Phys. Rev. D",
    volume = "91",
    number = "2",
    pages = "024043",
    year = "2015"
}

@article{Shaik:2019dym,
    author = "Shaik, Feroz H. and Lange, Jacob and Field, Scott E. and O'Shaughnessy, Richard and Varma, Vijay and Kidder, Lawrence E. and Pfeiffer, Harald P. and Wysocki, Daniel",
    title = "{Impact of subdominant modes on the interpretation of gravitational-wave signals from heavy binary black hole systems}",
    eprint = "1911.02693",
    archivePrefix = "arXiv",
    primaryClass = "gr-qc",
    doi = "10.1103/PhysRevD.101.124054",
    journal = "Phys. Rev. D",
    volume = "101",
    number = "12",
    pages = "124054",
    year = "2020"
}

@article{Schutz:2011tw,
    author = "Schutz, Bernard F.",
    title = "{Networks of gravitational wave detectors and three figures of merit}",
    eprint = "1102.5421",
    archivePrefix = "arXiv",
    primaryClass = "astro-ph.IM",
    reportNumber = "AEI-2011-008",
    doi = "10.1088/0264-9381/28/12/125023",
    journal = "Class. Quant. Grav.",
    volume = "28",
    pages = "125023",
    year = "2011"
}

@article{Siegel:2023lxl,
    author = "Siegel, Harrison and Isi, Maximiliano and Farr, Will M.",
    title = "{Ringdown of GW190521: Hints of multiple quasinormal modes with a precessional interpretation}",
    eprint = "2307.11975",
    archivePrefix = "arXiv",
    primaryClass = "gr-qc",
    reportNumber = "LIGO-P2300214",
    doi = "10.1103/PhysRevD.108.064008",
    journal = "Phys. Rev. D",
    volume = "108",
    number = "6",
    pages = "064008",
    year = "2023"
}

@article{Steinle:2020xej,
    author = "Steinle, Nathan and Kesden, Michael",
    title = "{Pathways for producing binary black holes with large misaligned spins in the isolated formation channel}",
    eprint = "2010.00078",
    archivePrefix = "arXiv",
    primaryClass = "astro-ph.HE",
    doi = "10.1103/PhysRevD.103.063032",
    journal = "Phys. Rev. D",
    volume = "103",
    number = "6",
    pages = "063032",
    year = "2021"
}

@article{Stevenson:2017dlk,
    author = "Stevenson, Simon and Berry, Christopher P. L. and Mandel, Ilya",
    title = "{Hierarchical analysis of gravitational-wave measurements of binary black hole spin\textendash{}orbit misalignments}",
    eprint = "1703.06873",
    archivePrefix = "arXiv",
    primaryClass = "astro-ph.HE",
    doi = "10.1093/mnras/stx1764",
    journal = "Mon. Not. Roy. Astron. Soc.",
    volume = "471",
    number = "3",
    pages = "2801--2811",
    year = "2017"
}

@article{Talbot:2017yur,
    author = "Talbot, Colm and Thrane, Eric",
    title = "{Determining the population properties of spinning black holes}",
    eprint = "1704.08370",
    archivePrefix = "arXiv",
    primaryClass = "astro-ph.HE",
    doi = "10.1103/PhysRevD.96.023012",
    journal = "Phys. Rev. D",
    volume = "96",
    number = "2",
    pages = "023012",
    year = "2017"
}

@ARTICLE{Teukolsky:1973,
       author = {{Teukolsky}, Saul A.},
        title = "{Perturbations of a Rotating Black Hole. I. Fundamental Equations for Gravitational, Electromagnetic, and Neutrino-Field Perturbations}",
      journal = {\apj},
         year = 1973,
        month = oct,
       volume = {185},
        pages = {635-648},
          doi = {10.1086/152444},
       adsurl = {https://ui.adsabs.harvard.edu/abs/1973ApJ...185..635T}
}

@article{Thomas:2020uqj,
    author = "Thomas, Lucy M. and Schmidt, Patricia and Pratten, Geraint",
    title = "{New effective precession spin for modeling multimodal gravitational waveforms in the strong-field regime}",
    eprint = "2012.02209",
    archivePrefix = "arXiv",
    primaryClass = "gr-qc",
    reportNumber = "LIGO-DCC P2000509",
    doi = "10.1103/PhysRevD.103.083022",
    journal = "Phys. Rev. D",
    volume = "103",
    number = "8",
    pages = "083022",
    year = "2021"
}

@article{Udall:2024ovp,
    author = "Udall, Rhiannon and Hourihane, Sophie and Miller, Simona and Davis, Derek and Chatziioannou, Katerina and Isi, Max and Deshong, Howard",
    title = "{Antialigned spin of GW191109: Glitch mitigation and its implications}",
    eprint = "2409.03912",
    archivePrefix = "arXiv",
    primaryClass = "gr-qc",
    doi = "10.1103/PhysRevD.111.024046",
    journal = "Phys. Rev. D",
    volume = "111",
    number = "2",
    pages = "024046",
    year = "2025"
}

@article{Unser:1984,
  author  = "Unser, Michael",
  title   = "On the approximation of the discrete Karhunen-Loeve transform for stationary processes",
  journal = "Signal Processing",
  year    = 1984,
  volume  = "7",
  number  = "3",
  pages   = "231--249", 
  doi = "10.1016/0165-1684(84)90002-1"
}

@article{Usman:2015kfa,
    author = "Usman, Samantha A. and others",
    title = "{The PyCBC search for gravitational waves from compact binary coalescence}",
    eprint = "1508.02357",
    archivePrefix = "arXiv",
    primaryClass = "gr-qc",
    reportNumber = "LIGO-P1500086",
    doi = "10.1088/0264-9381/33/21/215004",
    journal = "Class. Quant. Grav.",
    volume = "33",
    number = "21",
    pages = "215004",
    year = "2016"
}

@article{Usman:2018imj,
    author = "Usman, Samantha A. and Mills, Joseph C. and Fairhurst, Stephen",
    title = "{Constraining the Inclinations of Binary Mergers from Gravitational-wave Observations}",
    eprint = "1809.10727",
    archivePrefix = "arXiv",
    primaryClass = "gr-qc",
    doi = "10.3847/1538-4357/ab0b3e",
    journal = "Astrophys. J.",
    volume = "877",
    number = "2",
    pages = "82",
    year = "2019"
}

@article{Varma:2014jxa,
    author = {Varma, Vijay and Ajith, Parameswaran and Husa, Sascha and Bustillo, Juan Calderon and Hannam, Mark and P\"urrer, Michael},
    title = "{Gravitational-wave observations of binary black holes: Effect of nonquadrupole modes}",
    eprint = "1409.2349",
    archivePrefix = "arXiv",
    primaryClass = "gr-qc",
    reportNumber = "LIGO-P1400095-V3",
    doi = "10.1103/PhysRevD.90.124004",
    journal = "Phys. Rev. D",
    volume = "90",
    number = "12",
    pages = "124004",
    year = "2014"
}

@article{Varma:2016dnf,
    author = "Varma, Vijay and Ajith, Parameswaran",
    title = "{Effects of nonquadrupole modes in the detection and parameter estimation of black hole binaries with nonprecessing spins}",
    eprint = "1612.05608",
    archivePrefix = "arXiv",
    primaryClass = "gr-qc",
    doi = "10.1103/PhysRevD.96.124024",
    journal = "Phys. Rev. D",
    volume = "96",
    number = "12",
    pages = "124024",
    year = "2017"
}

@article{Varma:2021csh,
    author = "Varma, Vijay and Isi, Maximiliano and Biscoveanu, Sylvia and Farr, Will M. and Vitale, Salvatore",
    title = "{Measuring binary black hole orbital-plane spin orientations}",
    eprint = "2107.09692",
    archivePrefix = "arXiv",
    primaryClass = "astro-ph.HE",
    doi = "10.1103/PhysRevD.105.024045",
    journal = "Phys. Rev. D",
    volume = "105",
    number = "2",
    pages = "024045",
    year = "2022"
}

@article{Veitch:2008,
  title = {Bayesian approach to the follow-up of candidate gravitational wave signals},
  author = {Veitch, J. and Vecchio, A.},
  journal = {Phys. Rev. D},
  volume = {78},
  issue = {2},
  pages = {022001},
  numpages = {10},
  year = {2008},
  month = {Jul},
  publisher = {American Physical Society},
  doi = {10.1103/PhysRevD.78.022001},
  url = {https://link.aps.org/doi/10.1103/PhysRevD.78.022001}
}

@article{Vitale:2014mka,
    author = "Vitale, Salvatore and Lynch, Ryan and Veitch, John and Raymond, Vivien and Sturani, Riccardo",
    title = "{Measuring the spin of black holes in binary systems using gravitational waves}",
    eprint = "1403.0129",
    archivePrefix = "arXiv",
    primaryClass = "gr-qc",
    doi = "10.1103/PhysRevLett.112.251101",
    journal = "Phys. Rev. Lett.",
    volume = "112",
    number = "25",
    pages = "251101",
    year = "2014"
}

@article{Vitale:2015tea,
    author = "Vitale, Salvatore and Lynch, Ryan and Sturani, Riccardo and Graff, Philip",
    title = "{Use of gravitational waves to probe the formation channels of compact binaries}",
    eprint = "1503.04307",
    archivePrefix = "arXiv",
    primaryClass = "gr-qc",
    reportNumber = "LIGO-DOCUMENT-P1500022, LIGO-P1500022",
    doi = "10.1088/1361-6382/aa552e",
    journal = "Class. Quant. Grav.",
    volume = "34",
    number = "3",
    pages = "03LT01",
    year = "2017"
}

@article{Vitale:2016avz,
    author = "Vitale, Salvatore and Lynch, Ryan and Raymond, Vivien and Sturani, Riccardo and Veitch, John and Graff, Philp",
    title = "{Parameter estimation for heavy binary-black holes with networks of second-generation gravitational-wave detectors}",
    eprint = "1611.01122",
    archivePrefix = "arXiv",
    primaryClass = "gr-qc",
    doi = "10.1103/PhysRevD.95.064053",
    journal = "Phys. Rev. D",
    volume = "95",
    number = "6",
    pages = "064053",
    year = "2017"
}

@book{Wigner:1959,
  author    = {Wigner, Eugene P.},
  title     = {Group Theory and its Application to the Quantum Mechanics of Atomic Spectra},
  year      = {1959},
  publisher = {Academic Press},
  note      = {Originally published in German as \textit{Gruppentheorie und ihre Anwendungen auf die Quantenmechanik der Atomspektren}, Braunschweig: Vieweg Verlag, 1931. English translation by J.~J.~Griffin. Reprinted by Elsevier in 2013. ISBN: 978-1-4832-7576-5},
  translator = {Griffin, J. J.}
}

@article{Wysocki:2017isg,
    author = "Wysocki, Daniel and Gerosa, Davide and O'Shaughnessy, Richard and Belczynski, Krzysztof and Gladysz, Wojciech and Berti, Emanuele and Kesden, Michael and Holz, Daniel E.",
    title = "{Explaining LIGO\textquoteright{}s observations via isolated binary evolution with natal kicks}",
    eprint = "1709.01943",
    archivePrefix = "arXiv",
    primaryClass = "astro-ph.HE",
    reportNumber = "LIGO-P1700174",
    doi = "10.1103/PhysRevD.97.043014",
    journal = "Phys. Rev. D",
    volume = "97",
    number = "4",
    pages = "043014",
    year = "2018"
}

@article{Xu:2022zza,
    author = "Xu, Yumeng and Hamilton, Eleanor",
    title = "{Measurability of precession and eccentricity for heavy binary-black-hole mergers}",
    eprint = "2211.09561",
    archivePrefix = "arXiv",
    primaryClass = "gr-qc",
    doi = "10.1103/PhysRevD.107.103049",
    journal = "Phys. Rev. D",
    volume = "107",
    number = "10",
    pages = "103049",
    year = "2023"
}

@article{Zevin:2020gbd,
    author = "Zevin, Michael and Bavera, Simone S. and Berry, Christopher P. L. and Kalogera, Vicky and Fragos, Tassos and Marchant, Pablo and Rodriguez, Carl L. and Antonini, Fabio and Holz, Daniel E. and Pankow, Chris",
    title = "{One Channel to Rule Them All? Constraining the Origins of Binary Black Holes Using Multiple Formation Pathways}",
    eprint = "2011.10057",
    archivePrefix = "arXiv",
    primaryClass = "astro-ph.HE",
    doi = "10.3847/1538-4357/abe40e",
    journal = "Astrophys. J.",
    volume = "910",
    number = "2",
    pages = "152",
    year = "2021"
}

@article{Zhang:2023fpp,
    author = "Zhang, Rachel C. and Fragione, Giacomo and Kimball, Chase and Kalogera, Vicky",
    title = "{On the Likely Dynamical Origin of GW191109 and Binary Black Hole Mergers with Negative Effective Spin}",
    eprint = "2302.07284",
    archivePrefix = "arXiv",
    primaryClass = "astro-ph.HE",
    doi = "10.3847/1538-4357/ace4c1",
    journal = "Astrophys. J.",
    volume = "954",
    number = "1",
    pages = "23",
    year = "2023"
}

@article{Zhu:2023fnf,
    author = "Zhu, Hengrui and others",
    title = "{Black hole spectroscopy for precessing binary black hole coalescences}",
    eprint = "2312.08588",
    archivePrefix = "arXiv",
    primaryClass = "gr-qc",
    doi = "10.1103/PhysRevD.111.064052",
    journal = "Phys. Rev. D",
    volume = "111",
    number = "6",
    pages = "064052",
    year = "2025"
}

@unpublished{Miller:2025,
  author       = {Miller, Simona and Isi, Maximiliano and Chatziioannou, Katerina and Hourihane, Sophie and Varma, Vijay},
  title        = {Misinterpreting the spins of heavy binary black holes: insights from time-domain morphology},
  note         = {In preparation},
  year         = {2026}
}

@article{Kang:2025nio,
    author = "Kang, Karen and Miller, Simona J. and Chatziioannou, Katerina and Ferguson, Deborah",
    title = "{Mapping parameter correlations in spinning binary black hole mergers}",
    eprint = "2502.17402",
    archivePrefix = "arXiv",
    primaryClass = "gr-qc",
    doi = "10.1103/4vbp-1jwl",
    journal = "Phys. Rev. D",
    volume = "112",
    number = "6",
    pages = "064020",
    year = "2025"
}

@article{Vallisneri:2007ev,
    author = "Vallisneri, Michele",
    title = "{Use and abuse of the Fisher information matrix in the assessment of gravitational-wave parameter-estimation prospects}",
    eprint = "gr-qc/0703086",
    archivePrefix = "arXiv",
    reportNumber = "LIGO-P070009-00-Z",
    doi = "10.1103/PhysRevD.77.042001",
    journal = "Phys. Rev. D",
    volume = "77",
    pages = "042001",
    year = "2008"
}

@misc{github_release,
  author       = {Simona Miller},
  title        = {\texttt{heavy-precessing-bbh-time-domain}},
  month        = may,
  year         = 2025,
  publisher    = {Github},
  url          = {https://github.com/simonajmiller/heavy-precessing-bbh-time-domain}
}

@misc{animation_figure01,
  author       = {Simona Miller},
  title        = {Animation of real data versus maximum likelihood GW190521 posteriors},
  month        = may,
  year         = 2025,
  publisher    = {Github},
  url          = {https://github.com/simonajmiller/heavy-precessing-bbh-time-domain/blob/main/gifs/gif_figure_01.gif}
}

@misc{animation_figure02,
  author       = {Simona Miller},
  title        = {Animation of spin-precession posteriors for different SNR signals},
  month        = may,
  year         = 2025,
  publisher    = {Github},
  url          = {https://github.com/simonajmiller/heavy-precessing-bbh-time-domain/blob/main/gifs/gif_figure_02.gif}
}

@misc{animation_figure04,
  author       = {Simona Miller},
  title        = {Animation of Animation of spin-precession posteriors for different total mass signals},
  month        = may,
  year         = 2025,
  publisher    = {Github},
  url          = {https://github.com/simonajmiller/heavy-precessing-bbh-time-domain/blob/main/gifs/gif_figure_04.gif}
}

@misc{zenodo_release,
  author       = {Simona Miller},
  title        = {Dataset for ``Measuring spin precession from massive black holes binaries with gravitational waves: insights from time-domain signal morphology"},
  month        = may,
  year         = 2025,
  publisher    = {Zenodo},
  url          = {https://zenodo.org/record/15474960},
  doi          = {10.5281/zenodo.15474960}
}

@misc{suppl_figs,
  author       = {Simona Miller},
  title        = {Supplemental Figures},
  month        = oct,
  year         = 2025,
  publisher    = {Github},
  url          = {https://github.com/simonajmiller/heavy-precessing-bbh-time-domain/blob/main/figures/supplemental_figures/}
}

\end{document}